\documentclass[11pt]{article}
\usepackage{harvard}
\usepackage{amssymb}
\usepackage{amsmath}
\usepackage{graphicx}
\usepackage{lscape}
\usepackage{pslatex}
\usepackage{times}

\newcommand{\be}{\begin{equation}}
\newcommand{\ee}{\end{equation}}
\newcommand{\bea}{\begin{eqnarray}}
\newcommand{\eea}{\end{eqnarray}}
\newcommand{\go}{\rightarrow}

\newcommand{\E}{{\rm E}}
\newcommand{\Var}{{\rm Var}}
\newcommand{\Cov}{{\rm Cov}}

\begin{document}

\title{Empirical Distributions of Log-Returns: \\
between the Stretched Exponential and the Power Law?}
\author{Y. Malevergne$^{1,2}$, V. Pisarenko$^3$, and
       D. Sornette$^{2,4}$\\
\\
$^1$ Institut de Science Financi\`ere et d'Assurances - Universit\'e Lyon I\\
43, Bd du 11 Novembre 1918, 69622 Villeurbanne Cedex, France\\
$^2$ Laboratoire de Physique de la Mati\`{e}re Condens\'{e}e,
CNRS UMR6622 and Universit\'{e} des Sciences\\
Parc Valrose, 06108 Nice Cedex 2, France \\
$^3$ International Institute of Earthquake Prediction Theory and
Mathematical Geophysics\\ Russian Ac. Sci. Warshavskoye sh., 79, kor. 2,
Moscow 113556, Russia\\
$^4$ Institute of Geophysics and
Planetary Physics and Department of Earth and Space Science\\
University of California, Los Angeles, California 90095\\
e-mails: Yannick.Malevergne@unice.fr, Vlad@sirus.mitp.ru  and
sornette@unice.fr
}

\date{}

\maketitle

\setcounter{page}{1}


\begin{abstract}

A large consensus now seems to take for granted that the distributions
of empirical returns of financial time series are regularly varying,
with a tail exponent $b$ close to $3$. First, we show by synthetic tests
performed on time series with time dependence in the volatility
with both Pareto and Stretched-Exponential distributions that for sample
of moderate size, the standard generalized extreme value (GEV)
estimator is quite
inefficient due to the possibly slow convergence toward the asymptotic
theoretical distribution and the existence of biases in presence of dependence
between data. Thus it cannot distinguish reliably between rapidly and regularly
varying classes of distributions. The Generalized Pareto distribution (GPD)
estimator works better, but still lacks power in the presence of
strong dependence.
Then, we use a parametric representation of the tail of the distributions
of returns of 100 years of daily return of the Dow Jones Industrial Average and
over 1 years of $5$-minutes returns of the Nasdaq Composite index,
encompassing both a regularly varying distribution in one limit of
the parameters and rapidly varying distributions of the class
of the Stretched-Exponential (SE) and Log-Weibull
distributions in other limits. Using the
method of nested hypothesis testing (Wilks' theorem), we conclude that
both the SE distributions and Pareto distributions provide reliable
descriptions of the data and cannot be distinguished for sufficiently high
thresholds. However, the exponent $b$ of the Pareto increases with the
quantiles and its growth does not seem exhausted for the highest quantiles
of three out of the four tail distributions investigated here. Correlatively,
the exponent $c$ of the SE model decreases and seems to tend to zero.
Based on the discovery that the SE distribution tends to the Pareto
distribution
in a certain limit such that the Pareto
(or power law) distribution can be approximated with any desired accuracy on
an arbitrary interval by a suitable adjustment of the
parameters of the SE distribution,
we demonstrate that Wilks' test of nested hypothesis
still works for the non-exactly nested comparison between the SE and
Pareto distributions.
The SE distribution is found significantly better over the whole quantile range
but becomes unnecessary beyond the $95\%$ quantiles compared with the
Pareto law.  Similar conclusions hold for the log-Weibull model with respect
to the Pareto distribution.
Summing up all the evidence
provided by our
battery of tests, it seems that the tails ultimately decay slower than any SE
but probably faster than power laws with reasonable exponents. Thus, from
a practical view point, the log-Weibull model, which provides a
smooth interpolation between
SE and PD, can be considered as an appropriate approximation of the
sample distributions. We finally discuss the implications of our results on the
``moment condition failure'' and for risk estimation and management.

\end{abstract}

\vspace{2cm}
{\bf key-words:} Extreme-Value Estimators, Non-Nested Hypothesis
Testing, Pareto distribution, Weibull distribution

\pagebreak
\section{Motivation of the study}

The determination of the precise shape of the tail of the distribution
of returns is a
major issue both from a practical and from an academic point of view.
For practitioners, it is crucial to accurately estimate the low
value quantiles of the distribution of
returns (profit and loss) because they are
involved in almost all the modern risk management methods. From an
academic perspective, many economic and financial theories rely on a specific
parameterization of the distributions whose parameters are intended to
represent the ``macroscopic'' variables the agents are sensitive to.

The distribution of returns is one of the most basic characteristics of the
markets and many papers have been devoted to it.
Contrarily to the average or expected return, for which economic theory
provides guidelines to assess them in relation with risk premium, firm size or
book-to-market equity (see for instance \citeasnoun{FF96}), the functional form
of the distribution of returns, and especially of extreme returns, is much
less constrained and still a topic of
active debate. Naively, the central limit theorem would lead
to a Gaussian distribution for sufficiently large time intervals over which
the return is estimated. Taking the continuous time limit
such that any finite time interval is seen as the sum of an infinite
number of increments thus leads to the paradigm of log-normal distributions
of prices and equivalently of Gaussian distributions of
returns, based on the pioneering work of \citeasnoun{Bachelier00}
later improved by \citeasnoun{Samuelson}. The log-normal paradigm has been
the starting point of many
financial theories such as \citeasnoun{Markovitz59}'s portfolio selection
method, \citeasnoun{Sharpe64}'s market equilibrium model or \citeasnoun{BS73}'s
rational option pricing theory. However, for real financial data,
the convergence in distribution to a
Gaussian law is very slow \cite[for instance]{Campbell,BP00}, much slower
than predicted for independent returns. As table \ref{table:Stat} shows,
the excess kurtosis (which is zero for a normal distribution) remains large
even for monthly returns,
testifying (i) of significant deviations from normality, (ii) of
the heavy tail behavior of the distributions of returns and (iii) of
significant dependences between returns \cite{Campbell}.

Another idea rooted in economic theory consists in invoking the
      ``Gibrat principle'' \cite{Simon}
initially used to account for the growth of cities and of wealth through
a mechanism combining stochastic multiplicative and additive noises
\cite{LSR96,SC97,Biham98,DS98}
leading to a Pareto distribution of sizes \cite{Champenowne,Gabaix}.
Rational bubble models a la \citeasnoun{Blanchardwat}
can also be cast in this mathematical framework of stochastic
recurrence equations and leads to distribution with power law tails,
albeit with a strong constraint on the tail exponent \cite{LuxSor,MalSor}.
These frameworks suggest that an alternative and natural way to capture
the heavy tail character of the distributions of returns is to use
distributions with power-like tails (Pareto, Generalized Pareto, stable laws)
or more generally, regularly-varying distributions \cite{BGT87}
\footnote{The general representation of a
regularly varying distribution is given by $\bar F(x) = {\cal L}(x) \cdot
x^{-\alpha}$, where ${\cal L}(\cdot)$ is a slowly varying function, that is,
lim$_{x \to \infty} {\cal L}(tx)/{\cal L}(x) = 1$ for any finite $t$.},
the later encompassing all the former.

In the early 1960s, \citeasnoun{Mandelbrot63} and \citeasnoun{Fama65}
presented evidence that distributions of returns can be well approximated by
a symmetric L\'evy stable law with tail index $b$ about $1.7$.
These estimates of the
power tail index have recently been confirmed by \citeasnoun{MRP98},
and slightly different indices of the stable law ($b =1.4$) were
suggested by \citename{MS95}~\citeyear{MS95,MS00}.
On the other hand,
there are numerous evidences of a larger value of the tail index $b \cong 3$
\cite{Longin96,GDDMOP97,GMAS98,GMAS98_2,Plerouetal99,MDP98,Farmer99,Lux00}. See also
the various alternative parameterizations in term of the Student distribution
\cite{Blattberg,kon}, or Pearson type-VII distributions \cite{Naga}, which
all have an asymptotic power law tail and are regularly varying.
Thus, a general
conclusion of this group of authors concerning tail fatness can be formulated
as follows: tails of the distribution of returns are heavier than
a Gaussian tail and heavier than an exponential tail;
they certainly admit the existence of a finite variance
($b>2$), whereas the existence of the third (skewness) and the fourth
(kurtosis) moments is questionable.

These apparent contradictory results actually do not apply to the
same quantiles of the distributions of returns.
Indeed, \citeasnoun{MS95} have shown that
the distribution of returns can be described accurately by a
L\'evy law only within a limited range of perhaps up to 4
standard deviations, while a faster decay
of the distribution is observed beyond. This almost-but-not-quite
L\'evy stable description explains (in part) the slow convergence
of the returns distribution to the Gaussian law under time aggregation
\cite{Sorbook}. And it
is precisely outside this range where the L\'evy law
applies that a tail index of about three have been
estimated. This can be seen from the fact that most authors who have
reported a tail index $b \cong 3$ have
used some optimality criteria for choosing the sample fractions
(i.e., the largest values) for
the estimation of the tail index. Thus, unlike the authors
supporting stable laws, they have used only a fraction of the
largest (positive tail) and smallest (negative tail) sample values.

It would thus seem that all has been said on the distributions of returns.
However, there are dissenting views in the literature. Indeed,
the class of regularly varying distributions is not the sole one
able to account for the large kurtosis and fat-tailness
of the distributions of returns.
Some recent works suggest alternative descriptions for the
distributions of returns. For instance, \citeasnoun{GJ98} claim
that the distribution of returns on the French stock market decays faster than
any power law. \citeasnoun{contdub} have proposed to use
exponentially truncated stable
distributions, \citeasnoun{Barndorff}, \citeasnoun{Eberlein} and
\citeasnoun{Prause} have respectively considered normal inverse Gaussian and
(generalized) hyperbolic distributions, which asymptotically decay as
$x^{\alpha} \cdot \exp(- \beta x)$, while \citeasnoun{LS99} suggest to fit the
distributions of stock returns by the Stretched-Exponential (SE) law. These
results, challenging the traditional hypothesis of power-like tail, offer a new
representation of the returns distributions and need to be tested rigorously on
a statistical ground.

\textit{A priori}, one could assert that \citeasnoun{Longin96}'s results
should rule out the exponential and
Stretched-Exponential hypotheses. Indeed, his results,
based on extreme value theory, show that the distributions of log-returns
belong to the maximum domain of attraction of the
Fr\'{e}chet distribution, so that they are necessarily regularly
varying power-like laws.
However, his study, like almost all others on this subject, has been
performed under the assumption that (1) financial time series are made
of independent and identically distributed returns and (2) the corresponding
distributions of returns belong to one of only three possible
maximum domains of attraction. However, these assumptions are not
fulfilled in general. While
\citeasnoun{Smith85}'s results indicate that the dependence of the data does
not constitute a major
problem in the limit of large samples, we shall see that it
can significantly bias standard statistical methods for samples of size
commonly used in extreme tails studies.
      Moreover, Longin's conclusions are essentially based on an
\textit{aggregation} procedure which stresses the central part of the
distribution while smoothing the characteristics of the tail, which are
essential in characterizing the tail behavior.

In addition, real financial time series exhibit GARCH effects
\cite{Bollerslev86,BEN94} leading to heteroscedasticity and
to clustering of high threshold
exceedances due to a long memory of the volatility. These rather
complex dependent structures
make difficult if not questionable
the blind application of standard statistical tools for data analysis.
In particular, the existence of significant dependence in the return
volatility leads to the existence of a significant bias and an
increase of the true standard deviation of the
statistical estimators of tail indices. Indeed, there are now many examples
showing that dependences and long memories as well as nonlinearities mislead
standard statistical tests \cite[for instance]{AEL99,GT99}. Consider the Hill's
and Pickand's estimators, which play an important role in the study 
of the tails
of distributions. It is often overlooked that, for dependent time series,
Hill's estimator remains only consistent but not asymptotically efficient
\cite{RLdH98}.
Moreover, for financial time series with a
dependence structure described by a IGARCH process,
\citeasnoun{KP97} have shown that the standard deviation of Hill's
estimator obtained by a bootstrap method can be seven to eight time larger than
the standard deviation derived under the asymptotic normality assumption. These
figures are even worse for Pickand's estimator.

The question then arises whether
the many results and seemingly almost consensus
obtained by ignoring the limitations of
usual statistical tools could have led to erroneous conclusions about the tail
behavior of the distributions of returns. Here, we propose
to investigate once more this delicate problem of the tail behavior of
distributions of returns in order to shed new lights
\footnote{\citeasnoun{Picoli} have also presented fits comparing the relative
merits of SE and so-called $q$-exponentials (which are similar to Student
distribution with power law tails) for the description of the frequency
distributions of basketball baskets, cyclone
victims, brand-name drugs by retail sales, and highway length.}.
To this aim, we investigate
two time series: the daily
returns of the Dow Jones Industrial Average (DJ) Index over a
century (kindly provided by
Prof. H.-C. G. Bothmer) and the five-minutes returns of the Nasdaq Composite index (ND)
over one year from April 1997 to May 1998 obtained from Bloomberg.
These two sets of data have been chosen since they are typical
of the data sets used in most previous studies. Their size (about
$20,000$ data points),
while significant compared with those used in investment and
portfolio analysis,
is however much smaller than recent data-intensive studies using ten
of millions
of data points \cite{GMAS98,GMAS98_2,Plerouetal99,matia,mizumo}.

First, we show by synthetic tests
performed on time series with time dependence in the volatility
with both Pareto and Stretched-Exponential distributions that for sample
of moderate size, the standard generalized extreme value (GEV)
estimator is quite
inefficient due to the possibly slow convergence toward the asymptotic
theoretical distribution and the existence of biases in presence of dependence
between data. Thus it cannot distinguish reliably between rapidly and regularly
varying classes of distributions. The Generalized Pareto distribution (GPD)
estimator works better, but still lacks power in the presence of
strong dependence.
Then, we use a parametric representation of the tail of the distributions
of returns of our two time series,
encompassing both a regularly varying distribution in one limit of
the parameters and rapidly varying distributions of the class
of the Stretched-Exponential (SE) and Log-Weibull
distributions in other limits. 

Using the
method of nested hypothesis testing (Wilks' theorem),
our second conclusion is that none of the
standard parametric family distributions (Pareto, exponential,
stretched-exponential,
incomplete Gamma and Log-Weibull) fits satisfactorily the DJ and ND
data on the whole range of
either positive or negative returns. While this is also true for
the family of stretched exponential and the log-Weibull
distributions, these families
appear to be the
best among the five considered parametric families, in so far as they are able
to fit the data over the largest interval. For the high quantiles (far in the
tails), both the SE distributions and Pareto distributions provide reliable
descriptions of the data and cannot be distinguished for sufficiently high
thresholds. However, the exponent $b$ of the Pareto increases with the
quantiles and its growth does not seem exhausted for the highest quantiles
of three out of the four tail distributions investigated here. Correlatively,
the exponent $c$ of the SE model decreases and seems to tend to zero

Based on the discovery presented here that the SE distribution tends to the Pareto
distribution in a certain limit such that the Pareto
(or power law) distribution can be approximated with any desired accuracy on
an arbitrary interval by a suitable adjustment of the
parameters of the SE distribution,
we demonstrate that Wilks' test of nested hypothesis
still works for the non-exactly nested comparison between the SE and
Pareto distributions.
The SE distribution is found significantly better over the whole quantile range
but becomes unnecessary beyond the $95\%$ quantiles compared with the
Pareto law. The log-Weibull model 
seems to be a good candidate since it provides
a smooth interpolation between the SE and PD models. 
The log-Weibull distribution is at least as good as the
Stretched-Exponential model, on a large range of data, but again, the Pareto
distribution is ultimately the most parsimonious.

Collectively, these results suggest that
the extreme tails of the true distribution of returns of our two data sets 
are fatter that any
stretched-exponential, strictly speaking -i.e., with a strickly
positive fractional exponent- but thinner than any power law. Thus,
notwithstanding our best efforts,
we cannot conclude on the exact nature of the far-tail of
distributions of returns.

As already mentioned, other works have proposed the so-called
inverse-cubic law ($b=3$)
based on the analysis of distributions of returns of high-frequency
data aggregated
over hundreds up to thousands of stocks. This aggregating procedure
leads to novel
problems of interpretation. We think that the relevant question for
most practical
applications is not to determine what is the true asymptotic tail but what is
the best effective description of the tails in the domain of useful
applications.
As we shall show below, it may be that the extreme
asymptotic tail is a regularly varying function
with tail index $b=3$ for daily returns, but this is not very useful if
this tail describes events whose recurrence time is a century or more.
Our present work must thus be gauged as an attempt to provide a simple
efficient effective description of the tails of distribution of returns
covering most of the range of interest for practical applications. We feel
that the efforts requested to go deeper in the tails
beyond the tails analyzed here, while of
great interest from a scientific point of view to potentially help unravel
market mechanisms, may be too artificial and unreachable to have significant
applications.

The paper is organized as follows.

The next section is devoted to the presentation of our two data sets and to
some of their basic statistical properties, emphasizing their fat tailed
behavior. We discuss, in particular, the importance of the so-called
``lunch effect'' for the tail properties of intra-day returns. We then
obtain the well-known presence of a significant temporal dependence
structure and
study the possible non-stationary character of these time series.

Section 3 attempts to account for the temporal dependence of our time
series and investigates its effect on the determination of the extreme behavior
of the tails of the distribution of returns. In this goal, we build a simple
long memory stochastic volatility process whose stationary
distributions are by construction either asymptotically regularly varying or
exponential. We show that, due to the time dependence on the volatility,
the estimation with standard statistical estimators may become unreliable due
to the significant bias and increase of the standard deviation of these
estimators. These results justify our re-examination of previous claims of
regularly varying
tails.

To fit our two data sets, section 4 proposes two general parametric
representations of the distribution of returns encompassing both a regularly
varying distribution in one limit of the parameters and rapidly varying
distributions of the class of stretched exponential and log-Weibull
distributions
in another limit.
The use of regularly varying distributions have been justified above. From a
theoretical view point, the class of stretched exponentials is
motivated in part
by the fact that the large deviations of multiplicative processes are
generically distributed with stretched exponential distributions
\cite{FrischSor}. Stretched exponential distributions are also parsimonious
examples of the important subset of sub-exponentials, that is, of the general
class of distributions decaying slower than an exponential.
This class of sub-exponentials share several important properties of
heavy-tailed distributions \cite{EKM97}, not shared by exponentials or
distributions decreasing faster than exponentials.
The interest of the log-Weibull comes from the smooth interpolation it
provides between any Stretched-Exponential and any Pareto distributions.

The descriptive power of these different  hypotheses are
compared in section 5. We first consider nested hypotheses and use
Wilks' test to compare each distribution
(Pareto, Exponential, Gamma and Stretched Exponential) with
the most general parameterization which encompasses all of them.
It appears that both the stretched-exponential and the
Pareto distributions are the best and most parsimonious models
compatible with the data
with a slight advantage in favor of the stretched exponential model.
Then, in order to directly compare the descriptive power of these two
models, we use the important remark that, in a certain limit where the
exponent $c$ of the stretched exponential pdf goes to zero,
the stretched exponential pdf tends to the Pareto distribution. Thus,
the Pareto
(or power law) distribution can be approximated with any desired accuracy on
an arbitrary interval by a suitable adjustment of the parameters of the
stretched exponential pdf. This allows us to demonstrate in Appendix
\ref{app:NN} that Wilks' test also applies to this non-exactly nested
comparison
between the SE and Pareto models. We find that
the SE distribution is significantly better over the whole quantile range
but becomes unnecessary beyond the $95\%$ quantiles compared with the
Pareto law.
Similar results are found for the comparison of the Log-Weibull versus the
Pareto distributions.

Section 6 summarizes our results and presents the conclusions of our
study for risk management purposes.

\section{Some basic statistical features}

\subsection{The data}

We use two sets of data. The first sample consists in the daily
returns\footnote{Throughout the paper, we will use compound returns, i.e.,
log-returns.} of the Dow Jones Industrial Average Index (DJ) over the time
interval from May 27, 1896 to May 31, 2000, which represents a sample size
$n=28415$.  The second data set contains the high-frequency (5
minutes) returns of
Nasdaq Composite (ND) index for the period from April 8, 1997 to May 29, 1998
which represents n=22123 data points.  The choice of these two data sets is
justified by their similarity with (1) the data set of daily returns used
by \citeasnoun{Longin96} particularly and (2) the high frequency data used by
\citeasnoun{GDDMOP97}, \citeasnoun{Lux00}, \citeasnoun{MDP98} among others.

\begin{figure}
\begin{center}
\includegraphics[width=15cm]{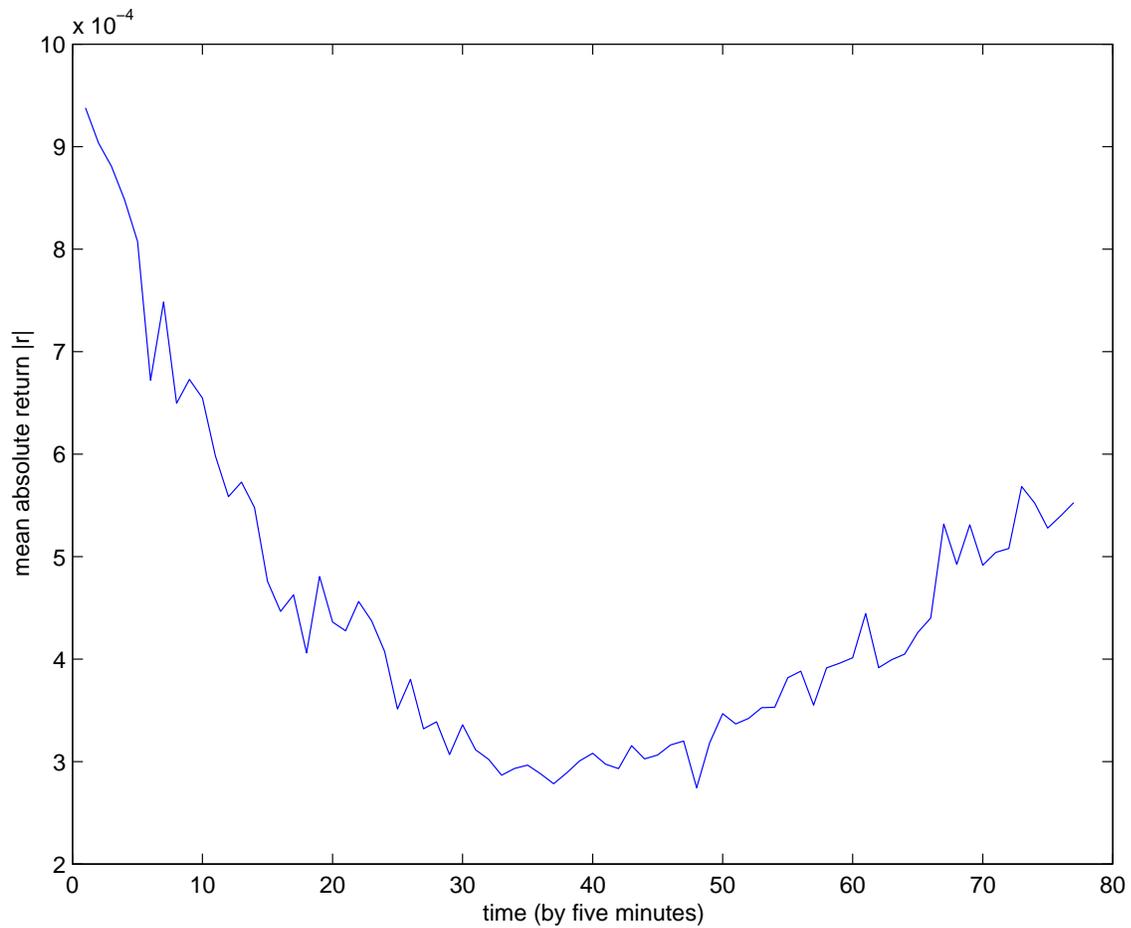}
\end{center}
\caption{\label{fig:Ushape} Average absolute return, as a function of time
within a trading day. The U-shape characterizes the so-called lunch effect.}
\end{figure}

For the intra-day Nasdaq data, there are two caveats that must be addressed.
First, in order to remove the effect of overnight price jumps, we have
determined the returns separately for each of 289 days contained in
the Nasdaq data
and have taken the union of all these 289 return data sets to obtain
a global return data set. Second, the volatility of intra-day data are
known to exhibit a U-shape, also called ``lunch-effect'', that is, an
abnormally
high volatility at the begining and the end of the trading day compared
with a low volatility at the approximate time of lunch. Such effect is
present in our data, as depicted on figure~\ref{fig:Ushape}, where the average
absolute returns are shown as a function of the time within a trading day. It
is desirable to correct the data from this systematic effect. This
has been performed
by renormalizing the 5 minutes-returns at a given moment of the
trading day by the
corresponding average absolute return at the same moment. We shall
refer to this
time series as the corrected Nasdaq returns in contrast with the raw
(incorrect)
Nasdaq returns and we shall examine both data sets for comparison.

The Dow Jones daily returns also exhibit some non-stationarity. Indeed, one
can observe a clear excess volatility roughly covering the time of
the bubble ending in the October 1929 crash following by the Great
Depression. To investigate the influence of such non-stationarity, time
interval, the statistical study exposed below has been performed twice: first
with the entire sample, and after having removed the
period from 1927 to 1936 from the sample.
The results are somewhat different, but on the whole,
the conclusions about the nature of the tail are the same. Thus, only the
results concerning the whole sample will be detailed in the paper.

Although the distributions of positive and negative returns are known
to be very
similar \cite[for instance]{JR01}, we have chosen to treat them separately.
For the Dow Jones, this gives us
$14949$ positive and $13464$ negative data points while, for the
Nasdaq, we have
$11241$ positive and $10751$ negative data points.

Table~\ref{table:Stat} summarizes the main statistical properties of these two
time series (both for the raw and for the corrected Nasdaq returns)
in terms of the average
returns, their standard deviations, the skewness and the excess kurtosis
for four time scales of five minutes, an hour, one day and one month.
The Dow Jones exhibits a
significantly negative skewness, which can be ascribed to the impact of the
market crashes. The raw Nasdaq returns are significantly positively skewed
while the returns corrected for the ``lunch effect'' are negatively skewed,
showing that the lunch effect plays an important role in the
shaping of the distribution of the intra-day
returns. Note also the important
decrease of the kurtosis after correction of the Nasdaq returns for
lunch effect,
confirming the strong impact of the lunch effect.
In all cases, the excess-kurtosis are high and remains significant
even after a time aggregation of one month.
The Jarque-Bera's test \cite{Cromwell}, a joint statistic using
skewness and kurtosis coefficients, is used to
reject the normality assumption for these time series.

\subsection{Existence of time dependence}

\begin{figure}
\begin{center}
\includegraphics[width=12cm]{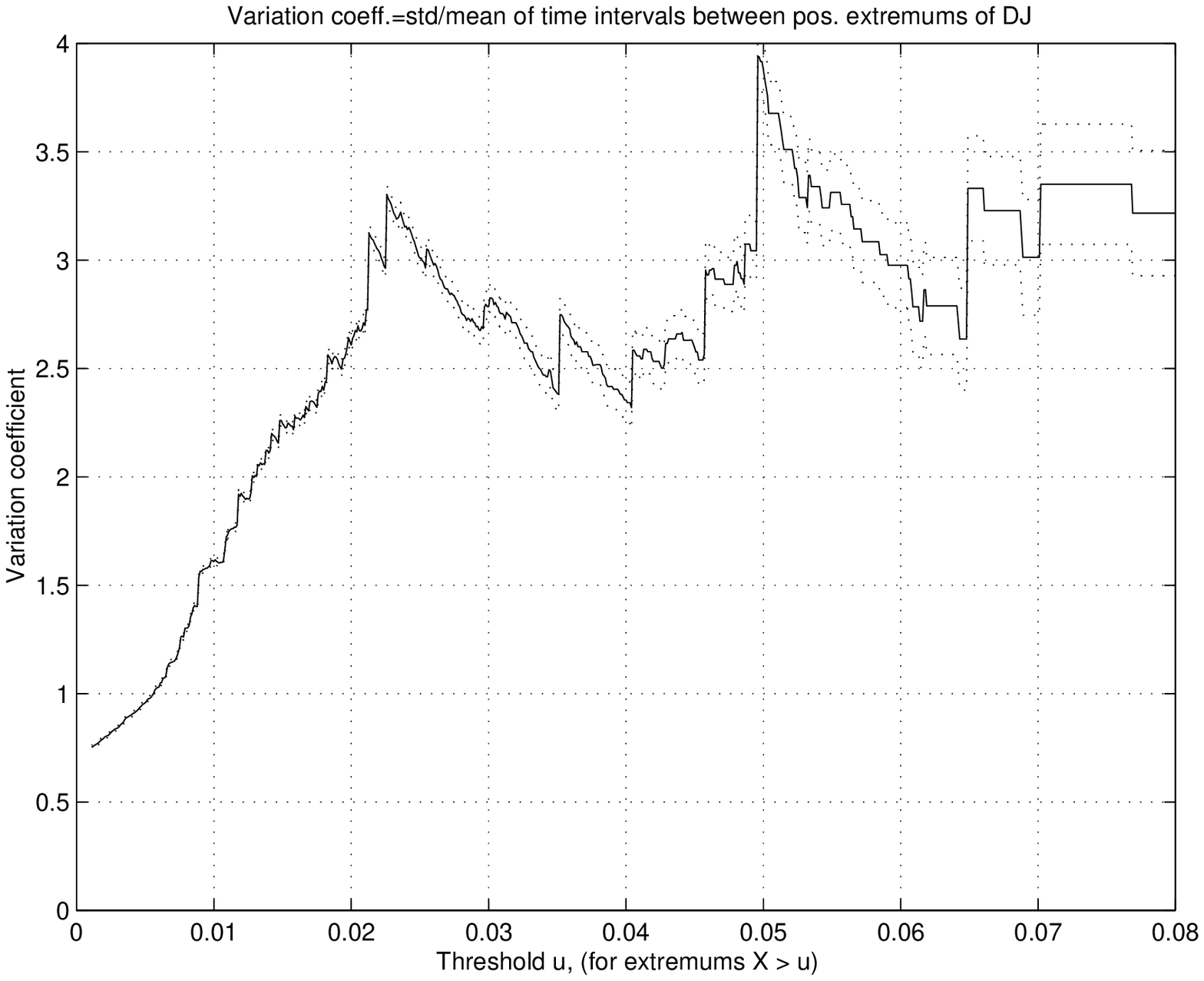}
\includegraphics[width=12cm]{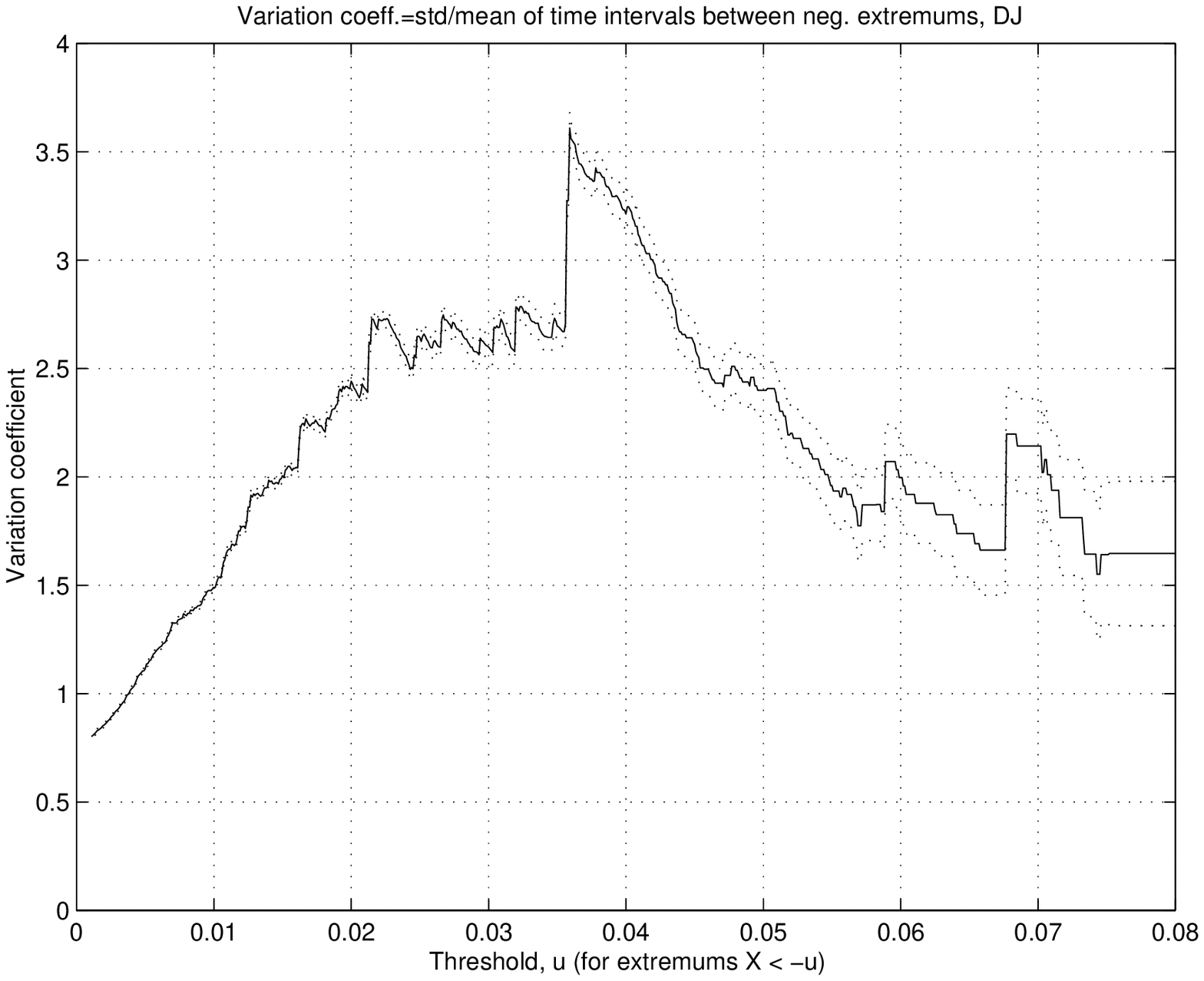}
\end{center}
\caption{\label{fig105-106} Coefficient of variation $V$ for the Dow
Jones daily
returns. An increase of $V$ characterizes the increase of ``clustering''.}
\end{figure}

It is well-known that financial time series exhibit complex dependence
structures like heteroscedasticity or non-linearities. These properties are
clearly observed in our two times series. For instance, we have estimated the
statistical characteristic $V$ (for positive random variables) called
coefficient
of variation
\be
V= \frac{{\rm Std}(X)}{{\rm E}(X)}~,
\ee
which is often used as a testing statistic of the
randomness property of a time series. It can be applied to a
sequence of points (or, intervals generated by these points on the line). If
these points are ``absolutely random,'' that is, generated by a
Poissonian flow,
then the intervals between them are distributed according to an exponential
distribution for which $V=1$. If $V<<1$, the process is
close to a periodic oscillation. Values $V>>1$ are associated with a clustering
phenomenon. We estimated $V=V(u)$ for extrema $X>u$ and $X<-u$ as function of
threshold $u$ (both for positive and for negative extrema). The results are
shown in figure~\ref{fig105-106} for the Dow Jones daily returns. As
the results are
essentially the same for the Nasdaq, we do not show them.
Figure~\ref{fig105-106} shows that, in the main
range $\vert X\vert <0.02$, containing $\sim95{\%}$ of the sample,
$V$ increases
with $u$, indicating that the ``clustering'' property
becomes stronger as the threshold $u$ increases.
The coefficient of variation has also been estimated for the Dow Jones when
the time interval from 1927 to 1936 is removed. Its maximum value decreases by
one, but it still significantly increases with the threshold $u$.

We have then applied several formal statistical tests of independence.
We have first performed
the Lagrange multiplier test proposed by \citeasnoun{Engle84} which leads to
the $T \cdot R^2$ test statistic, where $T$ denotes the sample size and $R^2$
is the determination coefficient of the regression of the squared centered
returns $x_t$ on a constant and on $q$ of their lags $x_{t-1}, x_{t-2}, \cdots,
x_{t-q}$. Under the null hypothesis of homoscedastic time series, $T
\cdot R^2$
follows a $\chi^2$-statistic with $q$ degrees of freedom. The test has been
performed up to $q=10$ and, in every case, the null hypothesis is
strongly rejected, at
any usual significance level. Thus, the time series are heteroskedastic and
exhibit volatility clustering. We have also
performed a BDS test \cite{BDS87} which allows us to detect not only volatility
clustering, like in the previous test, but also departure from iid-ness due to
non-linearities. Again, we strongly rejects the null-hypothesis of iid data, at
any usual significance level, confirming the Lagrange multiplier test.

\section{Can long memory processes lead to misleading measures of
extreme properties?}

Since the descriptive statistics given in the previous section have clearly
shown the existence of a significant temporal dependence structure, it is
important to consider the possibility that it can lead to erroneous
conclusions on the estimated parameters as previously shown by
\citeasnoun{KP97} for integrated GARCH processes. We first briefly recall the
standard procedures used to investigate extremal properties, stressing the
problems and drawbacks arising from the existence of temporal dependence. We
then perform a numerical simulation to study the behavior of the estimators in
presence of dependence. We put particular emphasis on the possible appearance
of significant biases due to dependence in the data set. Finally,
we present the results on the
extremal properties of our two DJ and ND data sets in the light of the
bootstrap results.

\subsection{Some theoretical results}

Two limit theorems allow one to
study the extremal properties and to determine the maximum domain of
attraction (MDA) of a distribution function in two forms.

First, consider a sample of $N$ iid realizations $X_1, X_2, \cdots, X_N$. Let
$X^{\wedge} $ denotes the maximum of this sample.
Then, the Gnedenko theorem states that, if, after an
adequate centering and normalization, the distribution of $X^\wedge$ converges
to a {\it non-degenerate} distribution as $N$ goes to infinity, this limit
distribution is then necessarily the Generalized
Extreme Value (GEV) distribution defined by
\be
H_\xi(x) = \exp \left[- (1+ \xi \cdot x)^{-1/\xi} \right]~.
\label{hjgjhsd}
\ee
When $\xi=0$, $H_\xi(x)$ should be understood as
\be
H_{\xi=0}(x) = \exp[- \exp(-x)].
\ee
Thus, for $N$ large enough
\be
\label{eq:Max}
\Pr \left\{X^\wedge < x \right\} = H_\xi \left( \frac{x-\mu}{\psi}\right),
\ee
for some value of the centering parameter $\mu$, scale factor $\psi$ and
tail index $\xi$. It should be noted that the existence of
non-degenerate limit distribution
of properly centered and normalized $X^\wedge$ is a rather strong limitation.
There are a lot of distribution functions that do not satisfy this limitation,
e.g., infinitely alternating functions between a power-like and an
exponential behavior.

The second limit theorem is called after Gnedenko-Pickands-Balkema-de
Haan (GPBH)
and its formulation is as follows.
In order to state the GPBH theorem, we define the right endpoint
$x_F$ of a distribution function $F(x)$ as $x_F  = {\rm sup}\{x: F(x) < 1 \}$.
Let us call the function
\be
\Pr \{ X - u  \ge x ~|~ X > u \} \equiv  {\bar F}_u(x)
\ee
the excess distribution function (DF).
Then, this DF ${\bar F}_u(x)$ belongs to the Maximum Domain of
Attraction of $H_\xi(x)$
defined by eq.(\ref{hjgjhsd}) if and only if there exists a positive
scale-function $s(u)$,
depending on the threshold $u$, such that
\be
\lim_{u \to x_F} ~~\sup_{0 \leq x \leq x_F-u}
|{\bar F}_u(x) - {\bar G}(x~|~\xi, s(u))| = 0~,
\ee
where
\be
G(x~|~\xi, s)  =  1 + \ln H_{\xi} \left( \frac{x}{s} \right)  = 1- \left(1+ \xi
\cdot \frac{x}{s} \right)^{-1/\xi}~. \label{eq:GPD}
\ee
By taking the limit
$\xi \to 0$, expression (\ref{eq:GPD}) leads to the exponential
distribution. The support of the distribution function (\ref{eq:GPD}) is
defined as follows:
\be
\begin{cases}
0 \leqslant x < \infty, & \text{ if $\xi \geqslant 0$}\\
0 \leqslant x \leqslant -d /\xi,& \text{ if $\xi < 0$}.
\end{cases}
\ee
Thus, the Generalized Pareto Distribution has a finite support for $\xi < 0$.

The form parameter $\xi$ is of paramount importance for the form of
the limiting distribution.
Its sign determines three possible limiting forms of the distribution
of maxima:
If $\xi>0$, then the limit distribution is the Fr\'echet power-like
distribution; If $\xi=0$, then the limit distribution is the Gumbel
(double-exponential) distribution; If $\xi<0$, then the limit
distribution has a
support bounded from above. All these three distributions are united in
eq.(\ref{hjgjhsd}) by this parameterization. The determination of the
parameter $\xi$ is the central problem of extreme value analysis. Indeed, it
allows one to determine the maximum domain of attraction of the underling
distribution. When $\xi>0$, the underlying distribution belongs to
the Fr\'echet
maximum domain of attraction and is regularly varying (power-like tail). When
$\xi=0$, it belongs to the Gumbel Maximum Domain of
Attraction and is rapidly varying
(exponential tail),
while if $\xi<0$ it belongs to the Weibull Maximum Domain of
Attraction and has a finite right endpoint.

\subsection{Examples of slow convergence to limit GEV and GPD distributions}

There exist two ways of estimating $\xi$. First, if there is a sample of maxima
(taken from sub-samples of sufficiently large size), then one can fit to this
sample the GEV distribution, thus estimating the parameters by
Maximum Likelihood method.
Alternatively, one can prefer the distribution of
exceedances over a large threshold given by the GPD~(\ref{eq:GPD}), whose tail
index can be estimated with Pickands' estimator or by Maximum Likelihood, as
previously. Hill's estimator cannot be used since it assumes $\xi>0$, while
the essence of extreme value analysis is, as we said, to test for
the class of limit distributions without excluding any possibility, and
not only to determine the quantitative value of an exponent.
Each of these methods has its advantages and drawbacks, especially
when one has to
study dependent data, as we show below.

Given a sample of size $N$, one considers the  $q$-maxima drawn from
$q$ sub-samples of size $p$ (such that $p \cdot q =N$) to estimate the
parameters $(\mu, \psi, \xi)$ in (\ref{eq:Max}) by Maximum Likelihood. This
procedure yields consistent and asymptotically Gaussian estimators, provided
that $\xi >-1/2$ \cite{Smith85}.
The properties of the estimators still hold approximately for dependent data,
provided that the interdependence of data is weak. However, it is difficult
to choose an optimal value of $q$ of the sub-samples. It depends both on
the size $N$ of the entire sample and on the
underlying distribution: the maxima drawn from an Exponential distribution
are known to converge very quickly to Gumbel's distribution \cite{HW79}, while
for the Gaussian law, convergence is particularly slow \cite{Hall79}.

The second possibility is to estimate the parameter $\xi$ from the distribution
of exceedances (the GPD). For this, one can use either the Maximum Likelihood
estimator or Pickands' estimator. Maximum Likelihood estimators are well-known
to be the most efficient ones (at least for $\xi > -1/2$ and
for independent data) but, in this
particular case, Pickands' estimator works reasonably well. Given an ordered
sample $x_1 \le x_2 \le \cdots x_N$ of size $N$,
Pickands' estimator is given by
\be
\label{eq:Pickands}
\hat \xi_{k,N} = \frac{1}{\ln 2} \ln \frac{x_k-x_{2k}}{x_{2k}-x_{4k}}~.
\ee
For independent and identically distributed data, this estimator is consistent
provided that $k$ is chosen so that $k \go \infty$ and $k/N \go 0$ as $N \go
\infty$. Moreover, $\hat \xi_{k,N}$ is asymptotically normal with variance
\be
\label{eq:PicSTD}
\sigma(\hat \xi_{k,N})^2 \cdot k= \frac{\xi^2
(2^{2\xi+1}+1)}{(2(2^\xi-1) \ln 2)^2}~.
\ee
In the presence of dependence between data, one can expect an increase
of the standard deviation, as reported by \citeasnoun{KP97}.
For time dependence of the GARCH class, \citeasnoun{KP97} have indeed
demonstrated
a significant increase of the standard deviation of the
tail index estimator, such as Hill's estimator, by a factor more than seven
with respect to their asymptotic properties for iid samples. This leads to
very inaccurate index estimates for time series with this kind of temporal
dependence.

Another
problem lies in the determination of the optimal threshold $u$ of the GPD,
which is in fact related to the optimal determination of the sub-samples
size $q$ in the case of the estimation of the parameters of the distribution of
maximum.

In sum, none of these methods seem really satisfying and each one
presents severe
drawbacks. The estimation of the parameters of the GEV distribution
and of the GPD may be less
sensitive to the dependence of the data, but this property is only asymptotic,
thus a bootstrap investigation is required to be able to compare the real
power of each estimation method for samples of moderate size.

As a first simple example illustrating the possibly very
slow convergence to the limit distributions of extreme value theory mentioned
above, let us consider a simulated sample of iid Weibull random variables (we
thus fulfill the most basic assumption of extreme values theory, i.e,
iid-ness). We take two values for the exponent
of the Weibull distribution: $c=0.7$ and $c=0.3$, with $d=1$ (scale parameter).
An estimation of $\xi$ by the distribution of the GPD of exceedance
should give estimated values of $\xi$ close to zero
in the limit of large $N$. In order to use the GPD,
we have taken the conditional Weibull distribution under condition $X > U_k,
k=1...15$, where the thresholds $U_k$ are chosen as: $U_1=0.1;~  U_2=0.3;~
U_3=1;~ U_4=3;~ U_5=10;~ U_6=30;~ U_7=100;~  U_8=300;~ U_9=1000;~
U_{10}=3000;~ U_{11}=10^4;~ U_{12}=3 \cdot 10^4;~ U_{13}=10^5;~ U_{14}=3 \cdot
10^5$ and $U_{15}=10^6$.

\begin{figure}
\begin{center}
\includegraphics[width=15cm]{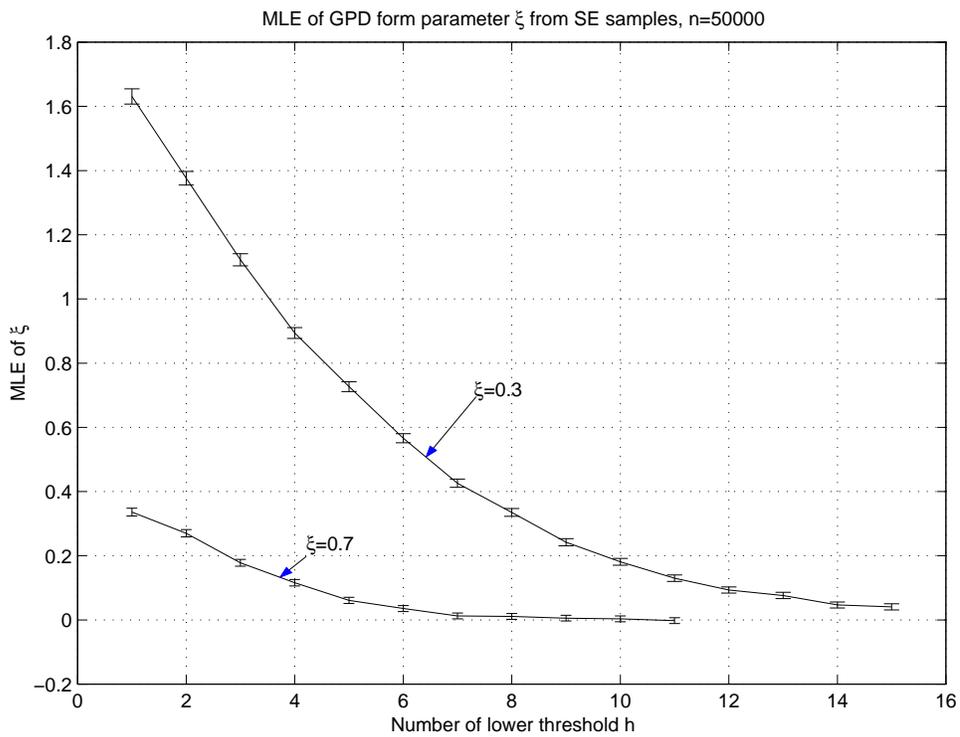}
\end{center}
\caption{\label{fig110} Maximum Likelihood estimates of the GPD form parameter
for Stretched-Exponentail samples of size 50,000. }
\end{figure}

For each simulation, the size of the sample above the considered
threshold $U_k$
is chosen equal to $50,000$ in order to get small standard deviations.
The Maximum-Likelihood estimates of the GPD form parameter $\xi$ are shown in
figure~\ref{fig110}. For $c=0.7$, the threshold $U_7$ gives an
estimate $\xi=0.0123$ with standard deviation equal to $0.0045$,
i.e., the estimate
for $\xi$ differs significantly from zero (recall that $\xi=0$ is the
correct theoretical limit value). This occurs notwithstanding the huge size of
the implied data set; indeed,
the probability $\Pr{X>U_7}$ for $c=0.7$ is about $10^{-9}$,
so that in order to obtain a data set of conditional samples from an
unconditional data set of the size studied here ($50,000$ realizations above
$U_7$), the size of
such unconditional sample should be approximately $10^9$ times larger
than the number of
``peaks over threshold'', i.e., it is practically impossible to have
such a sample.
For $c=0.3$, the
convergence to the theoretical value zero is even slower.
Indeed, even the largest financial datasets for a single
asset, drawn from high frequency data, are no larger than or of the order of
one million points\footnote{One year of data sampled at the $1$
minute time scale
gives approximately $1.2 \cdot 10^5$ data points}. The situation does not
change even for data sets one or two orders of magnitudes larger as
considered in \cite{GMAS98,GMAS98_2,Plerouetal99}, obtained by aggregating thousands of
stocks\footnote{In this
case, another issue arises concerning the fact that the aggregation of
returns from different assets may distort the information and the very
structure of the tails of the probability density functions (pdf),
if they exhibit some intrinsic variability
\cite{matia}.}. Thus, although the GPD form
parameter should be zero theoretically in the limit
of large sample for the Weibull distribution, this limit cannot
be reached for any available sample sizes.

This is a clear illustration that a rapidly varying distribution,
like the Weibull
distribution with exponent smaller than one, i.e., a Stretched-Exponential
distribution, can be mistaken for a Pareto or any other regularly varying
distribution for any practical applications.

\subsection{Generation of a long memory process with a well-defined stationary
distribution}

In order to study the performance of the various estimators of the tail index
$\xi$ and the influence of interdependence of sample values,
we have generated several samples with distinct properties. The first
three samples are
made of iid realizations drawn respectively from an asymptotic power-law
distribution with tail index $b=3$ and from a Stretched-Exponential 
distribution
with exponent $c=0.3$ and $c=0.7$. The other samples contain realizations
exhibiting different degrees of time dependence with the same three
distributions as for the first three samples: a regularly varying
distribution with tail index $b=3$ and a Stretched-Exponential
distribution with
exponent $c=0.3$ and $c=0.7$. Thus, the three first samples are the iid
counterparts of the later ones. The sample with regularly varying iid
distributions
converges to the Fr\'echet's maximum domain of attraction with
$\xi=1/3 = 0.33$,
while the iid Stretched-Exponential distribution converges to Gumbel's maximum
domain of attraction with $\xi=0$. We now study how well can one distinguish
between these two distributions belonging to two different maximum
domains of attraction.

For the stochastic processes with temporal dependence, we use a simple
stochastic volatility model. First, we construct a Markovian Gaussian
process $\{X_t\}_{t \ge 1}$ whose correlation function is
\be
\label{eq:CorrFunc}
C(t) = a^{|t|}, \quad a <1.
\ee
Varying $a$ allows us to change the strength of the time dependence,
characterized by the correlation length $\tau = - \frac{1}{\ln a}$.  When
$a=0$, the iid case is retrieved. In the following, we have chosen
$a=0.95$ and $0.99$, which correspond to correlation lengths of about 
20 and 100
lags respectively. For simplicity, we will refer to the first case as the
``short-memory'' process, while the second one will be called ``long-memory''
process. This denomination is only for convenience and does not refer to the
conventional distinction between processes with short and long range memory
\cite{Beran94}.

The next step consists in building the process $\{U_t\}_{t \ge 1}$, defined by
\be
U_t = \Phi (X_t)~,  \label{mgkled}
\ee
where $\Phi(\cdot)$ is the Gaussian distribution function. The process
$\{U_t\}_{t \ge 1}$ exhibits also a dependence qualitatively
similar to that of the process $\{X_t\}_{t \ge 1}$.
The precise nature of  the temporal dependence of the process $\{U_t\}_{t \ge
1}$ is revealed differently by different tools Indeed, if one quantifies
dependence by copulas, then the process $\{ U_t\}_{t \ge 1}$ has the same
dependence as $\{ X_t \}_{t \ge 1}$ because copulas are invariant under a
strickly increasing change of variables. Let us recall that a copula is the
mathematical embodiment of the dependence structure between different random
variables \cite{Joe97,Nelsen98}. The process $\{U_t\}_{t \ge 1}$ thus possesses
a Gaussian copula dependence structure with long memory and uniform marginals.
In contrast, if one quantifies the dependence by the correlation coefficient or
the correlation ratio or other reasonable standard measures of dependence,
the monotonous change of variable (\ref{mgkled}) is no more innocuous as
the correlation may become as small as one
wants under an suitable choice
of a strickly increasing transformation (see for instance \cite{Malconcept}
for a detailed discussion of the effect of conditioning on
correlation measures). However, in the present case, we can
calculate  exactly the correlation function of the process $\{ U_t 
\}_{t \ge 1}$, which
is nothing but the rank (or Spearman) correlation function of the process  $\{
X_t \}_{t \ge 1}$, so that
\bea
C_U(t) &=& \frac{6}{\pi} \arcsin \left( \frac{1}{2} ~C(t) \right),\\
&=& \frac{6}{\pi} \arcsin \left( \frac{a^{|t|}}{2} \right),\\
& \simeq & \frac{3}{\pi} a^{|t|}, \quad {\rm as}~ t\rightarrow \infty.
\eea
For our purpose, the important
point is to obtain a process with the correct asymptotic distribution
tails together
with some dependence: this allows us to probe some impact of the dependence
on estimators and show that
standard statistical estimators may become unreliable.

In the last step, we define the volatility process
\be
\sigma_t = \sigma_0 \cdot U_t^{-1/b}~,
\label{nmghjk}
\ee
which ensures that the stationary distribution of the volatility is a Pareto
distribution with tail index $b$. Such a distribution of the volatility is not
realistic in the bulk which is found to be approximately
a lognormal distribution for not too large volatilities \cite{SSA00}, but is
in agreement with the hypothesis of an asymptotic regularly varying
distribution.
A change of variable more complicated than (\ref{nmghjk}) can provide
a more realistic behavior of the volatility on
the entire range of the distribution but our
main goal is not to provide a realistic stochastic
volatility model but only to exhibit a stochastic process with
time dependence and well-defined prescribed marginals in order to test the
influence of the dependence structure.

The return process is then given by
\be
r_t = \sigma_t \cdot \epsilon_t~,
\label{hyje}
\ee
where the $\epsilon_t$ are Gaussian random variables independent from
$\sigma_t$. The construction (\ref{hyje})
ensures the de-correlation of the returns at every time lag.
The stationary distribution of $r_t$ admits the
density
\be
p(r)= \frac{2^{\frac{b}{2}-1}}{\sqrt{\pi}} \cdot \Gamma \left( \frac{b+1}{2},
\frac{r^2}{2 \sigma_0^2} \right) \frac{b \cdot \sigma_0^b}{|r|^{b+1}}~,
\ee
which is regularly varying at infinity since $\Gamma \left(
\frac{b+1}{2}, \frac{r^2}{2\cdot \sigma_0^2} \right)$ goes to $\Gamma \left(
\frac{b+1}{2} \right)$.
This completes the construction and characterization of
our long memory process with regularly varying stationary
distribution.

In order to obtain a process with Stretched-Exponential
distribution with long range dependence,
we apply to $\{r_t\}_{t \ge 1}$ the following increasing mapping
$G : r \go y$
\be
\label{eq:transfo}
G(r) =
\begin{cases}
(x_0 + \ln \frac{r}{r_0})^{1/c} & r > r_0\\
{\rm sgn}(r) \cdot |r|^{1/c} & |r| \le r_0\\
-(r_0 + \ln |r/r_0|)^{1/c} & r < -r_0~.
\end{cases}
\ee
This transformation gives a
stretched exponential of index $c$ for all values of the return larger than the
scale factor $r_0$. This derives from the fact that the process
$\{r_t\}_{t \ge 1}$
admits a regularly varying distribution function, characterized by
$\bar F_r(r) =1- F_r(r)= {\cal L}(r) |r|^{- b}$, for some slowly varying
function ${\cal L}$. As a consequence, the stationary distribution of
$\{Y_t\}_{t \ge1}$ is given by \bea
\bar F_Y(y) &=& {\cal L} \left( r_0 e^{-x_0} \exp \left( y^c \right)  \right)
\frac{e^{b r_0}}{x_0^b} \cdot e^{- b |y|^c}, \quad \forall |y| >
r_0,    \\
&=&  {\cal L}' (y) \cdot e^{- b |y|^c}, \quad \text{${\cal L}'$ is slowly
varying at infinity} ,
\eea
which is a Stretched-Exponential distribution.

To summarize, starting with a Markovian Gaussian process, we have defined
a stochastic process characterized by a stationary distribution function of
our choice, thanks to
the invariance of the temporal dependence structure (the copula) under strictly
increasing change of variable. In particular, this approach gives stochastic
processes with a regularly varying marginal distribution and with a
stretched-exponential distribution. Notwithstanding the difference in their
marginals, these two processes possess by construction exactly the same time
dependence. This allows us to compare the impact of the same 
dependence on these
two classes of marginals.

\subsection{Results of numerical simulations}
\label{sec:boot}

We have generated $1000$ replications of each process presented in the
previous section, i.e., iid Stretched-Exponential, iid Pareto, short and long
memory processes with a Pareto distribution and with a Stretched-Exponential
distribution. Each sample contains $10,000$ realizations,
which is approximately the number of points in each tail of our real samples.

Panel (a) of table~\ref{table:MaxLikelihood} presents the mean values and
standard deviations of the Maximum Likelihood estimates of $\xi$, using the
Generalized Extreme Value distribution and the Generalized Pareto Distribution
for the three samples of iid data. To estimate the parameters of the GEV
distribution and study the influence of the sub-sample size, we have grouped
the data in clusters of size $q=10, 20, 100$ and $200$.
For the analysis in terms of the GPD, we have considered four
different large thresholds $u$, corresponding to the quantiles $90\%$, $95\%$,
$99\%$ and $99.5\%$.
The estimates of $\xi$ obtained from the distribution of maxima are
compatible (at the
$95\%$ confidence level) with the expected value for the Stretched-Exponential
with $c=0.7$ for all cluster sizes and for the Pareto distribution for clusters
of size larger than 10. For the Stretched-Exponential with fractional
exponent $c=0.3$, we obtain an average value $\xi$ larger than $0.2$ over the
four different sizes of sub-samples. Except for the largest cluster, this value
is significantly different from the theoretical value $\xi=0.0$. This clearly
shows that the distribution of the maximum drawn from a Stretched-Exponential
distribution with $c=0.7$ converges very quickly toward the theoretical
asymptotic GEV distribution, while for $c=0.3$ the convergence is extremely
slow. Such a fast convergence for $c=0.7$ is not surprising since, for this
value of the fractional index, the Stretched-Exponential distribution remains
close to the Exponential distribution, which is known to converge very quickly
to the GEV distribution \cite{HW79}. For $c=0.3$, the
Stretched-Exponential distribution behaves, over a wide range, like the power
law - as we shall see in the next section - thus it is not
surprising to obtain an
estimate of $\xi$ which remains significantly positive.

Overall, the results are slightly better for the Maximum
Likelihood estimates obtained from the GPD. Indeed, the bias observed for the
Stretched-Exponential with $c=0.3$ seems smaller for large quantiles than the
smallest biases reached by the GEV method. Thus, it appears that the
distribution of exceedance converges faster to its asymptotic distribution than
the distribution of maximum. However, while in line with the 
theoretical values,
the standard deviations are found almost always larger than in the previous
case, which testifies of the higher variability of this estimator.
Thus,  for such sample sizes, the GEV and GPD Maximum Likelihood estimates
should be handled with care and there results interpreted with caution
due to possibly important bias and statistical fluctuations.
If a small value of $\xi$ seems to allow one to reliably conclude in favor of a
rapidly varying distribution, a positive estimate does not appear informative,
and in particular does not allow one to reject the rapidly varying
behavior of a
distribution.

Panel (b) and (c) of table~\ref{table:MaxLikelihood}
presents the same results for data with short and long memory,
respectively. We note the presence of a significant downward bias (with
respect to the iid case) in almost every cases for the GPD estimates: the
stronger the dependence, the more important is the bias.  At the same 
time, the empirical
values of the standard deviations remain comparable with those obtained in the
previous case for iid data. The downward bias can
be ascribed to the dependence between data. Indeed, positive dependence yields
important clustering of extremes and accumulation of realizations around some
values, which -- for small samples -- could (misleadingly) appear as the
consequence of the compactness of the support of the underlying distribution.
This rationalizes the negative $\xi$ estimates obtained for the
Stretched-Exponential distribution with $c=0.7$. In other words, for finite
sample, the dependence prevents the full exploration of the tails and create
clusters that mimics a thinner tail (even if the clusters are occurring all at
large values since what is important is the range of exploration of the tail in
order to control the value of $\xi$).

The situation is different for the GEV estimates which show either an 
upward  or
downward bias (with respect to the iid case).  Here two effects are competing.
On the one hand, the dependence creates a downward bias, as explained above,
while, on the other hand, the lack of convergence of the distribution of maxima
toward its GEV asymptotic distribution results in an upward bias, as 
observed on
iid data. This last phenomemon is strengthened by the existence of time
dependence which leads to decrease the ``effective'' sample size ( the actual
size divided by the correlation length $\lambda=\sum C(t)=(1-a)^{-1}$) and thus
slows down the convergence rate toward the asymptotic distribution even more.
Interestingly, both the GEV and GPD estimators for the Pareto
distribution may be utterly wrong in presence of long range dependence for any
cluster sizes.

To summarize, two opposite effects are competing. On the one hand,
non-asymptotic effects due to the slow convergence toward the asymptotic GEV or
GPD distributions yield an upward or downward bias. This effect seems more
pronounced for GEV distributions and becomes more important when the 
correlation
length increases since the ``effective'' sample size decreases. On the other
hand, the presence of dependence in the data induces a downward bias and
sometimes an increase of the standard deviation of the estimated values. The
qualitative effect can be described as follows: {\it the larger $a$ 
is, the smaller
is the $\xi$-estimate}, provided - of course - that the ``effective'' 
sample size is
kept constant, everything being otherwise taken equal.

These two entangled effects, which sometimes compete and sometimes 
oppose each other,
have also
been observed for non-Markovian processes drawn from Gaussian processes with
long range correlation. Thus, the existence of an important bias and the
increase in the scattering of estimates is a general and genuine 
progeny of the time
dependence. It leads us to the conclusion that the Maximum Likelihood 
estimators
derived from the GEV or GPD distributions are not very efficient for the
investigation of the financial data whose sample sizes are moderate and which
exhibit complicated serial dependence. The only positive note is that the GPD
estimator correctly recovers the range of
the index $\xi$ with an uncertainty smaller than $20\%$ for data with a pure
Pareto distribution while it is cannot reject the hypothesis that $\xi=0$ when
the data is generated with a Stretched-Exponential distribution, albeit with a
very large uncertainty, in other words with little power.

Table~\ref{tableBootstrap} focuses on the results given by Pickands' estimator
for the tail index of the GPD. For each thresholds $u$, corresponding to the
quantiles $90\%$, $95\%$, $99\%$ and $99.5\%$ respectively, the results of our
simulations are given for two particular values of $k$  (defined in
(\ref{eq:Pickands})) corresponding to $N/k=4$, which is the largest admissible
value, and $N/k=10$ corresponding to be sufficiently far in the tail of the
GPD. Table~\ref{tableBootstrap} provides the mean value
and the numerically estimated
as well as the theoretical (given by (\ref{eq:PicSTD}))
standard deviation of $\hat \xi_{k,N}$. Panel (a) gives the result for
iid data. The mean values do not exhibit a significant bias for the Pareto
distribution and the Stretched-Exponential with $c=0.7$, but are utterly wrong
in the case $c=0.3$ since the estimates are comparable with those given for
the Pareto distribution. In each case, we note a very good agreement
between the empirical and theoretical standard deviations, even for the larger
quantiles (and thus the smaller samples). Panels~(b-c) present the results for
dependent data. The estimated standard deviations remains of the same order as
the theoretical ones, contrarily to results reported by \citeasnoun{KP97} for
IGARCH processes. However, like these authors, we find that the bias, either
positive or negative, becomes very significant and leads one to misclassify a
Stretched-Exponential distribution with $c=0.3$ for a Pareto distribution with
$b=3$. Thus, in presence of dependence, Pickands' estimator is
unreliable.

To summarize, the determination of the maximum domain of attraction with usual
estimators does not appear to be a very efficient way to study the extreme
properties of dependent times series. Almost all the previous studies
which have
investigated the tail behavior of asset returns distributions have focused on
these methods (see the influential works of \citeasnoun{Longin96} for instance)
and may thus have led to spurious results on the determination of the tail
behavior. In particular, our simulations show that rapidly varying function may
be mistaken for regularly varying functions. Thus, according to our
simulations, this
casts doubts on the strength of the conclusion of previous works that
the distributions of returns are regularly varying as seems to have been
the consensus until now and suggests to re-examine the possibility that
the distribution of returns may be rapidly varying
as suggested by \citeasnoun{GJ98} or
\citeasnoun{LS99} for instance. We now turn to this question
using the framework of GEV and GDP estimators just described.

\subsection{GEV and GPD estimators of the Dow Jones and Nasdaq data sets}

We have applied the same analysis as in the previous section on the
real samples of the Dow Jones and Nasdaq (raw and corrected) returns.
In order to estimate the standard deviations of Pickands'
estimator for the GPD derived from the upper quantiles of these distributions,
and of ML-estimators for the distribution of maximum and for the GPD, we have
randomly generated one thousand sub-samples, each sub-sample being constituted
of ten thousand data points in the positive or negative parts of the samples
respectively (with replacement).  It should be noted that the ML-estimates
themselves were derived from the full samples. The results are given in
tables~\ref{table:ML_RealData} and \ref{tableRealData}.

These results confirm the confusion about the tail behavior of the returns
distributions and it seems impossible to exclude a rapidly varying
behavior of their tails. Indeed, even the estimations performed by Maximum
Likelihood with the GPD tail index, which have appeared as the least unreliable
estimator in our previous tests, does not allow us to clearly reject
the hypothesis
that the tails of the empirical distributions of returns are rapidly varying,
in particular for large quantile values.
For the Nasdaq dataset, accounting for the lunch effect does not
yield any significant change in the estimations. This observation will be
confirmed by the other tests presented in the next sections.

\begin{figure}
\begin{center}
\includegraphics[width=12cm]{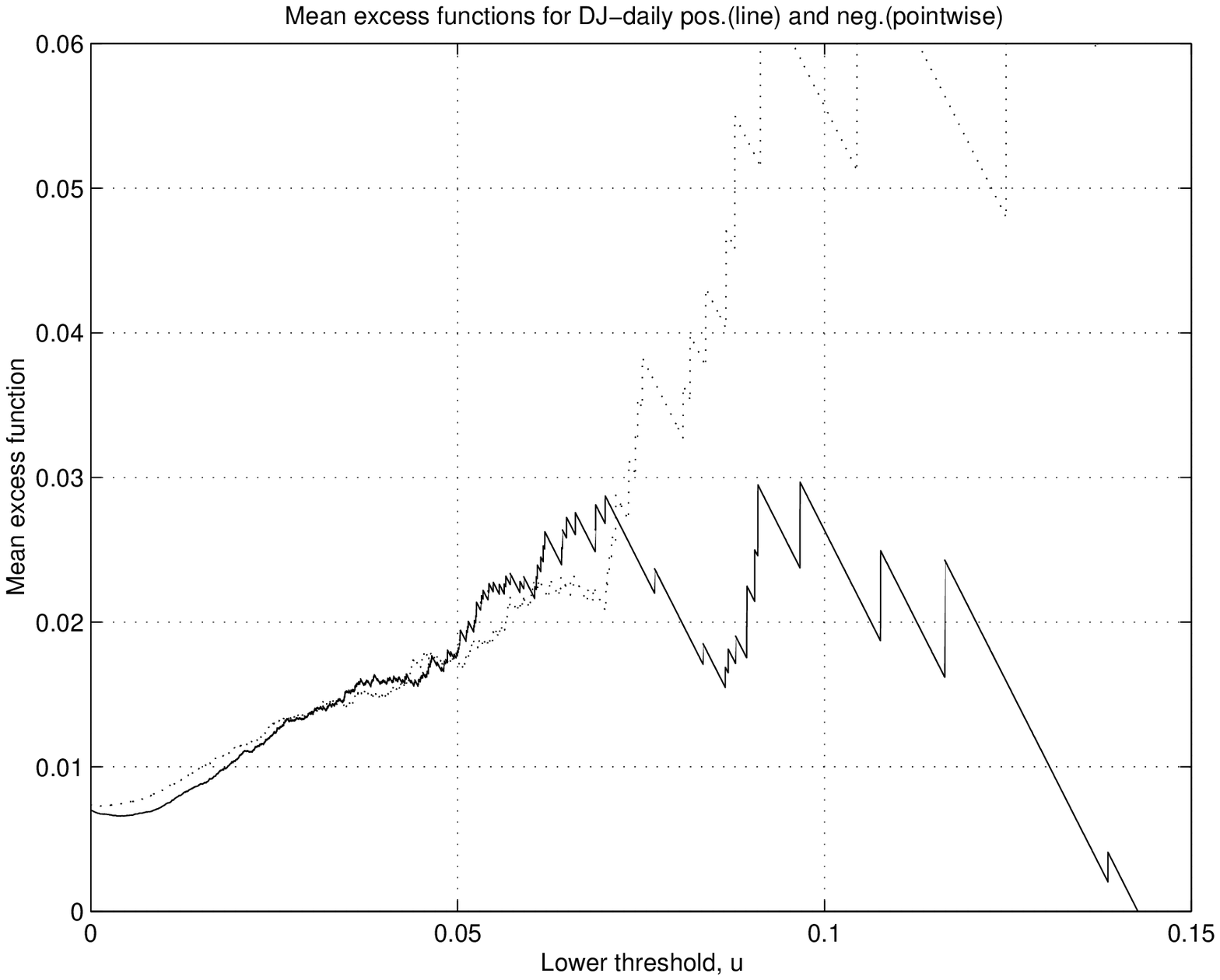}
\includegraphics[width=12cm]{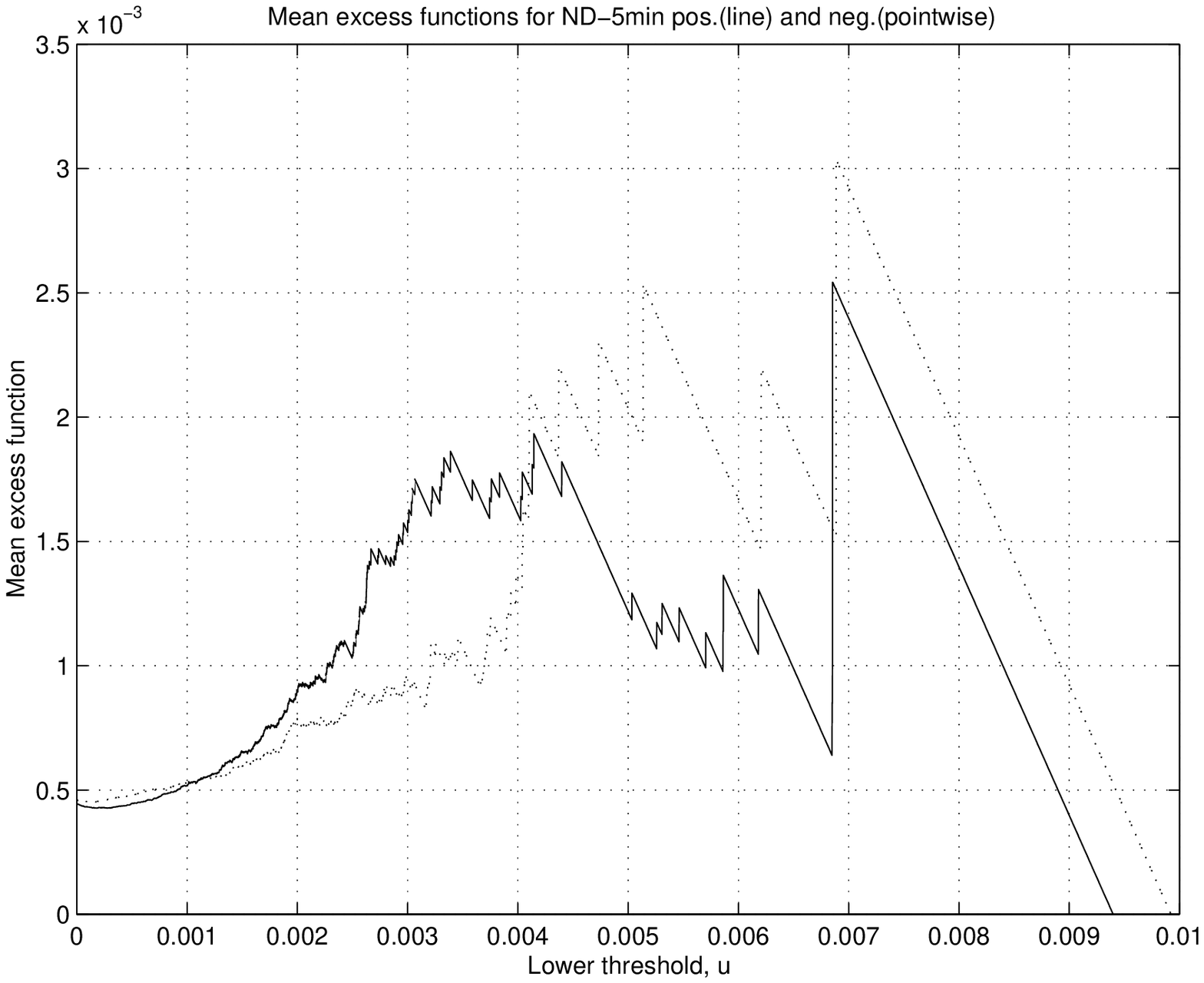}
\end{center}
\caption{\label{fig25} Mean excess functions for the Dow Jones daily returns
(upper panel) and the Nasdaq five minutes returns (lower panel). The plain
line represents the positive returns and the dotted line the negative ones }
\end{figure}

As a last non-parametric attempt to distinguish between a regularly
varying tail and a rapidly varying tail of the exponential or
Stretched-Exponential
families, we study the \textit{Mean Excess
Function} which is one of the known methods that often can help in
deciding what parametric
family is appropriate for approximation (see for details \citeasnoun{EKM97}).
The Mean Excess Function $MEF(u)$ of a random value $X$
(also called ``shortfall'' when applied
to negative returns in the context of financial risk management) is defined as
\be
MEF(u) = \E (X -u \vert X >u)~.
\ee
The Mean Excess Function $MEF(u)$ is obviously related to the GPD for
sufficiently large
threshold $u$ and its behavior can be derived in this limit for the three
maximum domains of attraction. In addition, more precise results can
be given for
particular random variables, even in a non-asymptotic regime. Indeed, for an
exponential random variable $X$, the $MEF(u)$ is just a constant. For a Pareto
random variable, the $MEF(u)$ is a straight increasing line, whereas for the
Stretched-Exponential and the Gauss distributions, the $MEF(u)$ is a decreasing
function. We evaluated the sample analogues of the $MEF(u)$ \cite[p.296]{EKM97}
which are shown in figure~\ref{fig25}. All attempts to find a constant  or a
linearly increasing behavior of the $MEF(u)$ on the main central part of the
range of returns were ineffective. In the central part of the range of negative
returns $(\vert X \vert > 0.002$; $q \cong  98\%$ for ND data, and
$\vert X \vert
> 0.025$ ; $q \cong  96\%$ for DJ data), the $MEF(u)$ behaves like a convex
function which exclude both exponential and power (Pareto)
distributions. Thus, the $MEF(u)$ tool does not support using
any of these two distributions.

\begin{figure}
\begin{center}
\includegraphics[width=12cm]{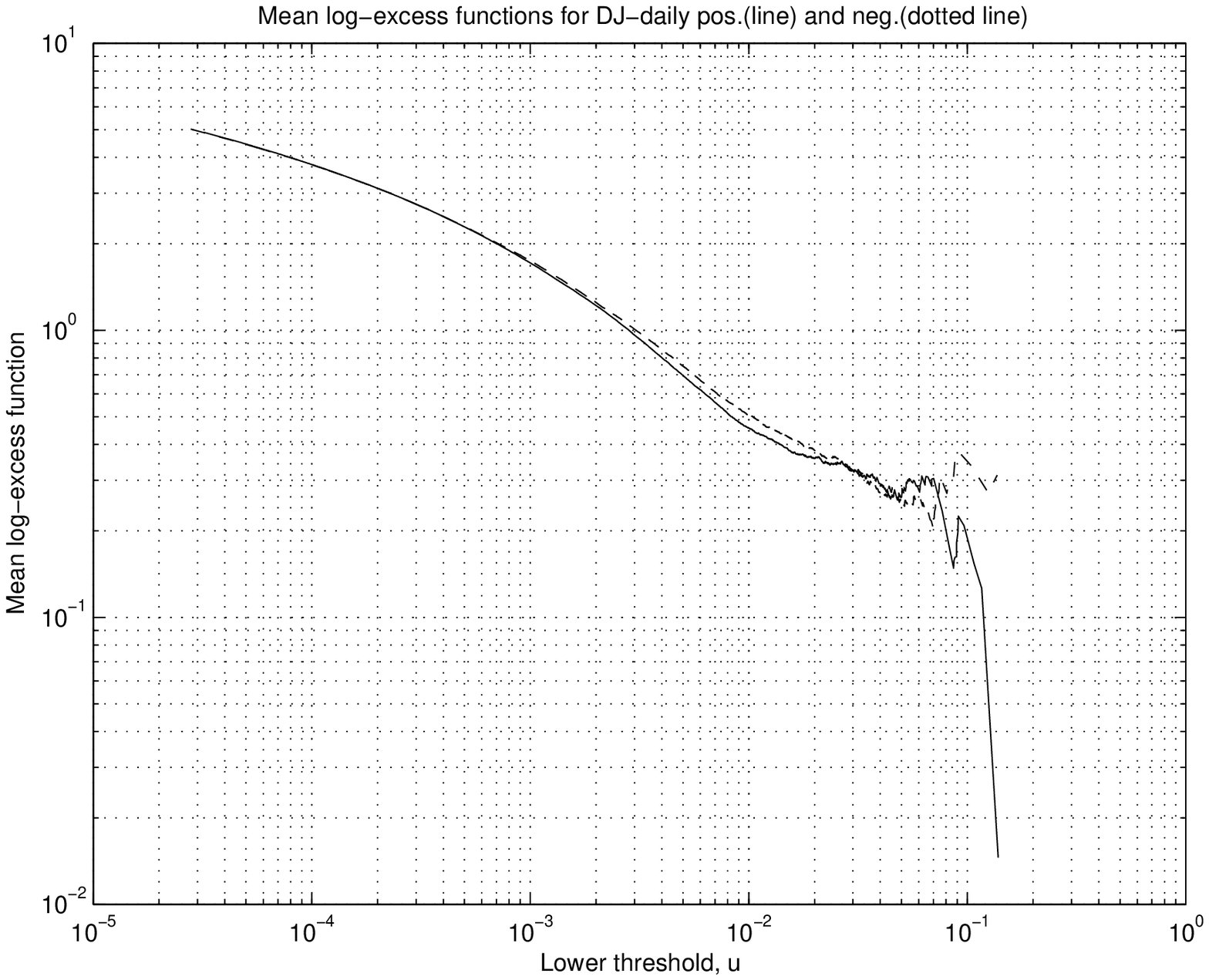}
\includegraphics[width=12cm]{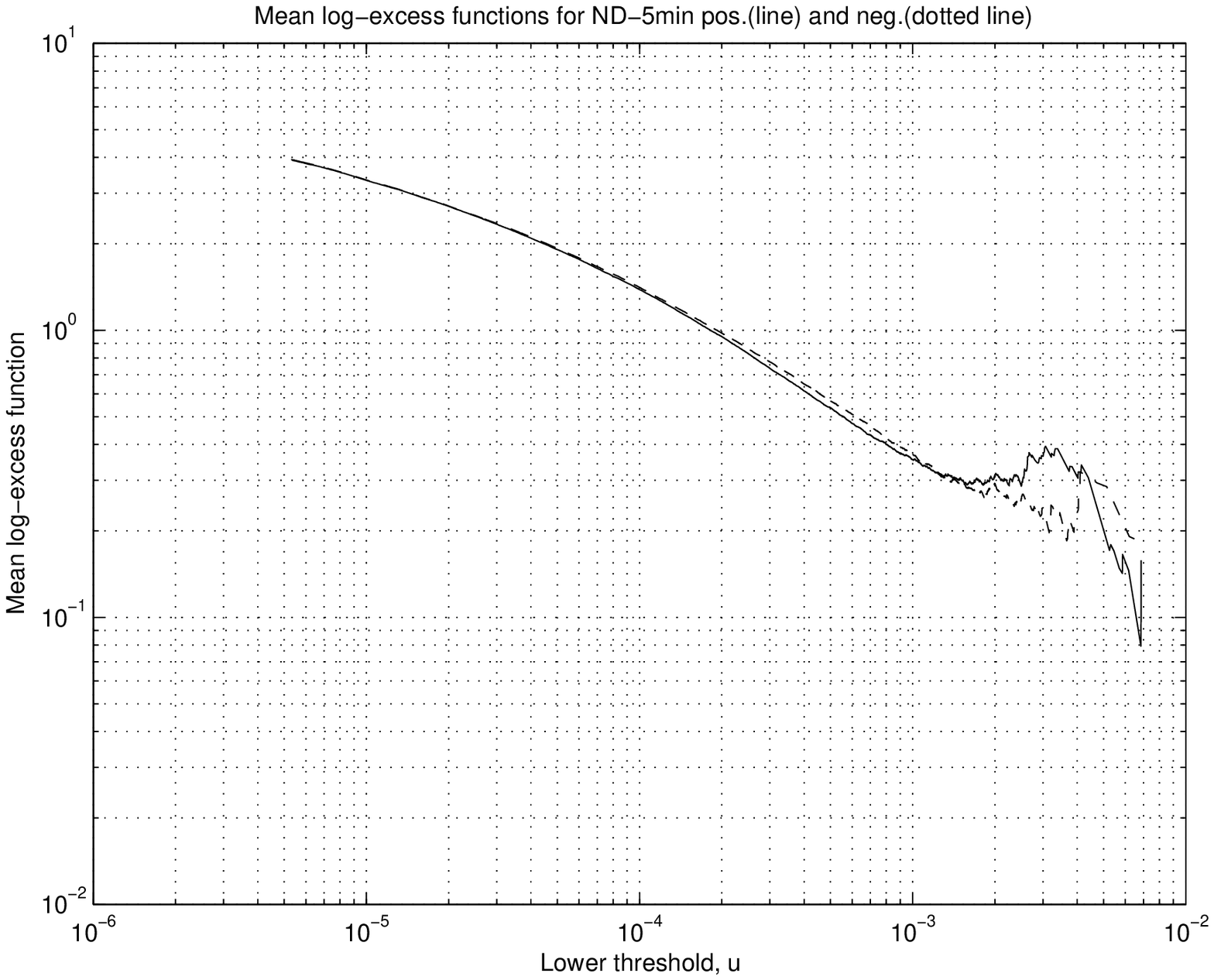}
\end{center}
\caption{\label{figLogExcess} Mean Log-excess functions for the Dow Jones daily
returns (upper panel) and the Nasdaq five minutes returns (lower panel). The
plain line represents the positive returns and the dotted line the negative ones
} \end{figure}

An alternative to the Mean Excess function is provided by the Mean Log-Excess
function:
\be
MLEF(u) = \E(\log(X/u) \vert X>u).
\ee
$MLEF(u)$ is again related to the GPD (of the variable $\log X$ instead of
$X$) for sufficiently large threhold $u$. In particular, when $X$ follows
asymtotically a power law, $\log X$ is asymptotically exponentially distributed,
so that $MLEF(u)$ goes to a constant equal to $\alpha^{-1}$, where $\alpha$
denotes the tail index of the distribution of $X$. For a Stretched-Exponential
variable $X$ with fractional exponent $c$, it turns out that $MLEF(u)$ behaves
like a regularly varying function whose tail index equals $-c$. Thus, in a
double logarithmic plot, such a behavior is characterized by a decreasing
straigth line with slope $-c$. Sample estimates of $MLEF(u)$ are shown in
figure~\ref{figLogExcess}. On about
$90\%$ of the range of the sample, the Mean Log-Excess functions behaves as expected for
Stretched-Exponentially distributed variables, while in the tail range (about
$10\%$ of the largest values), the results are very confusing, due to the
importance of the statistical fluctuations. Such behavior of  
$MLEF(u)$ in the tails cannot be
attributed definitely to a regularly varying or to a Stretched-Exponentially
distributed random variable. Therefore, a change of regime cannot be excluded
in the extreme tail of the distributions.

In view of the stalemate reached with the above non-parametric
approaches and in particular with the standard extreme value estimators,
the sequel of this paper is devoted to
the investigation of a parametric approach in order to decide
which class of extreme value distributions, rapidly versus regularly varying,
accounts best for the empirical distributions of returns.

\section{Fitting distributions of returns with parametric densities}

Since our previous results lead to doubt the validity of the
rejection of the hypothesis that the distribution of returns are
rapidly varying, we now propose to pit a parametric champion for this class of
functions against the Pareto champion of regularly varying functions.
To represent the class of rapidly varying functions, we propose
the family of Stretched-Exponentials. As discussed in the introduction,
the class of stretched
exponentials is motivated in part from a theoretical view point by
the fact that the
large deviations of multiplicative processes are generically distributed
with stretched exponential distributions \cite{FrischSor}.
Stretched exponential distributions are also parsimonious examples of
sub-exponential distributions with fat tails for instance in the
sense of the asymptotic
probability weight of the maximum compared with the sum of large
samples \cite{Feller}.
Notwithstanding their fat-tailness, Stretched Exponential
distributions have all
their moments finite\footnote{However, they do not admit an exponential
moment, which leads to problems in the reconstruction of the distribution
from the knowledge of their moments \cite{SO94}.}, in contrast with regularly
varying distributions for which moments of order equal to or larger than the
index $b$ are not defined. This property may provide a substantial advantage to
exploit in generalizations of the mean-variance portfolio theory using
higher-order moments \cite[for instance
]{R73,FL97,HS99,SSA00,AS01,JM02,MSmm02}.
Moreover, the existence of all moments is an important property allowing for an
efficient estimation of any high-order moment, since it ensures that the
estimators are asymptotically Gaussian. In particular, for
Stretched-Exponentially distributed random variables, the variance,
skewness and
kurtosis can be well estimated, contrarily to random variables with regularly
varying distribution with tail index in the range $3-5$.

\subsection{Definition of two parametric families}
\subsubsection{A general $3$-parameters family of distributions}
\label{sec:pdf}

We thus consider a general $3$-parameters family of distributions and its
particular restrictions corresponding to some fixed value(s) of two (one)
parameters. This family is defined by its density function given by:
\be \label{eq:CDF}
f_{u}(x \vert b,c,d) =
\begin{cases}
A(b,c,d,u)~ x^{-(b+1)} \exp \left[- \left( \frac{x}{d} \right)^{c} \right] &
\text{if $x\geqslant u >0$}\\ 0 & \text{if $x < u$}.
\end{cases}
\ee
Here, $b, c, d$ are unknown parameters, $u$ is a known lower threshold
that will be varied for the purposes of our analysis and $A(b,c,d,u)$
is a normalizing
constant given by the expression:
\be
A(b,c,d,u) = \frac{d^{b}~c}{\Gamma(-b/c,(u/d)^{c})},
\ee
where $\Gamma ( a , x )$ denotes the (non-normalized) incomplete Gamma
function. The parameter $b$ ranges from minus infinity to infinity while $c$
and $d$ range from zero to infinity. In the particular case where $c = 0$, the
parameter $b$ also needs to be positive to ensure the normalization of the
probability density function (pdf).
The interval of definition of this family is
the positive semi-axis. Negative log-returns will be studied
by taking their absolute values. The family~(\ref{eq:CDF})
includes several well-known pdf's often used in different
applications. We enumerate them.
\begin{enumerate}
\item
The Pareto distribution:
\be
\label{eq:Pareto}
F_{u}(x)=1-(u/x)^{b},
\ee
which corresponds to the set of parameters $(b>0, c=0)$ with
$A(b,c,d,u)=b\cdot u^{b}$. Several works have attempted to derive
or justified the existence of a power tail of the distribution
of returns from agent-based models \cite{chalmar}, from optimal
trading of large funds with sizes distributed according to the Zipf law
\cite{gabaixstan} or from stochastic processes \cite[2002]{SF02,Biham98}.

\item The Weibull distribution:
\be
\label{eq:Weibull}
      F_{u}(x)=1- \exp \left[- \left( \frac{x}{d} \right)^{c}+
\left( \frac{u}{d}\right)^{c}\right],
\ee
with parameter set $(b= -c, c>0,d>0)$ and normalization constant
$A(b,c,d,u)=\frac{c}{d^{c}} \exp \left[ \left(\frac{u}{d} \right)^{c}\right]$.
This distribution is said to be a ``Stretched-Exponential''
distribution when the
exponent $c$ is smaller than $1$, namely when the distribution decays
more slowly than an
exponential distribution.

\item The exponential distribution:
\be
\label{eq:Exponential}
F_{u}(x)=1- \exp \left( -\frac{x}{d}+\frac{u}{d} \right),
\ee
with parameter set $(b=-1,~ c=1,~ d>0)$ and normalization constant
$A(b,c,d,u)=\frac{1}{d} \exp \left(-\frac{u}{d} \right)$.
For sufficiently high quantiles, the exponential behavior can for instance
derive, from the hyperbolic model introduced by
\citeasnoun{Eberlein} or from a simple model
where stock price dynamics is governed by a
geometrical (multiplicative) Brownian motion with stochastic
variance. \citeasnoun{Dragu} have found an excellent fit of
the Dow-Jones index for time lags from 1 to
250 trading days with a model
with an asymptotic exponential tail of the distribution of log-returns.

\item The incomplete Gamma distribution:
\be
\label{eq:Gamma}
F_u(x)= 1 - \frac{\Gamma( -b , x/d )}{\Gamma( -b , u/d )}
\ee
with parameter set $(b,~ c=1,~ d>0)$ and normalization
$A(b,c,d,u)=\frac{d^{b}}{\Gamma( -b , u/d )}$.  Such an asymptotic tail
behavior can, for instance, be observed for the generalized hyperbolic models,
whose description can be found in \citeasnoun{Prause}.
\end{enumerate}

Thus, the Pareto distribution (PD) and exponential distribution (ED) are
one-parameter families, whereas the stretched exponential (SE) and the
incomplete Gamma distribution (IG) are two-parameter families. The
comprehensive distribution (CD) given by equation (\ref{eq:CDF}) contains
three unknown parameters.

Interesting links between
these different models reveal themselves under
specific asymptotic conditions.
Very interesting for our present study is the behavior of the (SE)
model when $c \to 0$ and $u>0$.
In this limit, and provided that
\be
c \cdot \left( \frac{u}{d}\right)^c \rightarrow \beta, \quad \text{as}~ c
\rightarrow 0~.
\label{jgjlke}
\ee
the (SE) model goes to the Pareto model. Indeed, we can write
\bea
\frac{c}{d^c} \cdot x^{c-1} \cdot \exp \left( - \frac{x^c-u^c}{d^c} \right) &=&
c \left( \frac{u}{d} \right)^c \cdot \frac{x^{c-1}}{u^c} \exp \left[-\left(
\frac{u}{d}\right)^c \cdot \left( \left(\frac{x}{u} \right)^c-1 \right)
\right]~,  \nonumber\\
& \simeq & \beta \cdot x^{-1} \exp \left[ - c\left( \frac{u}{d}\right)^c  \cdot
\ln \frac{x}{u} \right] , \quad \text{as}~ c \rightarrow 0   \nonumber \\
& \simeq & \beta \cdot x^{-1} \exp \left[ - \beta  \cdot \ln
\frac{x}{u} \right]
~,  \nonumber \\
& \simeq& \beta \frac{u^\beta}{x^{\beta+1}}~,   \label{jgjkwlw}
\eea
which is the pdf of the (PD) model with tail index $\beta$. The condition
(\ref{jgjlke}) comes naturally from the properties of the maximum-likelihood
estimator of the scale parameter $d$ given by equation (\ref{eq:dhat}) in
Appendix \ref{app:MLE}.
It implies that, as $c \to 0$, the characteristic scale $d$ of the (SE) model
must also go to zero with $c$ to ensure the convergence of the (SE)
model towards the (PD) model.

This shows that the Pareto model can be approximated with any desired accuracy
on an arbitrary interval $(u>0,U)$ by the (SE) model
with parameters $(c,d)$ satisfying equation~(\ref{jgjlke}) where
the arrow is replaced by an equality. Although the value
$c=0$ does not give strickly speaking a Stretched-Exponential distribution, the
limit $c \go 0$ provides any desired approximation to the
Pareto distribution, uniformly on any finite interval $(u,U)$.
This deep relationship between the SE and PD models allows us to
understand why it can be very
difficult to decide, on a statistical basis, which of these models fits
the data best.

Another interesting behavior is obtained in the limit $b \to +\infty$,
where the Pareto model tends to  the Exponential model \cite{BP00}. Indeed,
provided that the scale parameter $u$ of the power law is simultaneously scaled
as $u^{b} = (b/\alpha)^b$, we can  write the tail of the cumulative
distribution
function of the PD as $u^b/(u+x)^b$ which is indeed of the form
$u^b/x^b$ for large $x$. Then, $u^b/(u+x)^b = (1+\alpha x/b)^{-b} \to
\exp(-\alpha x)$ for $b \to +\infty$.
This shows that the Exponential model can be approximated with any
desired accuracy
on intervals $(u,u+A)$ by the (PD) model
with parameters $(\beta, u)$ satisfying $u^{b} =
(b/\alpha)^b$, for any positive constant A. Although the value
$b\to +\infty$ does not give strickly speaking a Exponential distribution, the
limit $u \propto b \go +\infty$ provides any desired approximation to the
Exponential distribution, uniformly on any finite interval $(u, u+A)$.
This limit is thus less general that the SE $\to$ PD limit since it is valid
only asymptotically for $u \go +\infty$ while $u$ can be finite
in the SE $\to$ PD limit.

\subsubsection{The log-Weibull family of distributions}

Let us also introduce the two-parameter log-Weibull family:
\be
     1-F(x)=\exp\left[-b \left(\ln(x/u)\right)^c\right]~,   ~~~~{\rm
for}~~ x \ge u ~. \label{mjgtkwl}
\ee
whose density is
\be
\label{eq:LW}
f_{u}(x \vert b,c,d) =
\begin{cases}
\frac{b \cdot c}{x} \left( \ln \frac{x}{u} \right)^{c-1} \exp \left[ -b
\left( \ln \frac{x}{u} \right)^c\right], & \text{if $x\geqslant u >0$}\\
0, & \text{if $x < u$}. \end{cases}
\ee
This family of pdf interpolates smoothly between the Stretched-Exponential and
Pareto classes. It recovers the Pareto family for $c=1$, in which case the
parameter $b$ is the tail exponent. For $c$ larger than $1$, the tail of the
log-Weibull is thinner than any Pareto distribution but heavier than any
Stretched-Exponential\footnote{A generalization of the SLE to the following
three-parameter family also contains the SE family in some formal limit.
Consider indeed $1-F(x)=\exp(-b (\ln(1+x/D))^c )$ for $x>0$, which has the same
tail as expression (\ref{mjgtkwl}). Taking $D \to +\infty$ together with $b =
(D/d)^c$ with $d$ finite yields $1-F(x)=\exp(-(x/d))^c)$.}. In
particular, when $c$ equals two, the
log-normal distribution is retrieved (above threshold $u$). For $c$ smaller
than $1$, the tails of the SLE are even heavier than any Pareto distributions.
This range of parameter is probably not useful except maybe to account of
``outliers'' in the spirit of \citeasnoun{JS02}; this will require a specific
investigation.

\subsection{Methodology  \label{mgmjlws}}

We start with fitting our two data sets (DJ and ND) by the five
distributions enumerated above
(\ref{eq:CDF}) and (\ref{eq:Pareto}-\ref{eq:Gamma}). Our first goal is
to show that no single parametric
representation among any of the cited pdf's
fits the \textit{whole range} of the data sets. Recall that we analyze
separately positive and negative returns (the later being converted
to the positive
semi-axis). We shall use in our analysis a \textit{movable} lower
threshold $u$, restricting by
this threshold our sample to observations satisfying to $x > u$.

In addition to estimating the parameters involved in each
representation (\ref{eq:CDF},\ref{eq:Pareto}-\ref{eq:Gamma}) by
maximum likelihood
for each particular threshold $u$\footnote{The estimators and their
asymptotic properties are derived in Appendix~\ref{app:MLE}.}, we need a
characterization of the goodness-of-fit. For this, we propose to use a distance
between the estimated distribution and the sample distribution. Many
distances can be used: mean-squared error, Kullback-Liebler distance\footnote{
This distance (or {\it divergence}, strictly speaking) is the natural
distance associated with maximum-likelihood estimation since it is for these
values of the estimated parameters that the distance between the true model and
the assumed model reaches its minimum.}, Kolmogorov distance, Sherman distance
(as in \citeasnoun{Longin96}) or Anderson-Darling distance, to cite a few. We
can also use one of these distances to determine the parameters of each pdf
according to the criterion of minimizing the distance between the estimated
distribution and the sample distribution. The chosen distance is thus useful
both for characterizing and for estimating the parametric pdf. In the later
case, once an estimation of the parameters of particular distribution
family has
been obtained according to the selected distance, we need to quantify the
statistical significance of the fit. This requires to derive the statistics
associated with the chosen distance. These statistics are known for most of the
distances cited above, in the limit of large sample.

We have chosen the Anderson-Darling distance to derive our estimated
parameters and perform our tests of goodness of fit. The Anderson-Darling
distance between a theoretical distribution function $F(x)$ and its
empirical analog $F_{N}(x)$, estimated from a sample of $N$ realizations, is
evaluated as follows:
      \bea
\label{eq:ADS}
{\rm ADS} &=&N \cdot \int \frac{ [F_{N}(x)-F(x)]^{2}}{F(x)(1-F(x))}~ dF(x)\\
\label{eq:ADSS}
&=& -N -2 \sum\limits_1^N {\{w_k \log (F(y_k)) + (1 - w_k)\log (1 -
       F(y_k))\}} ,
\eea
where $w_{k}= 2k /(2N+1)$, $k=1\ldots N$ and
$y_{1}\leqslant\ldots\leqslant y_{N}$ is its ordered sample. If
the sample is drawn from a population with distribution function $F(x)$,
the Anderson-Darling statistics (ADS)
has a standard AD-distribution \textit{free of the theoretical df F(x)}
\cite{AD52}, similarly to the $\chi ^{2}$ for the $\chi^{2}$-statistic, or
the Kolmogorov distribution for the Kolmogorov statistic. It should be noted
that the ADS weights the squared difference in eq.(\ref{eq:ADS}) by
$1/F(x)(1-F(x))$
which is nothing but the inverse of the variance of the
difference in square brackets. The AD distance thus emphasizes more
the tails of the distribution than, say, the Kolmogorov
distance which is determined by the \textit{maximum absolute}
deviation of $F_{n}(x)$
from $F(x)$  or the mean-squared error, which is mostly controlled by
the middle
of range of the distribution. Since we have to insert the
estimated parameters into the ADS, this statistic
does not obey any more the standard AD-distribution: the ADS decreases because
the use of the fitting parameters ensures a better fit to the sample
distribution. However, we can still use the standard quantiles of the
AD-distribution as \textit{upper} boundaries of the ADS. If the
observed ADS is larger
than the standard quantile with a high significance level $(1 - \varepsilon )$,
we can then conclude that the null hypothesis $F(x)$ is rejected with
significance level larger than $(1 - \varepsilon)$. If we wish
to estimate the real significance level of the ADS in the case where
it does not
exceed the standard quantile of a high significance level, we are forced to use
some other method of estimation of the significance level of the ADS,
such as the
bootstrap method.

In the following,
the estimates minimizing the Anderson-Darling distance will be refered to
as AD-estimates. The maximum likelihood estimates (ML-estimates) are
asymptotically more efficient than AD-estimates for independent data and
under the condition that
the null hypothesis (given by one of the four distributions
(\ref{eq:Pareto}-\ref{eq:Gamma}), for instance) corresponds to the true
data generating model. When this is
not the case, the AD-estimates provide a \textit{better practical tool}
for approximating sample distributions compared with the ML-estimates.

We have determined the AD-estimates for $18$ standard significance levels
$q_{1}\ldots q_{18}$ given in table~\ref{table1}. The corresponding
\textit{sample quantiles} corresponding to these  significance levels or
thresholds $u_{1} \ldots u_{18}$ for our samples are also shown in
table~\ref{table1}.
Despite the fact that thresholds $u_{k}$ vary from sample to sample, they
always corresponded to the same fixed set of significance levels $q_{k}$
throughout the paper and allows us to compare the goodness-of-fit
for samples of different sizes.

\subsection{Empirical results  \label{mgmjlws2}}

The Anderson-Darling statistics (ADS) for six parametric distributions
(Weibull or Stretched-Exponential, Generalized Pareto, Gamma, Exponential,
Pareto and Log-Weibull) are shown in table~\ref{table2} for two
quantile ranges,
the first top half of the table corresponding to the $90\%$ lowest thresholds
while the second bottom half corresponds to the $10\%$ highest ones. For the
lowest thresholds, the ADS rejects all distributions, except the
Stretched-Exponential for the Nasdaq. Thus, none of the considered
distributions
is really adequate to model the data over such large ranges. For the $10\%$
highest quantiles, only the exponential model is rejected at the $95\%$
confidence level. The Log-Weibull and the Stretched-Exponential distributions
are the best, just above the Pareto distribution and the Incomplete Gamma that
cannot be rejected. We now present an analysis of each case in more details.

\subsubsection{Pareto distribution \label{pwerooa}}

\begin{figure}
\begin{center}
\includegraphics[width=12cm]{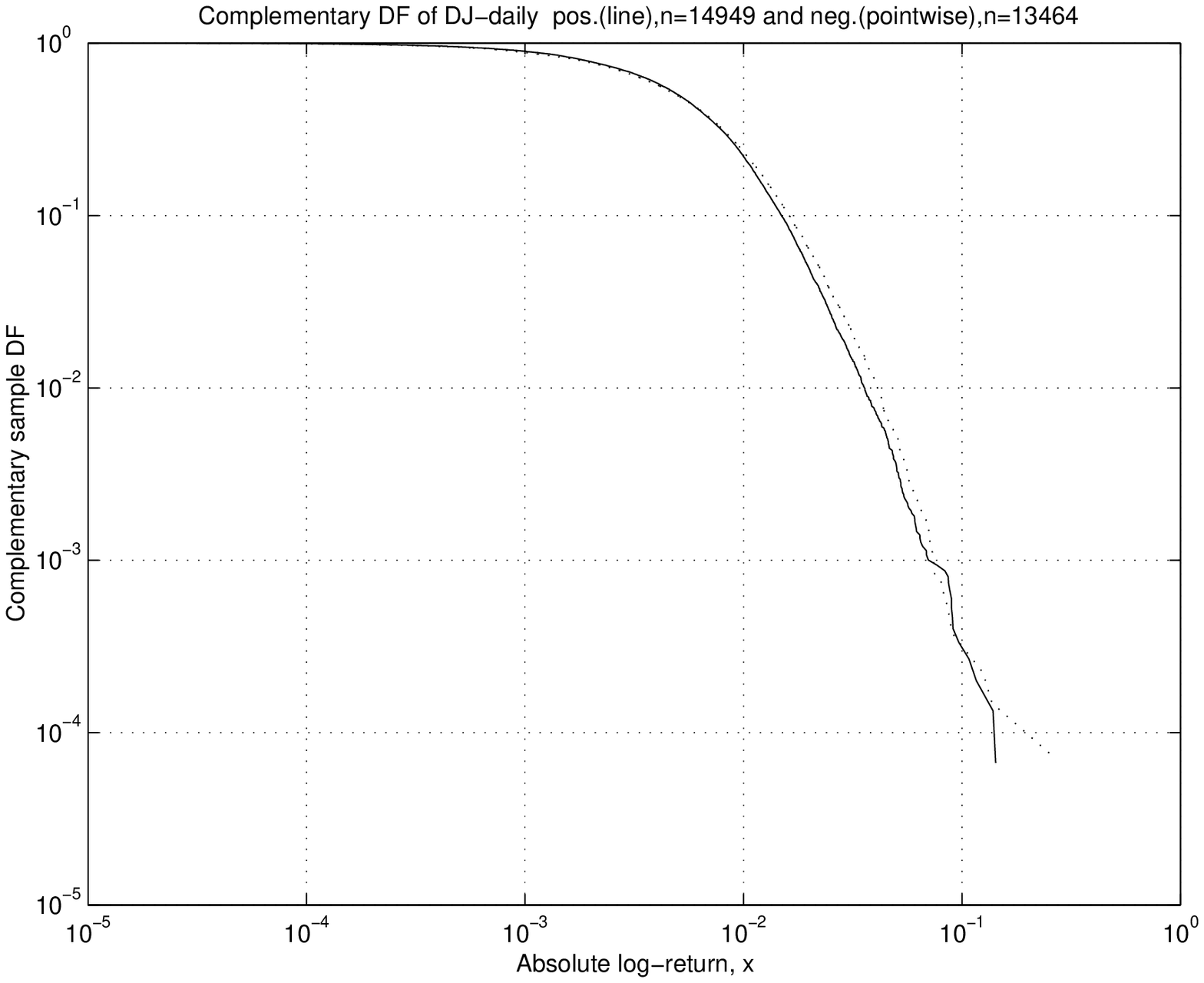}
\includegraphics[width=12cm]{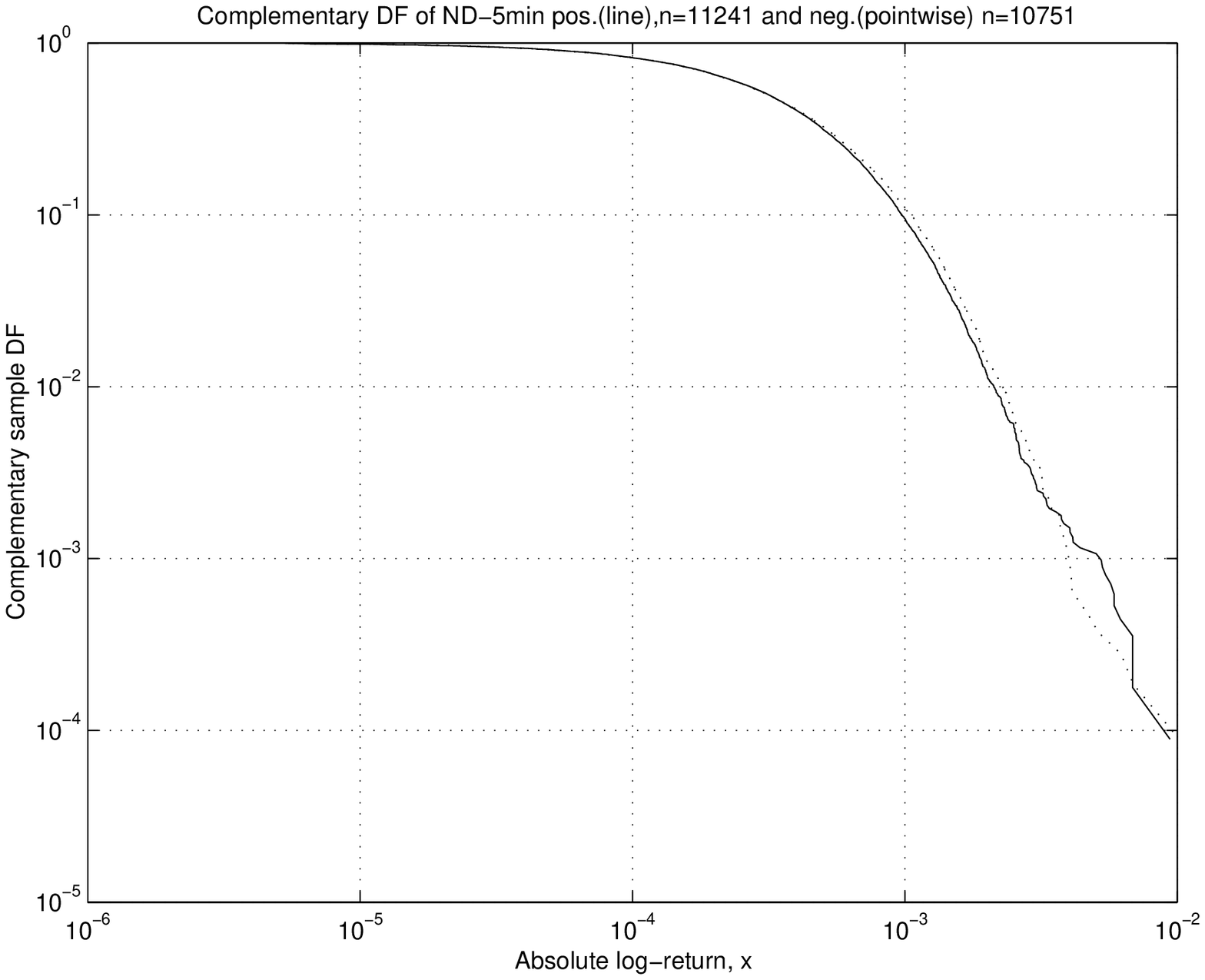}
\end{center}
\caption{\label{fig1ab} Cumulative sample distributions for the
Dow Jones (a) and for the Nasdaq (b) data sets.
      }
\end{figure}

Figure \ref{fig1ab}a shows the cumulative sample distribution
function $1- F(x)$
for the Dow Jones Industrial Average index, and in figure
\ref{fig1ab}b the cumulative sample
distribution function for the Nasdaq Composite index.
The mismatch between the Pareto distribution and the data can be
seen with the naked eye: if samples were taken from a Pareto population, the
graph in double log-scale should be a straight line. Even in the tails, this
is doubtful. To formalize this impression, we calculate the Hill and
AD estimators
for each threshold $u$. Denoting $y_{1}\geqslant \ldots \geqslant
y_{n_u}$ the ordered sub-sample of values exceeding $u$ where $N_{u}$
is the size of
this sub-sample, the Hill maximum likelihood estimate of parameter
$b$ is \cite{Hill75}
\be
\label{eq:Hill}
\hat {b}_{u} =\left[\frac{1}{N_{u}}~\sum\limits_1^{N_u} {\log (y_k / u)}
\right]^{-1}~.
\ee
The standard deviations of $\hat {b}_{u}$ can be estimated as
\be
{\rm Std}(\hat {b}_{u}) = \hat {b}_{u} /\sqrt {N_u},
\ee
under the assumption of iid data, but very severely underestimate the true
standard deviation when samples exhibit dependence, as reported by
\citeasnoun{KP97}.

\begin{figure}
\begin{center}
\includegraphics[width=12cm]{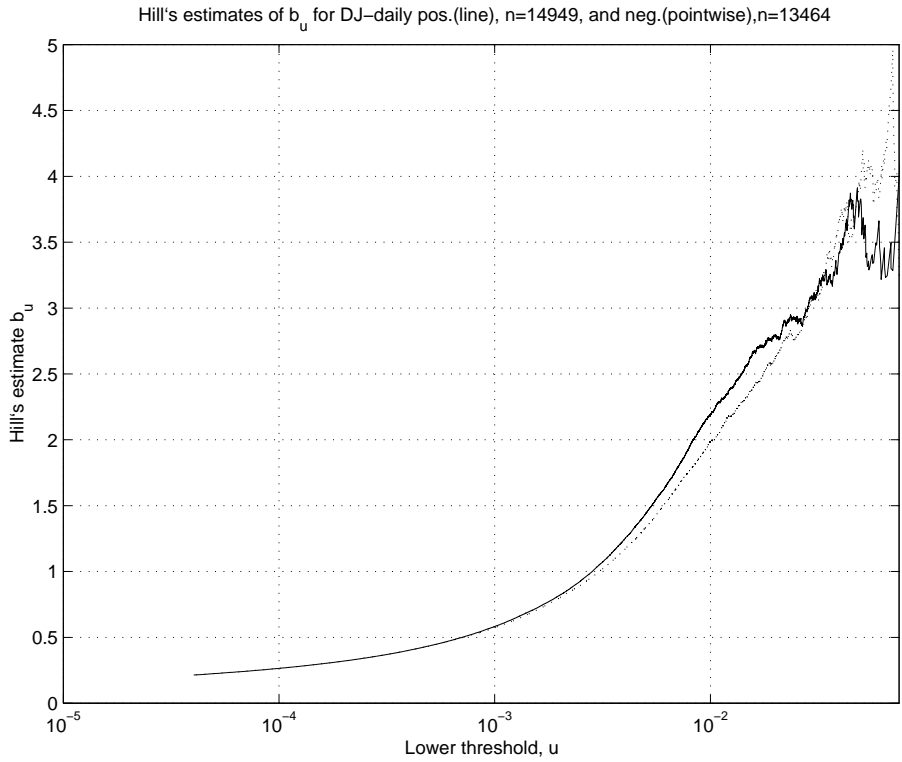}
\includegraphics[width=12cm]{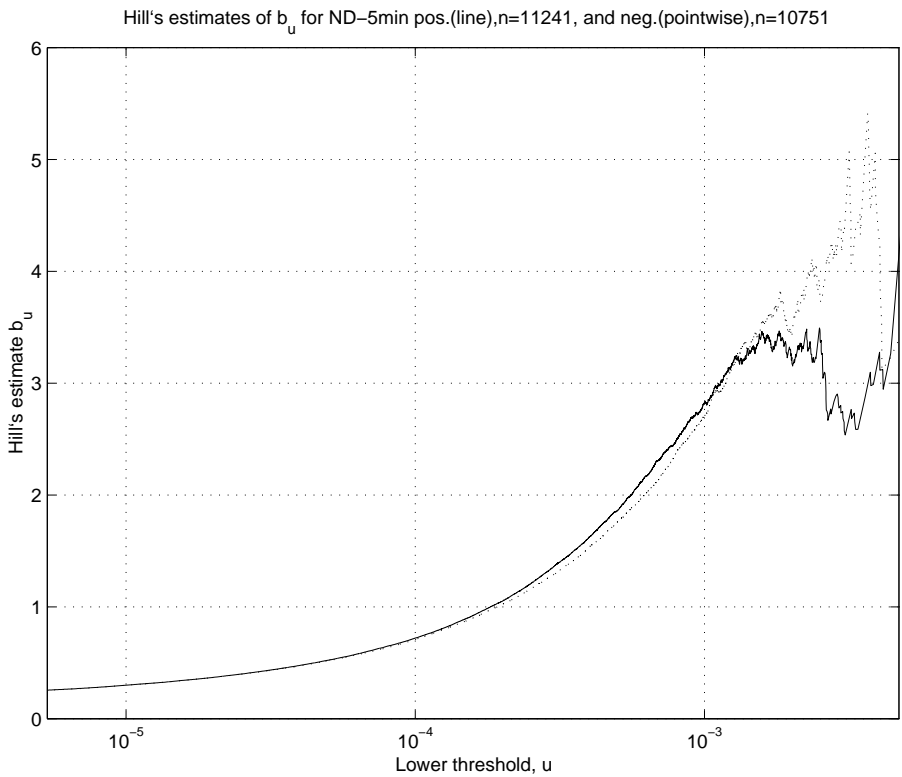}
\end{center}
\caption{\label{FIG2Apis}  Hill estimates $\hat {b}_{u}$ as a function of
the threshold $u$ for the Dow Jones (a) and for the Nasdaq (b).
      }
\end{figure}

Figure \ref{FIG2Apis}a and \ref{FIG2Apis}b
shows the Hill estimates $\hat {b}_{u}$ as a function of $u$ for
the Dow Jones and for the Nasdaq. Instead of
an approximately constant exponent (as would be the case for
true Pareto samples), the tail index estimator increases
until $u \cong 0.04$, beyond which it seems to slow its growth
and oscillates
around a value $\approx 3-4$ up to the threshold $u \cong .08$. It should be
noted that the interval $[0, 0.04]$ contains $99.12\%$ of the sample whereas
the interval $[0.04, 0.08]$ contains only $0.64\%$ of the sample. The behavior
of $\hat {b}_{u}$ for the ND shown in figure \ref{FIG2Apis}b is similar: Hill's
estimate $\hat {b}_{u}$ seems to slow its growth already at $u \cong 0.0013$
corresponding to the $95\%$ quantile.
Are these slowdowns of the growth
of $\hat {b}_{u}$ genuine signatures of a possible constant well-defined
asymptotic value that would qualify a regularly varying function?

As a first answer to this question,
table~\ref{table3} compares the AD-estimates of the tail exponent $b$ with
the corresponding
maximum likelihood estimates for the 18 intervals $u_{1} \ldots u_{18}$.
Both maximum
likelihood and Anderson-Darling estimates of $b$ steadily increase with the
threshold $u$ (except for the highest quantiles of the positive tail of the
Nasdaq). The corresponding figures for positive and negative returns are very
close to each other and almost never significantly different at the
usual $95\%$
confidence level. Some slight non-monotonicity of the increase for the highest
thresholds can be explained by small sample sizes. One can observe
that both MLE
and ADS estimates continue increasing as the interval of estimation is
contracting to the extreme values. It seems that their growth potential has not
been exhausted even for the largest quantile $u_{18}$, except for the positive
tail of the Nasdaq sample. This statement might not be very strong
as the standard deviations of the tail index estimators also grow
when exploring
the largest quantiles. However, the non-exhausted growth is observed for three
samples out of the four tails. Moreover, this effect is seen for
several threshold values
while random fluctuations would distort the $b$-curve in a random
manner rather than according to the increasing trend observed in three out
of four tails.

\begin{figure}
\begin{center}
\includegraphics[width=12cm]{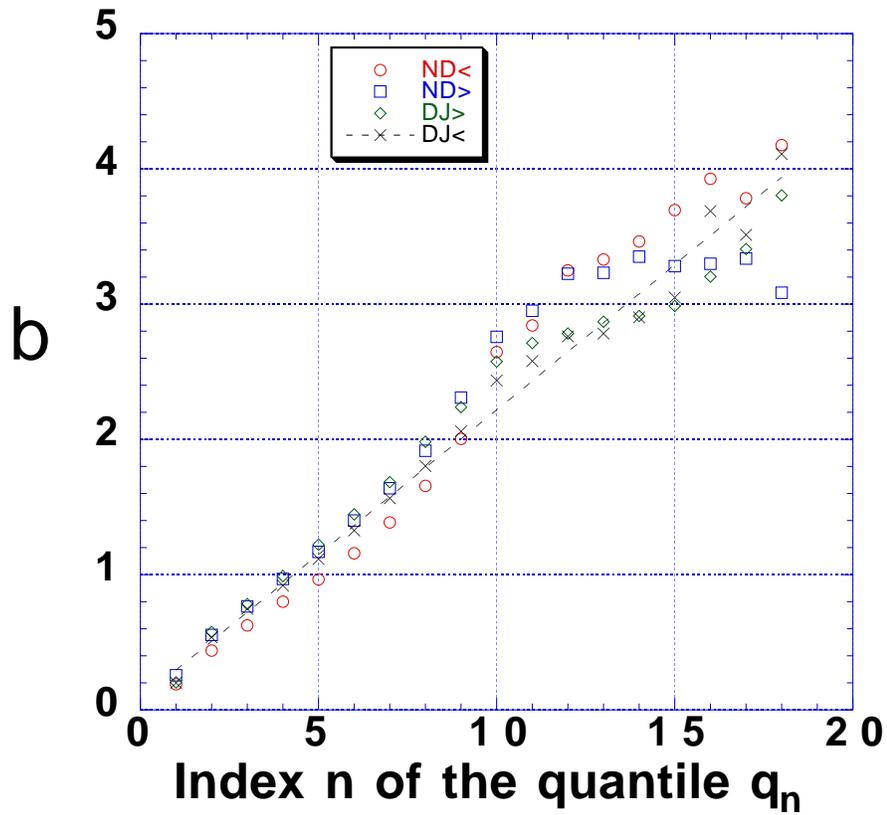}
\end{center}
\caption{\label{FigbofQUANTILE} Hill estimator $\hat {b}_{u}$
for all four data sets (positive and negative branches of the distribution
of returns for the DJ and for the ND) as a function of the index $n=1, ..., 18$
of the $18$ quantiles or standard significance levels
$q_{1}\ldots q_{18}$ given in table~\ref{table1}. The dashed line is
expression (\ref{bpredict}) with
$1- q_n = 3.08 ~e^{-0.342 n}$ given by (\ref{mgjer}).
}
\end{figure}

Assuming that the observation, that the sample distribution can
be approximated by a Pareto distribution with a growing index $b$,
is correct, an important
question arises: how far beyond the sample this growth will continue? Judging
from table~\ref{table3}, we can think this growth is still not exhausted.
Figure \ref{FigbofQUANTILE}
suggests a specific form of this growth, by plotting the hill
estimator $\hat {b}_{u}$
for all four data sets (positive and negative branches of the distribution
of returns for the DJ and for the ND) as a function of the index $n=1, ..., 18$
of the $18$ quantiles or standard significance levels
$q_{1}\ldots q_{18}$ given in table~\ref{table1}. Similar results
are obtained with the AD estimates. Apart from the positive
branch of the ND data set, all other three branches suggest a continuous
growth of the Hill estimator $\hat {b}_{u}$ as a function of
$n =1 ,..., 18$. Since the quantiles $q_{1}\ldots q_{18}$ given in
table~\ref{table1}
have been chosen to converge to $1$ approximately exponentially as
\be
1- q_n = 3.08 ~e^{-0.342 n}~,
\label{mgjer}
\ee
the linear fit of $\hat {b}_{u}$ as a function of $n$ shown as the dashed line
in figure \ref{FigbofQUANTILE} corresponds to
\be
\hat {b}_{u}(q_n) = 0.08 + 0.626 \ln \frac{3.08}{1- q_n}~.
\label{bpredict}
\ee
Expression (\ref{bpredict}) suggests
an unbound logarithmic growth of $\hat {b}_{u}$ as the quantile approaches $1$.
For instance, for a quantile $1-q=0.1\%$, expression (\ref{bpredict}) predicts
$\hat {b}_{u}(1-q=10^{-3}) = 5.1$. For a quantile $1-q=0.01\%$,
expression (\ref{bpredict}) predicts
$\hat {b}_{u}(1-q=10^{-4}) = 6.5$, and so on. Each time the quantile
$1-q$ is divided
by a factor $10$, the apparent exponent $\hat {b}_{u}(q)$ is increased by the
additive constant $\cong 1.45$: $\hat {b}_{u}((1-q)/10)= \hat
{b}_{u}(1-q) + 1.45$.
This very slow growth uncovered here may be an explanation for the
belief and possibly mistaken conclusion
that the Hill and other estimators of the tail index tends to a constant
for high quantiles. Indeed, it is now clear that the slowdowns of the growth
of $\hat {b}_{u}$ seen in
figures \ref{FIG2Apis} decorated by large fluctuations due
to small size effects is mostly the result of a
dilatation of the data expressed in terms of threshold $u$. When recast in the
more natural logarithm scale of the quantiles $q_{1}\ldots q_{18}$, this
slowdown disappears. Of course, it is impossible to know how long this growth
given by (\ref{bpredict}) may go on as the quantile $q$ tends to~$1$.
In other words, how can we escape from the sample range when
estimating quantiles? How can we estimate the so-called ``high
quantiles'' at the level $q > 1-1/T$ where $T$ is the total
number of sampled points. \citeasnoun{EKM97} have summarized the
situation in this way: ``there is no free lunch when
it comes to high quantiles estimation!'' It is possible
that $\hat {b}_{u}(q)$ will grow without limit as would be the case if
the {\it true} underlying distribution was rapidly varying.
Alternatively, $\hat {b}_{u}(q)$ may saturate to a large value, as
predicted for
instance by the traditional GARCH model which yields tails indices which can
reach $10-20$ \cite{EP01,SP99} or by the recent multifractal random walk (MRW)
model which gives an asymptotic tail exponent in the range $20-50$
\cite{Muzyetal,musor}. According to (\ref{bpredict}), a value $\hat {b}_{u}
\approx 20$ (respectively $50$) would be attained for $1-q \approx 10^{-13}$
(respectively $1-q \approx 10^{-34}$)! If one believes in the prediction of the
MRW model, the tail of the distribution of returns is regularly
varying but this
insight is completely useless for all practical purposes due to the
astronomically high statistics that would be needed to sample this regime. In
this context, we cannot hope to get access to the true nature of the pdf of
returns but only strive to define the best effective or apparent most
parsimonious and robust model. By comparing distributions
of aggregated returns with their corresponding reshuffled counterparts,
\citeasnoun{Viswanathan} suggest that the
fat tail nature of the returns result mainly from the existence of
long-range dependence, in agreement with the construction of GARCH and MRW
processes.

\subsubsection{Weibull distributions}

Let us now fit our data with the Weibull (SE) distribution
(\ref{eq:Weibull}). The Anderson-Darling statistics (ADS) for this
case are shown in
table~\ref{table2}. The ML-estimates and
AD-estimates of the form parameter $c$ are represented in table~\ref{table5}.
Table~\ref{table2} shows that, for the highest quantiles, the ADS for the
Stretched-Exponential is the smallest of all ADS, suggesting that the SE is the
best model of all.
Moreover, for the lowest quantiles, it is the sole model not systematically
rejected at the $95\%$ level.

The $c$-estimates are found to decrease when increasing
the order $q$ of the threshold $u_{q}$ beyond which the estimations
are performed. In addition, the $c$-estimate
is identically zero for $u_{18}$. However, this does not automatically imply
that the SE model is not the correct model for the data even for these highest
quantiles. Indeed, numerical simulations show that, even for
synthetic samples drawn from genuine
Stretched-Exponential distributions with exponent $c$ smaller than $0.5$
and whose size is comparable with that of our data, in about one case out
of three (depending on the exact value of $c$) the estimated value of $c$
is zero. This {\it a priori} surprising result
comes from condition (\ref{eq:condSE}) in appendix~\ref{app:MLE}
which is not fulfilled with certainty even for samples drawn for SE
distributions.

Notwithstanding this cautionary remark,
note that the $c$-estimate of the positive tail of the Nasdaq data equal zero
for all quantiles higher than $q_{14}=0.97\%$. In fact, in every
cases, the estimated $c$ is not significantly different from zero - at the
$95\%$ significance level - for quantiles higher than $q_{12}$-$q_{14}$.
In addition, table~\ref{table6} gives the values of the
estimated scale parameter $d$, which are found very small - particularly for
the Nasdaq - beyond $q_{12}=95\%$. In contrast, the Dow Jones keeps
significant scale factors until $q_{16} - q_{17}$.

These evidences taken all together provide
a clear indication on the existence of a change of behavior of the true pdf
of these four distributions: while the bulks of the distributions seem rather
well approximated by a SE model, a fatter tailed
distribution than that of the (SE) model is required for the highest quantiles.
Actually, the fact that both $c$ and $d$ are extremely small may be interpreted
according to
the asymptotic correspondence given by (\ref{jgjlke}) and (\ref{jgjkwlw})
as the existence of a possible power law tail.

\subsubsection{Exponential and incomplete Gamma distributions}

Let us now fit our data with the exponential distribution
(\ref{eq:Exponential}). The average ADS for this case are shown in
table~\ref{table2}. The maximum likelihood- and
Anderson-Darling estimates of the scale parameter $d$ are given in
table~\ref{table4}. Note that they always decrease as the
threshold $u_q$ increases. Comparing the mean ADS-values of table~\ref{table2}
with the standard AD quantiles, we can conclude that, on the whole, the
exponential distribution (\textit{even with moving scale parameter d)} does not
fit our data: this model is systematically rejected at the $95\%$ confidence
level for the lowest and highest quantiles - excepted for the negative tail of
the Nasdaq.

Finally, we fit our data by the IG-distribution (\ref{eq:Gamma}). The mean
ADS for this class of functions are shown in table~\ref{table2}.
The Maximum likelihood and Anderson Darling estimates of the power
index $b$ are
represented in table~\ref{tableXX}. Comparing the mean ADS-values
of table~\ref{table2} with the standard AD quantiles, we can again
conclude that, on
the whole, the IG-distribution does not fit our data. The model is rejected at
the $95\%$ confidence level excepted for the negative tail of the Nasdaq for
which it is not rejected marginally (significance level: $94.13\%$). However,
for the largest quantiles, this model becomes again relevant since it cannot be
rejected at the $95\%$ level.

\subsubsection{Log-Weibull distributions}

The parameters $b$ and $c$ of the log-Weibull defined by
(\ref{mjgtkwl}) are estimated
with both the Maximum Likelihood and Anderson-Darling methods
for the $18$ standard significance levels
$q_{1}\ldots q_{18}$ given in table~\ref{table1}. The results of these
estimations are given in table~\ref{tableSL}. For both positive and
negative tails of the Dow Jones, we find very stable results for all
quantiles lower than $q_{10}$: $c=1.09 \pm 0.02$ and
$b=2.71 \pm 0.07$. These results reject the Pareto distribution
degeneracy $c=1$
at the $95\%$ confidence level. Only for the quantiles higher than or equal to
$q_{16}$, we find an estimated value $c$ compatible with the Pareto
distribution. Moreover both for the positive and negative Dow Jones tails, we
find that $c \approx 0.92 $ and $b \approx 3.6-3.8$, suggesting a possible
change of regime or a sensitivity to ``outliers'' or a lack of
robustness due to
the small sample size. For the positive Nasdaq tail,  the exponent $c$ is found
compatible with $c=1$ (the Pareto value), at the $95\%$ significance level,
above $q_{11}$ while $b$ remains almost stable at $b \simeq 3.2$. For
the negative Nasdaq tail, we find that $c$ decreases almost systematically from
$1.1$ for $q_{10}$ to $1$ for $q_{18}$ for both estimators
while $b$ regularly increases from about $3.1$ to about $4.2$. The
Anderson-Darling distances are not worse but not significantly better than for
the SE and this statistics cannot be used to conclude neither in favor of nor
against the log-Weibull class.

\subsection{Summary}

At this stage, two conclusions can be drawn. First, it appears that none of
the considered distributions fit the data over the entire range, which is not a
surprise. Second, for the highest quantiles, four models seem to be
able to represent to data, the Gamma model, the Pareto model, the
Stretched-Exponential model and the log-Weibull model. The two last ones have
the lowest Anderson-Darling statistics and thus seems to be the most reasonable
models among the four models compatible with the data. For all the samples,
their Anderson-Darling statistic remain so close to each other
for the quantiles higher than
$q_{10}$ that the descriptive power of these two models cannot be
distinguished.

\section{Comparison of the descriptive power of the different families}

As we have seen by comparing the Anderson-Darling statistics corresponding
to the five parametric families (\ref{eq:Pareto}-\ref{eq:Gamma})
and~(\ref{eq:LW}), the best models in the sense of minimizing the
Anderson-Darling distance are the Stretched-Exponential and the Log-Weibull
distributions.

We now compare the four distributions (\ref{eq:Pareto}-\ref{eq:Gamma})
with the comprehensive distribution (\ref{eq:CDF}) using Wilks' theorem
\cite{Wilks38} of nested hypotheses to check whether or not some
of the four distributions are sufficient compared with the comprehensive
distribution to describe the data. It will appear that the Pareto and the
Stretched-Exponential models are the most parsimonious.
We then turn to a direct comparison of the best two parameter models
(the SE and
log-Weibull models) with the best one parameter model (the Pareto
model), which will
require an extension of Wilks' theorem derived in Appendix~\ref{app:NN} that
will allow us to directly test the SE model against the Pareto model.

\subsection{Comparison between the four parametric
families~(\ref{eq:Pareto}-\ref{eq:Gamma}) and the comprehensive
distribution~(\ref{eq:CDF})}

According to Wilks' theorem, the doubled generalized log-likelihood
ratio $\Lambda $:
\be
\label{eq:LogLike}
\Lambda = 2~ \log \frac{\max {\cal L} (CD,X,\Theta)}{ \max {\cal L}(
       z ,X,\theta )} ,
\ee
has asymptotically (as the size $N$ of the sample $X$ tends to
infinity) the $\chi^{2}$-distribution. Here ${\cal L}$ denotes the
likelihood function, $\theta$ and $\Theta$ are parametric spaces
corresponding to hypotheses $z$ and $CD$ correspondingly
(hypothesis $z$ is one of the four hypotheses
(\ref{eq:Pareto}-\ref{eq:Gamma}) that are particular cases of the $CD$
under some parameter relations). The statement of the theorem is valid
under the condition that the \textit{sample $X$ obeys hypothesis $z$ for some
      particular value of its parameter belonging to the space $\theta$.}
The number of degrees of freedom of the $\chi^{2}$-distribution equals to
the difference of the dimensions of the two spaces $\Theta$ and
$\theta$. We have
$\dim(\Theta) = 3, \dim(\theta) = 2$ for the Stretched-Exponential and
for the Incomplete Gamma distributions while $\dim(\theta) = 1$ for
the Pareto and
the Exponential
distributions. This corresponds to one degree of freedom for the two former
cases and two degrees of
freedom for the later pdf's. The maximum of the likelihood in the numerator of
(\ref{eq:LogLike}) is taken over the space $\Theta$, whereas the maximum of the
likelihood in the denominator of (\ref{eq:LogLike}) is taken over the
space $\theta$.
Since we have always $\theta \subset\Theta $, the likelihood ratio is
always larger
than $1$, and the log-likelihood ratio is non-negative. If the
observed value of
$\Lambda$ does not exceed some high-confidence level (say, $99\%$ confidence
level) of the $\chi^{2}$, we then reject the hypothesis CD in favor of
the hypothesis $z$, considering the space $\Theta$ redundant. Otherwise, we
accept the hypothesis CD, considering the space $\theta$ insufficient.

\begin{figure}
\begin{center}
\includegraphics[width=12cm]{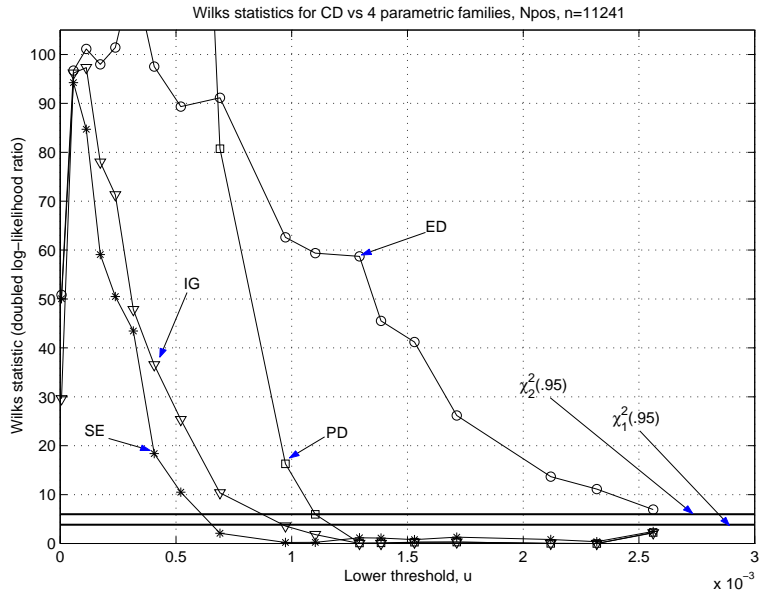}
\includegraphics[width=12cm]{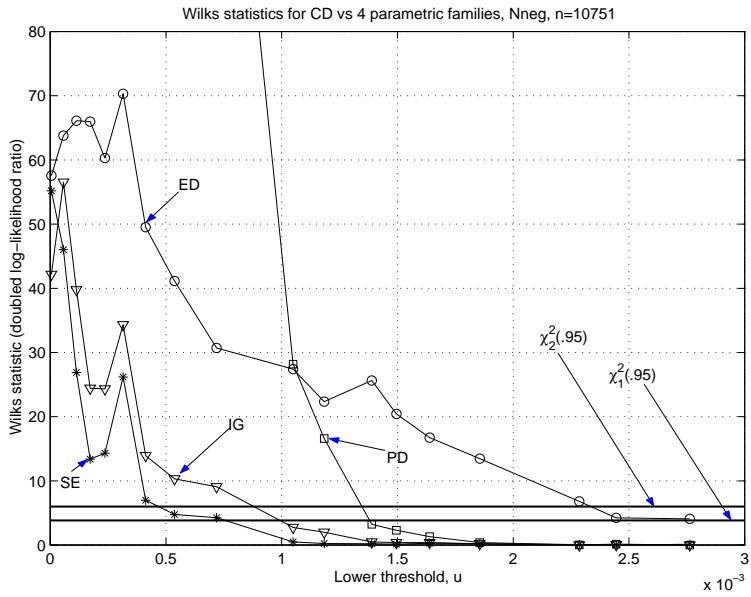}
\end{center}
\caption{\label{fig6a-b} Wilks statistic for the comprehensive distribution
versus the four parametric distributions : Pareto (PD), Weibull (SE),
Exponential (ED) and Incomplete Gamma (IG) for the Nasdaq five minutes
returns. The upper panel refers to the positive returns and lower panel to
the negative ones.
}
\end{figure}

\begin{figure}
\begin{center}
\includegraphics[width=12cm]{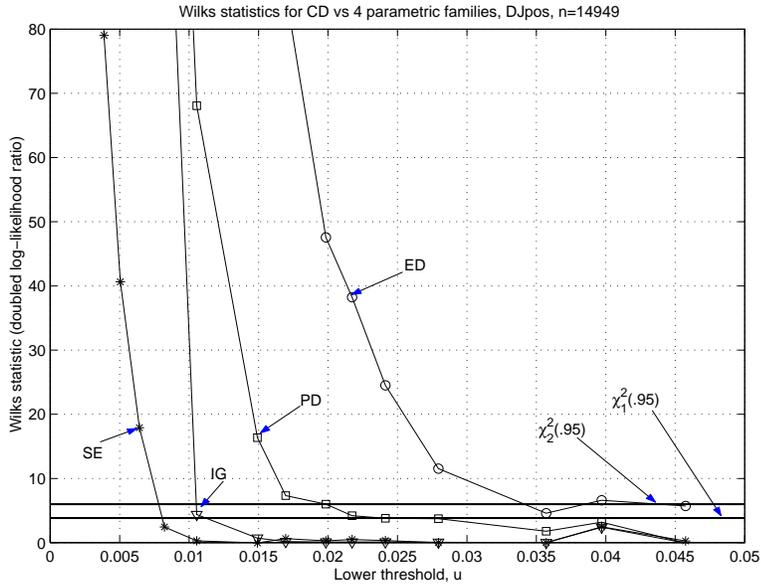}
\includegraphics[width=12cm]{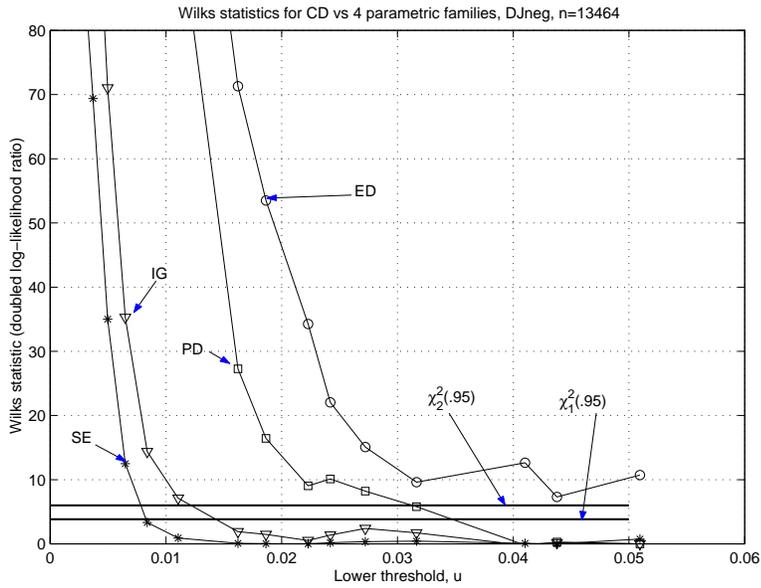}
\end{center}
\caption{\label{fig6c-d} Wilks statistic for the comprehensive distribution
versus the four parametric distributions : Pareto (PD), Weibull (SE),
Exponential (ED) and Incomplete Gamma (IG) for the Dow Jones daily
returns. The upper panel refers to the positive returns and the lower panel to
the negative ones.
}
\end{figure}

The doubled log-likelihood ratios (\ref{eq:LogLike}) are shown in
figures~\ref{fig6a-b}
for the positive and negative branches of the distribution of returns
of the Nasdaq
and in figures \ref{fig6c-d} for the Dow Jones.
The $95\%$ $\chi^{2}$ confidence levels for $1$ and
$2$ degrees of freedom are given by the horizontal lines.

For the Nasdaq data, figure \ref{fig6a-b} clearly shows that
Exponential distribution is completely insufficient: for all
lower thresholds, the Wilks log-likelihood ratio exceeds the $95\%$
$\chi_{1}^{2}$ level $3.84$. The Pareto distribution is insufficient for
thresholds $u_{1 } - u_{11}$ ($92.5\%$ of the ordered sample) and becomes
comparable with the Comprehensive distribution in the tail $u_{12} - u_{18}$
($7.5\%$ of the tail probability). It is natural that two-parametric families
Incomplete Gamma and Stretched-Exponential have higher goodness-of-fit than the
one-parametric Exponential and Pareto distributions. The Incomplete
Gamma distribution is
comparable with the Comprehensive distribution starting with $u_{10}$ ($90\%$),
whereas the Stretched-Exponential is somewhat better ($u_{9}$ or
$u_{8}$ , i.e.,
$70\%$). For the tails representing $7.5\%$ of the data, all
parametric families except for
the Exponential distribution fit the sample distribution with almost the same
efficiency.
The results obtained for the Dow Jones data shown in figure~\ref{fig6c-d}
are similar. The Stretched-Exponential is comparable with the Comprehensive
distribution starting with $u_{8}$ ($70\%$). On the whole, one can say that the
Stretched-Exponential distribution performs better than the three
other parametric
families.

We should stress that each log-likelihood ratio represented in
figures~\ref{fig6a-b} and~\ref{fig6c-d}, so-to say ``acts on its own ground,''
that is, the corresponding $\chi^{2}$-distribution is valid \textit{under
the assumption of the validity of each particular hypothesis whose
likelihood stands in
the numerator of the double log-likelihood (\ref{eq:LogLike}).}
It would be desirable
to compare all combinations of pairs of hypotheses directly, in
addition to comparing
each of them with the comprehensive distribution. Unfortunately, the Wilks
theorem can not be used in the case of pair-wise comparison because the problem
is not more that of comparing nested hypothesis (that is, one hypothesis is
a particular case of the comprehensive model). As a consequence,
our results on the comparison of the relative merits of each of the
four distributions
using the generalized log-likelihood ratio should be interpreted with
a care, in
particular, in a case of contradictory conclusions. Fortunately, the
main conclusion of the comparison (an advantage of the Stretched-Exponential
distribution over the three other distribution) does not contradict our earlier
results discussed above.

\subsection{Pair-wise comparison of the Pareto model with
the Stretched-Exponential and Log-Weibull models \label{wilsmmfl}}

We now want to compare formally the descriptive power of the
Stretched-Exponential
distribution and the Log-Weibull distribution (the two best two-parameter
models) with that of the Pareto distribution (the best one-parameter
model). For the comparison of the Log-Weibull model versus the Pareto model,
Wilks' theorem can still be applied since the Log-Weibull distribution
encompasses the Pareto distribution. {\it A contrario}, the comparison of the
Stretched-Exponential versus the Pareto distribution should in
principle require
that we use the methods for testing non-nested hypotheses \cite{GM94}, such as
the Wald encompassing test or the Bayes factors \cite{Kass}.
Indeed, the Pareto model and the
(SE) model are not, strictly speaking, nested. However, as exposed in section
\ref{sec:pdf}, the Pareto distribution is a limit case of the
Stretched-Exponential distribution, as the fractional exponent $c$
goes to zero.
Changing the parametric representation of the (SE) model into
\be
f(x|b,c)=    b~ u^c~ x^{c-1} \exp \left[ -\frac{b}{c} \left( \left(\frac{x}{u}
\right)^c -1 \right) \right], \quad x>u,
\ee
i.e., setting $b= c\cdot \left(\frac{u}{d} \right)^c$, where the parameter $d$
refers to the former (SE) representation (\ref{eq:Weibull}), we show in
Appendix~\ref{app:NN} that the doubled log-likelihood ratio
\be
W= 2 \log \frac{\max_{b,c} {\cal L}_{SE}}{\max_{b}{\cal L}_{PD}}
\ee
still follows
Wilks' statistic, namely is asymptotically distributed according to a
$\chi^2$-distribution, with one degree of freedom in the present case.
Thus, even in this case of non-nested hypotheses, Wilks' statistic still allows
us to test the null hypothesis $H_0$ according to which the Pareto model is
sufficient to describe the data.

The results of these tests are given in tables~\ref{table:ParSE}
and~\ref{table:ParSWeib}. The $p$-value (figures within
parentheses) gives the significance with which one can reject the null
hypothesis $H_0$ that the Pareto distribution is sufficient to accurately
describe the data. Table~\ref{table:ParSE} compares the
Stretched-Exponential with Pareto distribution.  $H_0$ is found to be
more often rejected for the Dow Jones than for the Nasdaq. Indeed, beyond
quantile $q_{12}=95\%$, $H_0$ cannot be rejected at the $95\%$ confidence level
for the Nasdaq data. For the Dow Jones, we must consider quantiles higher than
$q_{16}=99\%$ -at least for the negative tail- in order not to reject $H_0$ at
the $95\%$ significance level. These results are in qualitative
agreement with what we could expect from the action of the central
limit theorem: the power-law regime (if it really exists) is pushed back to
higher quantiles due to time aggregation (recall that the Dow Jones data is at
the daily scale while the Nasdaq data is at the $5$ minutes time scale).

Table \ref{table:ParSWeib} shows Wilks' test for the Pareto
distribution versus the log-Weibull distribution.  For quantiles
above $q_{12}$,
the Wilks' statistic is mostly insignificant, that is, the Pareto distribution
cannot be rejected in favor of of the Log-Weibull. This parallels the lack of
rejection of the Pareto distribution  against the Stretched-Exponential beyond
the significance level $q_{12}$.

In summary, Stretched-Exponential and Log-Weibull models encompass the Pareto
model as soon as one considers quantiles higher than $q_6 = 50\%$. The null
hypothesis that the true distribution is the Pareto distribution is
strongly rejected until quantiles $90\%-95\%$ or so. Thus, within this range,
the (SE) and (SLE) models seem the best and
the Pareto model is insufficient to describe the data. But,
for the very highest quantiles (above
$95\%-98\%$), we cannot reject any more the hypothesis that the Pareto model is
sufficient compared with the (SE) and (SLE) model. These two parameter models
can then be seen as a redundant parameterization for the extremes compared with
the Pareto distribution.

\section{Discussion and Conclusions}

\subsection{Is there a best model of tails?}

We have presented a statistical analysis of the tail behavior of the
distributions of the daily log-returns of
the Dow Jones Industrial Average and of the $5$-minutes log-returns of the
Nasdaq Composite index. We have emphasized
practical aspects of the application of
statistical methods to this problem.
Although the application of statistical methods to the study of empirical
distributions of returns seems to be an obvious approach, it
is necessary to keep in mind the existence of necessary conditions
that the empirical data must obey for the conclusions of the
statistical study to be valid. Maybe the most important condition
in order to speak meaningfully about distribution functions
is the stationarity of the data, a difficult issue that we have
barely touched upon
here.  In particular, the importance of regime switching is
now well established \cite{RS98,AB01} and its possible
role should be accounted for.

Our purpose here has been to revisit
a generally accepted fact that the tails of the distributions of returns
present a power-like behavior. Although there are some disagreements concerning
the exact value of the power
indices (the majority of previous workers accepts index values
between $3$ and $3.5$,
depending on the particular asset and the investigated time
interval), the power-like
character of the tails of distributions of returns
is not subjected to doubts. Often, the conviction of the existence of
a power-like tail is based on the Gnedenko theorem stating
the existence of only three possible types
of limit distributions of normalized maxima (a finite maximum value, an
exponential tail, and a power-like tail) together with the exclusion of the
first two types by experimental evidence. The power-like character of
the log-return tail $\bar {F}(x)$ follows then simply from the
power-like distribution
of maxima. However, in this chain of arguments, the conditions needed for the
fulfillment of the corresponding mathematical theorems are often
omitted and not
discussed properly. In addition, widely used arguments in favor of
power law tails
invoke the \textit{self-similarity} of the data but
are often \textit{assumptions} rather than
experimental evidence or consequences of economic and financial laws.

Here, we have shown that standard statistical estimators
of heavy tails are much less efficient that often assumed and cannot in
general clearly distinguish between a power law tail and a Stretched
Exponential tail
{even in the absence} of long-range dependence in the volatility.
In fact, this can be rationalized
by our discovery that, in a certain limit where the
exponent $c$ of the stretched exponential pdf goes to zero (together
with condition (\ref{jgjlke}) as seen in the derivation (\ref{jgjkwlw})),
the stretched exponential pdf tends to the Pareto distribution. Thus,
the Pareto
(or power law) distribution can be approximated with any desired accuracy on
an arbitrary interval by a suitable adjustment of the pair $(c,d)$ of the
parameters of the stretched exponential pdf. We have then
turned to parametric tests which indicate that the class of
Stretched Exponential and log-Weibull distributions provide a significantly
better fit to empirical returns than the Pareto, the exponential or
the incomplete Gamma distributions. All our tests are consistent
with the conclusion that these two model provide the best effective
apparent and parsimonious models to account for the empirical data on the
largest possible range of returns.

However, this does not mean that the stretched exponential (SE) or the
log-Weibull model is the correct
description of the tails of empirical distributions of returns. Again, as
already mentioned, the strength of these models come from the fact that
they encompass the Pareto model in the tail and offers a better description
in the bulk of the distribution. To see where the problem arises,
we report in table \ref{sumse} our best ML-estimates for the SE
parameters $c$ (form parameter) and $d$ (scale
parameter) restricted to the quantile level $q_{12}=95\%$, which
offers a good compromise between a sufficiently
large sample size and a restricted tail range leading to an accurate
approximation in this range.

One can see that $c$ is very small (and all the more so
for the scale parameter $d$) for the tail of positive returns of the
Nasdaq data
suggesting a convergence to a power law tail. The exponents $c$ for the three
other tails are an order of magnitude larger but our tests show that they are
not incompatible with an asymptotic power tail either. Indeed, we have shown
in section \ref{wilsmmfl}
that, for the very highest quantiles (above
$95\%-98\%$), we cannot reject the hypothesis that the Pareto model is
sufficient compared with the (SE) model.

Note also that the exponents $c$ seem larger for the daily DJ data than for
the $5$-minutes ND data, in agreement with an expected (slow)
convergence to the Gaussian law according to the central limit theory
\footnote{see \citeasnoun{SSA00} and figures 3.6-3.8 pp. 68 of
\citeasnoun{Sorbook} where
it is shown that SE distributions are approximately stable in family and the
effect of aggregation can be seen to slowly increase the exponent
$c$.  See also
\citeasnoun{Drozdzconv} which studies specifically this convergence
to a Gaussian law
as a function of the time scale level.}.
However, a $t$-test does not allow us to reject the hypotheses that the
exponents $c$ remains the same for a given tail (positive or
negative) of the Dow Jones data.
Thus, we confirm previous results \cite[for
instance]{Lux96,JR01} according to which the extreme tails can be considered as
symmetric, at least for the Dow Jones data. In contrast, we find a very
strong asymmetry for the $5$-minute sampled Nasdaq data.

These are the evidence in favor of the existence of an asymptotic
power law tail.
Balancing this, many of our tests have shown that the power law model is not as
powerful compared with the SE and SLE models, even arbitrarily far in the tail
(as far as the available data allows us to probe). In addition, our
attempts for
a direct estimation of the exponent $b$ of a possible power law tail has
failed to confirm the existence of a well-converged asymptotic value (except
maybe for the positive tail of the Nasdaq). In contrast, we have found
that the exponent $b$ of the power law model systematically increases when
going deeper and deeper in the tails, with no visible sign of exhausting this
growth. We have proposed a parameterization of this growth of the
apparent power law exponent. We note again that this behavior is expected
from models such as the GARCH or the Multifractal Random Walk models which
predict asymptotic power law tails but with exponents of the order of $20$ or
larger, that would be sampled at unattainable quantiles.

Attempting to wrap up the different results obtained by the battery of tests
presented here, we can offer the following
conservative conclusion: it seems that the four tails examined here
are decaying faster than any (reasonable) power law
but slower than any stretched exponentials. Maybe log-normal or log-Weibull
distributions could offer a better effective description of the distribution
of returns\footnote{Let us stress that we are speaking of a log-normal
distribution of returns, not of price! Indeed, the standard Black and Scholes
model of a log-normal distribution of prices is equivalent to a Gaussian
distribution of returns. Thus, a log-normal distribution of returns
is much more fat tailed, and in fact bracketed by power law tails and
stretched exponential tails.}. Such a model has already been suggested by
\cite{lognoserva}.

In sum, the PD is sufficient above quantiles $q_{12}=95\%$
but is not stable enough to ascertain with strong confidence
a power law asymptotic nature of the pdf. Other studies
using much larger database of up to tens of millions
of data points \cite{GMAS98,GMAS98_2,Plerouetal99,matia,mizumo} seem to confirm
an asymptotic power law with exponent close to $3$ but
the effect of aggregation of
returns from different assets may distort the information and the very
structure of the tails of pdf
if they exhibit some intrinsic variability \cite{matia}.

\subsection{Implications for risk assessment}

The correct description of the
distribution of returns has important implications for the assessment
of large risks not
yet sampled by historical time series. Indeed, the whole purpose
of a characterization of the functional form of the distribution of
returns is to extrapolate currently available historical time series
beyond the range provided by the empirical reconstruction of the
distributions.
For risk management, the determination of the tail of the distribution
is crucial. Indeed, many risk measures,
such as the Value-at-Risk or the
Expected-Shortfall, are based on the properties of the tail of the
distributions of returns. In order to assess risk at
probability levels of $95\%$ or more, non-parametric methods have merits.
However, in order to estimate risks at high probability level such as
$99\%$ or larger,
non-parametric estimations fail by lack of data and parametric models become
unavoidable. This shift in strategy has a cost and replaces sampling errors
by model errors. The considered distribution can be too
thin-tailed as when using normal laws, and risk will be underestimated,
or it is too fat-tailed and risk
will be over estimated as with L\'evy law and possibly with Pareto tails
according to the present study.
In each case, large amounts of money are at stake and
can be lost due to a too conservative or too optimistic risk measurement.

In order to bypass these problems, some authors
\cite[among many others]{Bali03,Longin00,MF00} have proposed to
estimate the extreme quantiles of the distributions in a semi-parametric way,
which allows one (i) to avoid the model errors and (ii) to limit the 
sampling errors
with respect to non-parametric methods and thus to keep a reasonable 
accuracy in
the estimation procedure. To this aim, it has been suggested to use the extreme
value theory\footnote{See, for instance,
http://www.gloriamundi.org for an overview of the extensive application of
EVT methods for VaR and Expected-Shortfall estimation.} . However, as 
emphasized
in section~\ref{sec:boot}, estimates of the parameters of such (GEV or GPD)
distributions can be very unreliable in presence of dependence, so that such
methods finally appears to be not very accurate and one cannot avoid a
parametric approach for the estimations of the highest quantiles.

Our present study suggests that the Paretian paradigm leads to an
overestimation of
the probability of large events and therefore leads to
the adoption of too conservative positions. Generalizing to larger time scales,
the overly pessimistic view of large risks deriving from the Paretian
paradigm should be all the more revised, due to the action of the
central limit theorem.
Our comparison between several models which turn out to be almost
undistinguishable
such as the stretched exponential, the Pareto and the log-Weibull
distributions,
offers the important possibility of developing scenarios that can
test the sensitivity
of risk assessment to errors
in the determination of parameters and even more interesting with respect
to the choice of models, often refered to as model errors.

Finally, an additional note of caution is in order. This study has
focused on the
marginal distributions of returns calculated at fixed time scales
and thus neglects the possible occurrence of runs of dependencies, such
as in cumulative drawdowns. In the presence of dependencies between returns,
and especially if the dependence is non stationary and increases
in time of stress, the characterization of
the marginal distributions of returns is not sufficient. As an example,
\citeasnoun{JS02} have recently shown that the recurrence time of
very large drawdowns
cannot be predicted from the sole knowledge of the distribution of returns
and that transient dependence effects occurring in time of stress make
very large drawdowns more frequent, qualifying them as abnormal ``outliers.''

\clearpage
\appendix

\section{Maximum likelihood estimators}
\label{app:MLE}

In this appendix, we give the expressions of the maximum likelihood estimators
derived from the four distributions (\ref{eq:Pareto}-\ref{eq:Gamma})

\subsection{The Pareto distribution}

According to expression (\ref{eq:Pareto}), the Pareto distribution is given by
\be
F_u(x) = 1- \left(\frac{u}{x}\right)^b, \quad x \ge u
\ee
and its density is
\be
f_u(x|b) = b \frac{u^b}{x^{b+1}}
\ee
Let us denote by
\be
L_T^{PD}(\hat b) = \max_b \sum_{i=1}^T \ln f_u(x_i |b)
\ee
the maximum of log-likelihood function derived under hypothesis (PD). $\hat b$
is the maximum likelihood estimator of the tail index $b$ under such
hypothesis.

The maximum of the likelihood function is solution of
\be
\frac{1}{b} + \ln u - \frac{1}{T} \sum \ln x_i = 0,
\ee
which yields
\be
\hat b = \left[ \frac{1}{T} \sum_{i=1}^T \ln x_i  - \ln u\right]^{-1}, \quad
\text{and} \quad
\frac{1}{T}~ L_T^{PD}(\hat b) = \ln \frac{\hat b}{u} - \left( 1+ \frac{1}{\hat
b} \right). \ee
Moreover, one easily shows that $\hat b$ is asymptotically normally
distributed:
\be
\sqrt{T} (\hat b -b) \sim {\cal N}(0,b).
\ee

\subsection{The Weibull distribution}

The Weibull distribution is given by equation (\ref{eq:Weibull}) and
its density
is
\be
f_u(x|c,d) = \frac{c}{d^c} \cdot e^{\left( \frac{u}{d}\right)^c} ~x^{c-1} \cdot
\exp\left[- \left( \frac{x}{d}\right)^c \right], \quad x \ge u.
\ee
The maximum of the log-likelihood function is
\be
L_T^{SE}(\hat c, \hat d) = \max_{c,d} \sum_{i=1}^T \ln f_u(x_i |c,d)
\ee
Thus, the maximum likelihood estimators $(\hat c, \hat d)$ are solution of
\bea
\frac{1}{c} &=& \frac{\frac{1}{T} \sum_{i=1}^T
       \left(\frac{x_i}{u}\right)^c \ln \frac{x_i}{u}}{\frac{1}{T}
\sum_{i=1}^T \left(\frac{x_i}{u}\right)^c-1}
- \frac{1}{T}\sum_{i=1}^T \ln \frac{x_i}{u},   \label{eq:chat} \\
d^c &=& \frac{u^c}{T} \sum_{i=1}^T \left(\frac{x_i}{u}\right)^c - 1.
\label{eq:dhat}
\eea
Equation (\ref{eq:chat}) depends on $c$ only and must be solved numerically.
Then, the resulting value of $c$ can be reinjected in (\ref{eq:dhat}) to get
$d$.
The maximum of the log-likelihood function is
\be
\frac{1}{T}~L_T^{SE}(\hat c,\hat d)=  \ln \frac{\hat c}{\hat d^{\hat c}} +
\frac{\hat c-1}{T} \sum_{i=1}^T \ln x_i -1.
\ee
Since $c>0$, the vector $\sqrt{N}(\hat c -c, \hat d- d)$ is
asymptotically normal, with a covariance matrix whose expression is
given in appendix~\ref{app:AsympVar}.

It should be noted that the maximum likelihood equations
(\ref{eq:chat}-\ref{eq:dhat}) do not admit a solution with positive $c$ for all
possible samples $(x_1, \cdots, x_N)$. Indeed, the function
\be
h(c) = \frac{1}{c} - \frac{\frac{1}{T} \sum_{i=1}^T
       \left(\frac{x_i}{u}\right)^c \ln \frac{x_i}{u}}{\frac{1}{T}
\sum_{i=1}^T \left(\frac{x_i}{u}\right)^c-1}
+ \frac{1}{T}\sum_{i=1}^T \ln \frac{x_i}{u},
\ee
which is the total derivative of $L_T^{SE}(c,\hat d(c))$, is a decreasing
function of $c$. It means, as one can expect, that the likelihood function is
concave. Thus, a necessary and sufficient condition for equation
(\ref{eq:chat})
to admit a solution is that $h(0)$ is positive. After some
calculations, we find
\be
h(0)=\frac{2\left(\frac{1}{T} \sum \ln \frac{x_i}{u} \right)^2 - \frac{1}{T}
\sum \ln^2 \frac{x_i}{u}}{\frac{2}{T} \sum \ln \frac{x_i}{u}},
\ee
which is positive if and only if
\be
\label{eq:condSE}
2\left(\frac{1}{T} \sum \ln \frac{x_i}{u} \right)^2 - \frac{1}{T}
\sum \ln^2 \frac{x_i}{u}>0.
\ee
However, the probability of occurrence of a sample leading to
a negative maximum-likelihood estimate of $c$ tends to zero (under the
Hypothesis of SE with a positive $c$) as
\be
\Phi \left( - \frac{c \sqrt{T}}{\sigma} \right) \simeq \frac{\sigma}{\sqrt{2
\pi ~ T }c} e^{-\frac{c^2T}{2 \sigma^2}},
\ee
i.e. exponentially with respect to $T$. $\sigma^2$ is the variance of
the limit Gaussian distribution of maximum-likelihood $c$-estimator that can be
derived explicitly.
If $h(0)$ is negative, $L_T^{SE}$ reaches its maximum at $c=0$ and in such a
case
\be
\frac{1}{T} L_T^{SE} (c=0) = -\ln \left( \frac{1}{T} \sum \ln \frac{x_i}{u}
\right) - \frac{1}{T} \sum \ln x_i -1~.
\ee
In contrast, if the maximum likelihood estimation based on the SE
assumption is applied to samples distributed differently from the SE, negative
$c$-estimate can then be obtained with some positive probability
not tending to zero with $N \go \infty$.
If the sample is distributed according to the Pareto distribution, for
instance, then the maximum-likelihood $c$-estimate converges in
probability to a
Gaussian random variable with zero mean, and thus the probability for negative
$c$-estimates converges to $0.5$.

\subsection{The Exponential distribution}

The Exponential distribution function is given by equation
(\ref{eq:Exponential}), and its density is
\be
f_u(x|d)=\frac{\exp \left[ \frac{u}{d} \right]}{d}~ \exp \left[- \frac{x}{d}
\right], \quad x \ge u.
\ee
The maximum of the log-likelihood function is reach at
\be
\hat d = \frac{1}{T} \sum_{i=1}^T x_i - u    ,
\ee
and is given by
\be
\frac{1}{T} L_T^{ED}(\hat d) = -(1+\ln \hat d).
\ee
The random variable $\sqrt{T} (\hat d - d)$ is asymptotically normally
distributed with zero mean and variance $d^2/T$.

\subsection{The Incomplete Gamma distribution}

The expression of the Incomplete Gamma distribution function is given by
(\ref{eq:Gamma}) and its density is
\be
f_u(x|b,d) = \frac{d^b}{\Gamma \left( -b, \frac{u}{d} \right)} \cdot x^{-(b+1)}
\exp \left[- \left( \frac{x}{d} \right) \right], \quad x\ge u.
\ee
Let us introduce the partial derivative of the logarithm of the
incomplete Gamma
function:
\be
\Psi(a,x) = \frac{\partial}{\partial a} \ln \Gamma(a,x)  =
\frac{1}{\Gamma(a,x)}
\int_x^\infty  dt ~ \ln t ~ t^{a-1}~ e^{-t }.
\ee
The maximum of the log-likelihood function is reached at the point $(\hat b,
\hat d)$ solution of
\bea
      \frac{1}{T} \sum_{i=1}^T \ln
\frac{x_i}{d} &=&  \Psi \left(- b, \frac{u}{d} \right) ,\\
      \frac{1}{T} \sum_{i=1}^T
\frac{x_i}{d} &=& \frac{1}{\Gamma \left( -b, \frac{u}{d} \right)}~
\left( \frac{u}{d}\right)^{-b} e^{-\frac{u}{d}} -b,
\eea
and is equal to
\be
\frac{1}{T}~ L_T^{IG}(\hat b, \hat d) =  -\ln \hat d - \ln \Gamma \left( -b,
\frac{u}{d} \right)  + (b+1) \cdot \Psi \left(- b, \frac{u}{d} \right) + b -
\frac{1}{\Gamma \left( -b, \frac{u}{d} \right)}~
\left( \frac{u}{d}\right)^{-b} e^{-\frac{u}{d}}.
\ee

\subsection{The Log-Weibull distribution}

The Log-Weibull distribution is given by equation (\ref{eq:LW}) and
its density
is
\be
f_u(x|b,c) = \frac{b \cdot c}{x} \cdot \left( \ln \frac{x}{u}\right)^{c-1}
\cdot \exp\left[-b \left(\ln \frac{x}{d}\right)^c \right], \quad x \ge u. \ee
The maximum of the log-likelihood function is
\be
L_T^{SE}(\hat b, \hat c) = \max_{b,c} \sum_{i=1}^T \ln f_u(x_i |b,c)
\ee
Thus, the maximum likelihood estimators $(\hat b, \hat b)$ are solution of
\bea
b^{-1} &=& \frac{1}{T} \sum_{i=1}^T \left(\ln \frac{x_i}{u}\right)^c,\\
\frac{1}{c} &=& \frac{\frac{1}{T} \sum_{i=1}^T
       \left(\ln \frac{x_i}{u}\right)^c \ln \left(\ln
\frac{x_i}{u}\right)}{\frac{1}{T} \sum_{i=1}^T
\left(\ln \frac{x_i}{u}\right)^c} - \frac{1}{T}\sum_{i=1}^T \ln \left( \ln
\frac{x_i}{u} \right).
\eea
The solution of these equations is unique and it can be shown that the vector
$\sqrt{T}(\hat b-b, \hat c-c)$ is asymptotically Gaussian with a covariance
which can be deduced from matrix~(\ref{eq:thry}) given in
appendix~\ref{app:AsympVar}.

\clearpage

\section{Asymptotic variance-covariance of  maximum likelihood
estimators of the SE parameters}
\label{app:AsympVar}

We consider the Stretched-Exponential (SE) parametric family with complementary
distribution function
\be
\bar F = 1-F(x)=\exp \left[-\left(\frac{x}{d}\right)^{c} +
\left(\frac{u}{d}\right)^{c} \right] \quad x \geqslant  u,
\ee
where $c,d$ are unknown parameters and $u$ is a known lower threshold.

Let us take a new parameterization of (SE) distribution, more appropriate for
the derivation of asymptotic variances. It should be noted that this
reparameterization does not affect asymptotic variance of the form parameter
$c$. In the new parameterization, the complementary distribution
function has form:
\be
\bar F(x) = \exp \left[ -v \left(\left(\frac{x}{u} \right)^{c}- 1 \right)
\right], \quad x \geqslant  u. \ee

Here, the parameter $v$ involves both unknown parameters $c,d$ and the known
threshold $u$:
\be
v = \left(\frac{u}{d} \right)^{c}.
\ee
The log-likelihood $L$ for sample $(x_{1}\ldots x_{N})$ has the form:
\be
L = N \ln v + N \ln c + (c-1) \sum_{i=1}^N  \ln \frac{x_{i}}{u}- v \sum_{i=1}^N
\left[ \left(\frac{x_i}{u}\right)^{c} - 1 \right].
\ee

Now we derive the Fisher matrix $\Phi $:
\be
\Phi = \left(
{\begin{array}{*{20}c}
   {E \left[ - \partial^2_v L \right]}  & {E \left[ - \partial ^2_{v,c} L
\right]}\\
   {E \left[ - \partial^2_{c,v} L \right]}  & {E \left[ - \partial ^2_c L
\right]}\\
\end{array} }
\right) .
\ee

We find:
\bea
\frac{\partial^{2} L}{ \partial v^2} &=& -  \frac{N}{v^{2}},\\
\frac{\partial ^{2} L}{\partial v \partial c} &=& -N \cdot \frac{1}{N}
\sum_{i=1}^N \left(\frac{x_{i}}{u}\right)^{c } \ln \frac{x_{i}}{u}
\stackrel{N \rightarrow \infty}{\longrightarrow}  - N
E \left[ \left( \frac{x}{u} \right)^{c} \ln \frac{x}{u} \right],\\
\frac{\partial ^{2}L}{\partial  c^{2} }&=& -\frac{N}{c^{2}}- N v \cdot
\frac{1}{N} \sum_{i=1}^N  \left(\frac{x_{i}}{u} \right)^{c} \ln^{2}
\frac{x_{i}}{u} \stackrel{N \rightarrow \infty}{\longrightarrow}
-\frac{N}{c^{2}}- N v \cdot E \left[ \left( \frac{x}{u} \right)^{c} \ln^2
\frac{x}{u} \right]. \eea

After some calculations we find:
\be
E \left[ \left(\frac{x}{u} \right)^{c } \ln \left(\frac{x}{u}
\right)\right]= \frac{1+E_{1}(v)}{c \cdot v},
\ee
where $E_{1}(v) $ is the integral exponential function:
\be
E_{1}(v) = \int_v^\infty \frac{e^{ - t}}{t}~ dt.
\ee

Similarly we find:
\be
E \left[ \left( \frac{x}{u} \right)^{c} \ln^{2} \frac{x}{u} \right] =
\frac{2 e^{v}}{v \cdot c^{2}} [ E_{1}(v) + E_{2}(v) - \ln(v) E_{1}(v)],
\ee
where $E_{2}(v)$ is the partial derivative of the incomplete Gamma function:
\be
E_{2}(v) = \left. \int_v^\infty \frac{\ln (t)}{t} \cdot e^{ - t} dt =
\frac{\partial}{\partial  a} \left. \int_v^\infty {t^{a - 1}} e^{ - t} dt
\right | _{a = 0}  = \frac{\partial}{\partial a } \Gamma (a,x) \right
| _{a = 0}.
\ee

Now we find the Fisher matrix (multiplied by $N$) :
\be
\label{eq:FisherMatrix}
N \Phi = \left(
\begin{array}{*{20}c}
\frac{1}{v^{2}} & \frac{1+ e^{v} E_{1}(v)}{c \cdot v}  \\
\frac{1+ e^{v} E_{1}(v)}{c \cdot v} &
\frac{1}{c^2} (1+2 e^{v} \left[ E_{1}(v)+ E_{2}(v) - \ln(v)E_{1}(v) \right] )
\end{array}
\right)
\ee

The covariance matrix $B$ of ML-estimates ($\tilde {v},\tilde {c})$ is equal to
the inverse of the Fisher matrix. Thus, inverting the Fisher matrix
$\Phi $ in equation (\ref{eq:FisherMatrix}) we find:
\be
\label{eq:CovarianceMatrix}
B= \left(
\begin{array}{*{20}c}
\frac{v^{2}}{N H(v)} [1+2e^{v} E_{1}(v)+2 e^{v}E_{2}(v)-\ln(v) e^{v}
E_{1} (v)] & -\frac{c v}{N H(v)} [1+ e^{v} E_{1}(v)] \\
-\frac{c v}{N H(v)} [1+ e^{v} E_{1}(v)] & \frac{c^{2}}{N H(v)}
\end{array}
\right)
\ee
where $H(v)$ has form:
\be
H(v)=2 e^{v}E_{2}(v) -2 \ln(v)e^{v} E_{1}(v) - (e^{v} E_{1}(v))^{2}.
\ee
Thus, the matrix (\ref{eq:CovarianceMatrix}) provides the desired covariance
matrix.

We present here as well the covariance matrix of the limit distribution of
ML-estimates for the SE distribution on the whole semi-axis $(0,\infty)$:
\be
1-F(x)=\exp(-g \cdot x^{c}), \quad x \geqslant 0.
\ee

After some calculations by the same scheme as above we find the covariance
matrix $B$ of the limit Gaussian distribution of ML-estimates
($\tilde{g},\tilde {c})$:
\be
\label{eq:thry}
B= \frac{6}{N \pi^2} \left(
\begin{array}{*{20}c}
g^{2} \left[ \frac{\pi^{2}}{6} +\gamma + \ln(g)-1)^{2} \right] & g \cdot
c~ [\gamma+\ln(g)-1]\\
g \cdot c~ [\gamma + \ln(g)-1] & c^{2}
\end{array}
\right)
\ee
where $\gamma$ is the Euler number: $\gamma \simeq 0.577~ 215 \ldots$

\clearpage
\section{Minimum Anderson-Darling Estimators}
We derive in this appendix the expressions allowing the calculation of the
parameters which minimize the Anderson-Darling distance between the assumed
distribution and the true distribution.

Given the ordered sample $x_1 \le x_2 \le \cdots \le x_N$, the AD-distance is
given by
\be
AD_N = -N- 2 \sum_{k=1}^N \left[ w_k \log F(x_k|\alpha) + (1-w_k)
\log(1-F(x_k|\alpha)) \right], \ee
where $\alpha$ represents the vector of parameters and $w_k=2k/(2N+1)$.  It
is easy to
show that the minimum is reached at the point $\hat \alpha$ solution of
\be
\label{eq:minADS}
\sum_{k=1}^N \left( 1- \frac{w_k}{F(x_k| \alpha)} \right)
\log(1-F(x_k|\alpha))=0.
\ee

\subsection{The Pareto distribution}
Applying equation (\ref{eq:minADS}) to the Pareto distribution yields
\be
\sum_{k=1}^N \frac{w_k}{1-\left( \frac{u}{x_k}\right)^b} \ln \frac{u}{x_k} =
\sum_{k=1}^N   \ln \frac{u}{x_k}.
\ee
This equation always admits a unique solution, and can easily be solved
numerically.

\subsection{Stretched-Exponential distribution}
In the Stretched-Exponential case, we obtain the two following equations
\bea
\sum_{k=1}^N \left(1-\frac{w_k}{F_k} \right) \left[ \ln \frac{u}{d}
\left(\frac{u}{d} \right)^c - \ln \frac{x_k}{d} \left(\frac{x_k}{d} \right)^c
\right]&=&0,\\
\sum_{k=1}^N \left(1-\frac{w_k}{F_k} \right) \left( u^c -x_k^c \right)&=&0,
\eea
with
\be
F_k = 1-\exp \left[ -  \frac{u^c-x_k^c}{d^c} \right].
\ee
After some simple algebraic manipulations, the first equation can be slightly
simplified, to finally yields
\bea
\sum_{k=1}^N \left(1-\frac{w_k}{F_k} \right) \ln \frac{x_k}{u}
\left(\frac{x_k}{u} \right)^c &=&0,\\
\sum_{k=1}^N \left(1-\frac{w_k}{F_k} \right) \left( \left(\frac{x_k}{u}
\right)^c -1 \right)&=&0.
\eea
However, these two equations remain coupled.  Moreover, we have not yet been
able to prove the unicity of the solution.

\subsection{Exponential distribution}
In the exponential case, equation(\ref{eq:minADS}) becomes
\be
\sum_{k=1}^N \left( \frac{w_k}{F_k}-1 \right) (u-x_k)=0,
\ee
with
\be
F_ k= 1-\exp \left[ - \frac{u-x_k}{d} \right].
\ee
Here again, we can show that this equation admits a unique solution.

\clearpage

\clearpage
\section{Testing the Pareto model versus the (SE) model using Wilks' test}
\label{app:NN}

Our goal is to test the (SE) hypothesis $f_{1}(x \vert c,b)$ versus the Pareto
hypothesis $f_{0}(x \vert  b)$ on a semi-infinite interval $(u, \infty ),~  u >
0$. Here, we use the parameterization
\be
f_{1}(x \vert c,b)= b~u^{c} x^{c - 1} \exp \left[ -\frac{b}{c} \left(
\left( \frac{x}{u} \right)^{c} - 1\right) \right]; \quad x  \ge  u
\ee
for the stretched-exponential distribution and
\be
\label{eq:P2}
f_{0}(x \vert  b)= b~ \frac{u^{b}}{x^{1 + b}}; \quad x \ge  u
\ee
for the Pareto distribution.

{\bf Theorem}:
Assuming that the sample $x_{1} \ldots x_{N}$ is generated from the
Pareto distribution (\ref{eq:P2}), and taking the
supremums of the log-likelihoods $L_{0}$ and $L_{1}$ of the Pareto and (SE)
models respectively
over the domains $(b > 0)$ for $L_{0}$ and $(b>0, c>0)$ for $L_{1}$,
then Wilks' log-likelihood ratio $W$:
\be
W = 2 \left[ \sup_{b,c} L_{1} - \sup_{b} L_{0} \right],
\ee
is distributed according to the $\chi^{2}$-distribution with one
degree of freedom, in the limit $N \to \infty$.

{\bf Proof}

The log-likelihood $L_{0}$ reads
\be
\label{eq:P4}
L_{0} = \sum \limits_{i=1}^N \log x_{i} + N \log(b) - b \sum \limits_{i=1}^N
\log \frac{x_{i }}{u}.
\ee
The supremum over $b$ of $L_{0}$ given by (\ref{eq:P4}) is reached at
\be
\tilde {b} = \left[ \frac{1}{N} \sum\limits_{i=1}^N \log \frac{x_{i }}{u}
\right]^{-1},
\ee
and is equal to
\be
\label{eq:P6}
\sup_{b} L_{0} = -N \left(1+\log u +\frac{1}{b} - \log \tilde {b} \right).
\ee

The log-likelihood $L_{1}$ is
\be
\label{eq:P7}
L_{1} = -N \left\{ \log u -(c-1)  \sum \limits_{i=1}^N  \log \frac{x_i}{u} +
\log b + \frac{b}{c} \sum \limits_{i=1}^N \left[ \left(\frac{x_i}{u}
\right)^{c}
- 1 \right] \right\} ~.
    \ee
The supremum over $b$ of $L_{1}$ given by (\ref{eq:P7}) is reached at
\be
\bar {b} = c \left(  \frac{1}{N} \sum\limits_{i=1}^N \left[
\left(\frac{x_i}{u} \right)^{c}- 1 \right] \right)^{-1}
\ee
and is equal to
\be
\label{eq:P9}
\sup_{b} L_{1} = -N \left(1+ \log u - (c-1) \frac{1}{N} \sum\limits_{i=1}^N
\log \frac{x_i}{u} - \log \bar {b} \right)~.
\ee
Taking the derivative of expression (\ref{eq:P9}) with respect to $c$
obtains the maximum
likelihood equation for the (SE) parameter $c$
\be
\label{eq:P10}
\frac{1}{c} = \frac{\frac{1}{N} \sum_{i=1}^N
       \left(\frac{x_i}{u}\right)^c \log \frac{x_i}{u}}{\frac{1}{N}
\sum_{i=1}^N \left(\frac{x_i}{u}\right)^c-1}
- \frac{1}{N}\sum_{i=1}^N \log \frac{x_i}{u}~.
\ee

If the sample $x_{1} \ldots x_{N}$ is generated by the Pareto
distribution (\ref{eq:P2}), then by the
strong law of large numbers, we have with probability $1$ as $N \to +\infty$
\bea
\frac{1}{N} \sum\limits_{i=1}^N \log \frac{x_i}{u} &\longrightarrow&
\frac{1}{b},\\
\frac{1}{N} \sum\limits_{i=1}^N \left[ \left(\frac{x_i}{u}
\right)^{c} -1 \right] &\longrightarrow& \frac{c}{b-c},\\
\frac{1}{N} \sum\limits_{i=1}^N \left( \frac{x_i}{u} \right)^{c} \log
\frac{x_i}{u} &\longrightarrow& \frac{b}{(b-c)^{2}}~.
\eea

Inserting these limit values into (\ref{eq:P10}), the only limit
solution of this equation is $c=0$. Thus, the solution of equation
(\ref{eq:P10}) for finite $N$, denoted as $c(N)$, converges with
probability $1$
to zero as $N \to +\infty$.

Expanding $(x_i/u)^c$ in power series in the neighborhood of $c=0$ gives
\be
\left(\frac{x_i}{u}\right)^{c} \cong  1 + c \cdot
\log\left(\frac{x_i}{u}\right) + \frac{c^{2}}{2} \cdot \log^{2}
\left(\frac{x_i}{u}\right) + \frac{c^{3}}{6} \cdot \log^{3}
\left(\frac{x_i}{u}\right)+ \cdots, \quad
\text{as}~ c \to 0~,
\ee
which yields
\bea
\frac{1}{N} \sum  \left(\frac{x_i}{u}\right)^{c} &\cong&  1 + c
\cdot S_{1} + \frac{c^{2}}{2} \cdot S_{2} +\frac{c^3}{3} S_3,\\
\frac{1}{N} \sum \left(\frac{x_i}{u}\right)^{c}
\log\left(\frac{x_i}{u}\right) &\cong& S_{1} + c \cdot S_{2}+
\frac{c^{2}}{2} S_{3}~,
\eea
where
\bea
S_{1} = \frac{1}{N} \sum_{i=1}^N \log\left(\frac{x_i}{u}\right),\\
S_{2} = \frac{1}{N} \sum_{i=1}^N \log^2\left(\frac{x_i}{u}\right),\\
S_{3} = \frac{1}{N} \sum_{i=1}^N \log^3\left(\frac{x_i}{u}\right).\\
\eea

Putting these expansions into (\ref{eq:P10}) and keeping only the terms of
lowest orders in $c$, the solution of equation (\ref{eq:P10}) reads
\be
\bar c \simeq \frac{\frac{1}{2}  S_{2}- S_{1}^{2}}{ \frac{1}{2}S_{1}S_{2} -
\frac{1}{3} S_{3}}~.
\label{mglke}
\ee
Inserting this solution (\ref{mglke}) for
$c$ into (\ref{eq:P9}) gives
$\sup_{b,c} L_{1}$. Using equation (\ref{eq:P6}) for $\sup_{b} L_{0}$,
we obtain the explicit formula
\bea
W &=& 2 \left[  \sup_{b,c} L_{1}- \sup_{b} L_{0} \right],\\
&=& 2 N [ \log S_{1} +  \bar {c} - 1].
\eea

Now, accounting for the fact that the variables $\xi_1=S_1 - b^{-1}$,
$\xi_2=S_2
- 2 b^{-2}$ and $\xi_3=S_3 - 6 b^{-3}$ are asymptotically Gaussian random
variables with zero mean and variance of order $N^{-1/2}$, at the
lowest order in $N^{-1/2}$, we obtain
\be
\bar c = b^2 \left(2 \xi_1 - \frac{b}{2} \xi_2 \right)~,
\ee
and
\be
W= N {\bar c} ^{2} / ( \tilde {b})^{2}~.
\ee
Thus, $\bar c$ converges in probability to a Gaussian random variable with
standard deviation $b/ \sqrt N$ since
\be
\Var(\xi _{1}) = \frac{1}{N b^{2}}, \quad \Var(\xi _{2}) = \frac{20}{N b
^{4}}, \quad \text{and}~ \Cov(\xi _{1}, \xi _{2}) = \frac{4}{N b ^{3}}~.
\ee
Since $\tilde {b}$ converges to $b$, the Wilks'
statistic $W$ converges to a $\chi^{2}$-random variable with one degree of
freedom.

\clearpage


\begin{table}
\begin{center}
\begin{tabular}{lccccc}
\hline
\hline
      &Mean & St. Dev. & Skewness & Ex. Kurtosis & Jarque-Bera\\
\hline
Nasdaq (5 minutes) $^\dagger$&1.80 $\cdot 10^{-6}$ & 6.61 $\cdot 10^{-4}$ &
0.0326 & 11.8535 & 1.30 $\cdot 10^5$ {\tiny (0.00)} \\
Nasdaq (1 hour) $^\dagger$& 2.40 $\cdot10^{-5} $& 3.30 $\cdot
10^{-3}$ & 1.3396&
23.7946 &4.40 $\cdot 10^4$ {\tiny (0.00)}\\

Nasdaq (5 minutes) $^\ddagger$&- 6.33 $\cdot 10^{-9}$ & 3.85 $\cdot 10^{-4}$ &
-0.0562 & 6.9641 & 4.50 $\cdot 10^4$ {\tiny (0.00)} \\
Nasdaq (1 hour) $^\ddagger$ & 1.05 $\cdot10^{-6} $& 1.90 $\cdot 10^{-3}$ &
-0.0374& 4.5250 &1.58 $\cdot 10^3$ {\tiny (0.00)}\\

Dow Jones (1 day)&   8.96$\cdot$10$^{-5}$
&4.70 $\cdot 10^{-3}$&-0.6101&22.5443 &  6.03 $\cdot$10$^5$ {\tiny (0.00)} \\
Dow jones (1 month)& 1.80 $\cdot 10^{-3}$  &2.54 $\cdot 10^{-2}$ &-0.6998&
5.3619 & 1.28 $\cdot$10$^3$ {\tiny (0.00)} \\
\hline
\hline
\end{tabular}
\end{center}
\caption{\label{table:Stat} Descriptive statistics for the Dow Jones
returns calculated over one day and one month
and for the Nasdaq returns calculated over five minutes and one hour.
The numbers within parenthesis represent the $p$-value of Jarque-Bera's
normality test. $^{(\dagger)}$ raw data, $^{(\ddagger)}$ data corrected for the
$U$-shape of the intra-day volatility due to the opening, lunch and
closing effects.}
\end{table}

\clearpage

\begin{landscape}
\begin{table}
{\small
\begin{center}
\begin{tabular}{lrrrrclrrrrclrrrrr}
\hline
\hline
\multicolumn{5}{c}{Stretched-Exponential c=0.7} &  &
\multicolumn{5}{c}{Stretched-Exponential c=0.3} &  &
\multicolumn{5}{c}{Pareto Distribution b=3} \\
\hline
(a) & \multicolumn{16}{c}{Independent Data} \\
\hline
  & \multicolumn{4}{c}{GEV} &  &  & \multicolumn{4}{c}{GEV} &  &  & 
\multicolumn{4}{c}{GEV} \\
cluster & 10 & 20 & 100 & 200 &  & cluster & 10 & 20 & 100 & 200 &  & 
cluster & 10 & 20 & 100 & 200 \\
\cline{1-5}  \cline{7-11} \cline{13-17}
mean & -0.050 & -0.001 & 0.032 & 0.023 &  & mean & 0.209 & 0.230 & 
0.229 & 0.208 &  & mean & 0.240 & 0.288 & 0.338 & 0.335 \\
Emp Std & 0.055 & 0.064 & 0.098 & 0.131 &  & Emp Std & 0.025 & 0.038 
& 0.085 & 0.132 &  & Emp Std & 0.027 & 0.040 & 0.095 & 0.149 \\
  &  &  &  &  &  &  &  &  &  &  &  &  &  &  &  &  \\
  & \multicolumn{4}{c}{GPD} &  &  & \multicolumn{4}{c}{GPD} &  &  & 
\multicolumn{4}{c}{GPD} \\
quantile  & 0.9 & 0.95 & 0.99 & 0.995 &  & quantile  & 0.9 & 0.95 & 
0.99 & 0.995 &  & quantile  & 0.9 & 0.95 & 0.99 & 0.995 \\
\cline{1-5}  \cline{7-11} \cline{13-17}
mean & 0.012 & 0.030 & 0.013 & -0.009 &  & mean & 0.226 & 0.231 & 
0.187 & 0.160 &  & mean & 0.236 & 0.296 & 0.314 & 0.295 \\
Emp Std & 0.032 & 0.048 & 0.122 & 0.175 &  & Emp Std & 0.037 & 0.055 
& 0.134 & 0.193 &  & Emp Std & 0.037 & 0.058 & 0.142 & 0.220 \\
Theor std & 0.032 & 0.046 &0.101 & 0.140 & & Theor std & 0.039 & 
0.055 & 0.119 & 0.164 &  & Theor std & 0.037 & 0.058 & 0.140 & 0.218 
\\
  &  &  &  &  &  &  &  &  &  &  &  &  &  &  &  &  \\
\hline
(b) & \multicolumn{16}{c}{Dependent Data, Correlation length $\tau=20$} \\
\hline
  & \multicolumn{4}{c}{GEV} &  &  & \multicolumn{4}{c}{GEV} &  &  & 
\multicolumn{4}{c}{GEV} \\
\cline{1-5}  \cline{7-11} \cline{13-17}
cluster & 10 & 20 & 100 & 200 &  & cluster & 10 & 20 & 100 & 200 &  & 
cluster & 10 & 20 & 100 & 200 \\
mean & -0.148 & -0.065 & 0.012 & 0.022 &  & mean & 0.206 & 0.216 & 
0.297 & 0.268 &  & mean & 0.136 & 0.223 & 0.361 & 0.364 \\
Emp Std & 0.031 & 0.038 & 0.047 & 0.053 &  & Emp Std & 0.036 & 0.046 
& 0.088 & 0.129 &  & Emp Std & 0.034 & 0.043 & 0.085 & 0.144 \\
  &  &  &  &  &  &  &  &  &  &  &  &  &  &  &  &  \\
  & \multicolumn{4}{c}{GPD} &  &  & \multicolumn{4}{c}{GPD} &  &  & 
\multicolumn{4}{c}{GPD} \\
quantile  & 0.9 & 0.95 & 0.99 & 0.995 &  & quantile  & 0.9 & 0.95 & 
0.99 & 0.995 &  & quantile  & 0.9 & 0.95 & 0.99 & 0.995 \\
\cline{1-5}  \cline{7-11} \cline{13-17}
mean & 0.000 & 0.018 & -0.024 & -0.080 &  & mean & 0.217 & 0.217 & 
0.144 & 0.085 &  & mean & 0.229 & 0.290 & 0.290 & 0.249 \\
Emp Std & 0.050 & 0.066 & 0.130 & 0.182 &  & Emp Std & 0.061 & 0.082 
& 0.151 & 0.206 &  & Emp Std & 0.061 & 0.084 & 0.168 & 0.230 \\
Theor std & 0.032 & 0.046 & 0.098 &0.130 &  &Theor std & 0.039 & 
0.054 & 0.114 & 0.153 & & Theor std &0.039 & 0.064 &0.134 &0.177 \\
  &  &  &  &  &  &  &  &  &  &  &  &  &  &  &  &  \\
\hline
(c) & \multicolumn{16}{c}{Dependent Data, Correlation length $\tau=100$} \\
\hline
& \multicolumn{4}{c}{GEV} &  &  & \multicolumn{4}{c}{GEV} &  &  & 
\multicolumn{4}{c}{GEV} \\
cluster & 10 & 20 & 100 & 200 &  & cluster & 10 & 20 & 100 & 200 &  & 
cluster & 10 & 20 & 100 & 200 \\
\cline{1-5}  \cline{7-11} \cline{13-17}
mean & -0.162 & -0.080 & 0.110 & 0.112 &  & mean & 0.197 & 0.186 & 
0.305 & 0.320 &  & mean & 0.131 & 0.196 & 0.372 & 0.439 \\
Emp Std & 0.043 & 0.046 & 0.094 & 0.128 &  & Emp Std & 0.052 & 0.059 
& 0.117 & 0.158 &  & Emp Std & 0.061 & 0.073 & 0.116 & 0.156 \\
  &  &  &  &  &  &  &  &  &  &  &  &  &  &  &  &  \\
  & \multicolumn{4}{c}{GPD} &  &  & \multicolumn{4}{c}{GPD} &  &  & 
\multicolumn{4}{c}{GPD} \\
quantile  & 0.9 & 0.95 & 0.99 & 0.995 &  & quantile  & 0.9 & 0.95 & 
0.99 & 0.995 &  & quantile  & 0.9 & 0.95 & 0.99 & 0.995 \\
\cline{1-5}  \cline{7-11} \cline{13-17}
mean & -0.026 & -0.019 & -0.089 & -0.157 &  & mean & 0.187 & 0.174 & 
0.072 & -0.019 &  & mean & 0.207 & 0.252 & 0.184 & 0.151 \\
Emp Std & 0.078 & 0.087 & 0.131 & 0.173 &  & Emp Std & 0.107 & 0.114 
& 0.153 & 0.182 &  & Emp Std & 0.108 & 0.131 & 0.184 & 0.222 \\
Theor std & 0.031 & 0.044 &0.091 &0.119 &  & Theor std & 0.038 & 
0.053 & 0.107 & 0.139 & & Theor std & 0.038 & 0.056 & 0.121 &0.163 \\
\hline
\hline
\end{tabular}
\end{center}
}
\caption{\label{table:MaxLikelihood} {\small Mean values and
standard deviations of the Maximum Likelihood estimates of the parameter $\xi$
(inverse of the Pareto exponent)
for the distribution of maxima (cf. equation~\ref{eq:Max}) when data are
clustered in samples of size $10, 20, 100$ and $200$ and for the
Generalized Pareto Distribution
(\ref{eq:GPD}) for thresholds $u$ corresponding to quantiles $90\%, 95\%, 99\%$
ans $99.5\%$. In panel (a),  we have used iid samples of size $10000$ drawn
from a Stretched-Exponential distribution with $c=0.7$ and $c=0.3$ and a Pareto
distribution with tail index $b=3$, while in panel (b) the samples are
drawn from a long memory process with Stretched-Exponential marginals and
regularly-varying marginal as explained in the text.}
}
\end{table}
\end{landscape}

\clearpage

\begin{landscape}
\begin{table}
{\small
\begin{center}
\begin{tabular}{lrrrrrlrrrrrlrrrr}
\hline
\hline
\multicolumn{5}{c}{Stretched-Exponential  c=0.7} &  & 
\multicolumn{5}{c}{Stretched-Exponential  c=0.3} &  & 
\multicolumn{5}{c}{Pareto Distribution  b=3} \\
\hline
(a) & \multicolumn{16}{c}{Independent data} \\
\hline
quantile & 0.9 & 0.95 & 0.99 & 0.995 &  & quantile & 0.9 & 0.95 & 
0.99 & 0.995 &  & quantile & 0.9 & 0.95 & 0.99 & 0.995 \\
\hline
N/k=4 &  &  &  &  &  & N/k=4 &  &  &  &  &  & N/k=4 &  &  &  &  \\
mean & -0.1846 & -0.0309 & 0.0436 & 0.0370 &  & mean & 0.2135 & 
0.2783 & 0.2782 & 0.2886 &  & mean & 0.1224 & 0.2489 & 0.3294 & 
0.3297 \\
emp. Std & 0.1157 & 0.1699 & 0.3713 & 0.5646 &  & emp. Std & 0.1230 & 
0.1723 & 0.3853 & 0.5354 &  & emp. Std & 0.1201 & 0.1772 & 0.3866 & 
0.5888 \\
th. Std & 0.1119 & 0.1607 & 0.3624 & 0.5121 &  & th. Std & 0.1173 & 
0.1675 & 0.3746 & 0.5307 &  & th. Std & 0.1158 & 0.1668 & 0.3778 & 
0.5343 \\
  &  &  &  &  &  &  &  &  &  &  &  &  &  &  &  &  \\
N/k=10 &  &  &  &  &  & N/k=10 &  &  &  &  &  & N/k=10 &  &  &  &  \\
mean & 0.0031 & 0.0620 & 0.0458 & 0.0329 &  & mean & 0.2946 & 0.3042 
& 0.2809 & 0.2608 &  & mean & 0.2802 & 0.3412 & 0.3413 & 0.3423 \\
emp. Std & 0.1757 & 0.2571 & 0.6228 & 0.8576 &  & emp. Std & 0.2004 & 
0.2699 & 0.5989 & 0.8702 &  & emp. Std & 0.1842 & 0.2683 & 0.6522 & 
0.8903 \\
th. Std & 0.1803 & 0.2568 & 0.5731 & 0.8093 &  & th. Std & 0.1878 & 
0.2660 & 0.5926 & 0.8354 &  & th. Std & 0.1874 & 0.2677 & 0.5985 & 
0.8466 \\
  &  &  &  &  &  &  &  &  &  &  &  &  &  &  &  &  \\
\hline
(b) & \multicolumn{16}{c}{Dependent data, Correlation length $\tau=20$} \\
\hline
quantile & 0.9 & 0.95 & 0.99 & 0.995 &  & quantile & 0.9 & 0.95 & 
0.99 & 0.995 &  & quantile & 0.9 & 0.95 & 0.99 & 0.995 \\
\hline
N/k=4 &  &  &  &  &  & N/k=4 &  &  &  &  &  & N/k=4 &  &  &  &  \\
mean & -0.1879 & -0.0267 & 0.0324 & 0.0364 &  & mean & 0.2204 & 
0.2705 & 0.2628 & 0.2464 &  & mean & 0.1194 & 0.2536 & 0.3154 & 
0.3274 \\
emp. Std & 0.1189 & 0.1674 & 0.3711 & 0.5334 &  & emp. Std & 0.1231 & 
0.1798 & 0.3741 & 0.5483 &  & emp. Std & 0.1261 & 0.1779 & 0.3914 & 
0.5600 \\
th. Std & 0.1119 & 0.1607 & 0.3619 & 0.5120 &  & th. Std & 0.1174 & 
0.1673 & 0.3737 & 0.5272 &  & th. Std & 0.1157 & 0.1669 & 0.3769 & 
0.5341 \\
  &  &  &  &  &  &  &  &  &  &  &  &  &  &  &  &  \\
N/k=10 &  &  &  &  &  & N/k=10 &  &  &  &  &  & N/k=10 &  &  &  &  \\
mean & 0.0077 & 0.0444 & 0.0132 & 0.0095 &  & mean & 0.2955 & 0.2982 
& 0.2336 & 0.1699 &  & mean & 0.2847 & 0.3226 & 0.3063 & 0.3118 \\
emp. Std & 0.1769 & 0.2678 & 0.5966 & 0.8306 &  & emp. Std & 0.1957 & 
0.2752 & 0.6018 & 0.8499 &  & emp. Std & 0.1885 & 0.2834 & 0.6236 & 
0.8722 \\
th. Std & 0.1804 & 0.2563 & 0.5709 & 0.8070 &  & th. Std & 0.1878 & 
0.2658 & 0.5882 & 0.8241 &  & th. Std & 0.1875 & 0.2668 & 0.5951 & 
0.8423 \\
  &  &  &  &  &  &  &  &  &  &  &  &  &  &  &  &  \\
\hline
(c) & \multicolumn{16}{c}{Dependent data, Correlation length $\tau=100$} \\
\hline
quantile & 0.9 & 0.95 & 0.99 & 0.995 &  & quantile & 0.9 & 0.95 & 
0.99 & 0.995 &  & quantile & 0.9 & 0.95 & 0.99 & 0.995 \\
\hline
N/k=4 &  &  &  &  &  & N/k=4 &  &  &  &  &  & N/k=4 &  &  &  &  \\
mean & -0.1946 & -0.0424 & 0.0220 & -0.0217 &  & mean & 0.2056 & 
0.2677 & 0.2479 & 0.1952 &  & mean & 0.1122 & 0.2386 & 0.2934 & 
0.2641 \\
emp. Std & 0.1295 & 0.1793 & 0.3729 & 0.5142 &  & emp. Std & 0.1455 & 
0.1968 & 0.3890 & 0.5676 &  & emp. Std & 0.1478 & 0.2005 & 0.3985 & 
0.5792 \\
th. Std & 0.1118 & 0.1605 & 0.3614 & 0.5086 &  & th. Std & 0.1172 & 
0.1673 & 0.3729 & 0.5231 &  & th. Std & 0.1156 & 0.1665 & 0.3756 & 
0.5286 \\
  &  &  &  &  &  &  &  &  &  &  &  &  &  &  &  &  \\
N/k=10 &  &  &  &  &  & N/k=10 &  &  &  &  &  & N/k=10 &  &  &  &  \\
mean & -0.0157 & 0.0138 & -0.0412 & -0.1016 &  & mean & 0.2793 & 
0.2694 & 0.1494 & 0.0807 &  & mean & 0.2639 & 0.2880 & 0.2228 & 
0.1682 \\
emp. Std & 0.1971 & 0.2676 & 0.5732 & 0.8222 &  & emp. Std & 0.2188 & 
0.2940 & 0.6115 & 0.8567 &  & emp. Std & 0.2230 & 0.3005 & 0.6252 & 
0.8707 \\
th. Std & 0.1799 & 0.2553 & 0.5674 & 0.7974 &  & th. Std & 0.1874 & 
0.2645 & 0.5810 & 0.8141 &  & th. Std & 0.1869 & 0.2653 & 0.5873 & 
0.8239 \\
\hline
\hline
\end{tabular}
\end{center}
}
\caption{\label{tableBootstrap} {\small Pickands estimates 
(\ref{eq:Pickands}) of the
parameter $\xi$ for the Generalized Pareto Distribution (\ref{eq:GPD}) for
thresholds $u$ corresponding to quantiles $90\%, 95\%, 99\%$ ans $99.5\%$ and
two different values of the ratio $N/k$ respectively equal to $4$ and $10$. In
panel (a), we have used iid samples of size $10000$ drawn from a
Stretched-Exponential distribution with $c=0.7$ and $c=0.3$ and a
Pareto distribution with
tail index $b=3$, while in panel (b) the samples are drawn from a
long memory process with Stretched-Exponential marginals and regularly-varying
marginal.}}
\end{table}
\end{landscape}

\clearpage

\begin{table}
\begin{center}
\begin{tabular}{lccccclcccc}
\hline
\hline
(a) & \multicolumn{10}{c}{Dow Jones} \\
\hline
\multicolumn{5}{c}{Positive Tail} &  & \multicolumn{5}{c}{Negative Tail} \\
\multicolumn{5}{c}{GEV} &  & \multicolumn{5}{c}{GEV} \\
cluster & 20 & 40 & 200 & 400 &  & cluster & 20 & 40 & 200 & 400 \\
\cline{1-5} \cline{7-11}
$\xi$ & 0.273 & 0.280 & 0.304 & 0.322 &  & $\xi$ & 0.262 & 0.295 & 
0.358 & 0.349 \\
Emp Std & 0.029 & 0.039 & 0.085 & 0.115 &  & Emp Std & 0.030 & 0.045 
& 0.103 & 0.143 \\
  &  &  &  &  &  &  &  &  &  &  \\
\multicolumn{5}{c}{GPD} &  & \multicolumn{5}{c}{GPD} \\
quantile  & 0.9 & 0.95 & 0.99 & 0.995 &  & quantile  & 0.9 & 0.95 & 
0.99 & 0.995 \\
\cline{1-5} \cline{7-11}
$\xi$ & 0.248 & 0.247 & 0.174 & 0.349 &  & $\xi$ & 0.214 & 0.204 & 
0.250 & 0.345 \\
Emp Std & 0.036 & 0.053 & 0.112 & 0.194 &  & Emp Std & 0.041 & 0.062 
& 0.156 & 0.223 \\
Theor Std & 0.032 & 0.046 & 0.096 & 0.156 &  & Theor Std & 0.033 & 
0.046 & 0.108 & 0.164 \\
  &  &  &  &  &  &  &  &  &  &  \\
\hline
(b) & \multicolumn{10}{c}{Nasdaq (Raw data)} \\
\hline
\multicolumn{5}{c}{Positive Tail} &  & \multicolumn{5}{c}{Negative Tail} \\
\multicolumn{5}{c}{GEV} &  & \multicolumn{5}{c}{GEV} \\
cluster & 20 & 40 & 200 & 400 &  & cluster & 20 & 40 & 200 & 400 \\
\cline{1-5} \cline{7-11}
$\xi$ & 0.209 & 0.193 & 0.388 & 0.516 &  & $\xi$ & 0.191 & 0.175 & 
0.292 & 0.307 \\
Emp Std & 0.031 & 0.115 & 0.090 & 0.114 &  & Emp Std & 0.030 & 0.038 
& 0.094 & 0.162 \\
  &  &  &  &  &  &  &  &  &  &  \\
\multicolumn{5}{c}{GPD} &  & \multicolumn{5}{c}{GPD} \\
quantile  & 0.9 & 0.95 & 0.99 & 0.995 &  & quantile  & 0.9 & 0.95 & 
0.99 & 0.995 \\
\cline{1-5} \cline{7-11}
$\xi$ & 0.200 & 0.289 & 0.389 & 0.470 &  & $\xi$ & 0.143 & 0.202 & 
0.229 & 0.242 \\
Emp Std & 0.040 & 0.058 & 0.120 & 0.305 &  & Emp Std & 0.040 & 0.057 
& 0.143 & 0.205 \\
Theor Std & 0.036 & 0.054 & 0.131 & 0.196 &  & Theor Std & 0.035 & 
0.052 & 0.118 & 0.169 \\
  &  &  &  &  &  &  &  &  &  &  \\
\hline
(c) & \multicolumn{10}{c}{Nasdaq (Corrected data)} \\
\hline
\multicolumn{5}{c}{Positive Tail} &  & \multicolumn{5}{c}{Negative Tail} \\
\multicolumn{5}{c}{GEV} &  & \multicolumn{5}{c}{GEV} \\
cluster & 20 & 40 & 200 & 400 &  & cluster & 20 & 40 & 200 & 400 \\
\cline{1-5} \cline{7-11}
$\xi$ & 0.090 & 0.175 & 0.266 & 0.405 &  & $\xi$ & 0.099 & 0.132 & 
0.138 & 0.266 \\
Emp Std & 0.029 & 0.039 & 0.085 & 0.187 &  & Emp Std & 0.030 & 0.041 
& 0.079 & 0.197 \\
  &  &  &  &  &  &  &  &  &  &  \\
\multicolumn{5}{c}{GPD} &  & \multicolumn{5}{c}{GPD} \\
quantile  & 0.9 & 0.95 & 0.99 & 0.995 &  & quantile  & 0.9 & 0.95 & 
0.99 & 0.995 \\
\cline{1-5} \cline{7-11}
$\xi$ & 0.209 & 0.229 & 0.307 & 0.344 &  & $\xi$ & 0.165 & 0.160 & 
0.210 & 0.054 \\
Emp Std & 0.039 & 0.052 & 0.111 & 0.192 &  & Emp Std & 0.039 & 0.052 
& 0.150 & 0.209 \\
Theor Std & 0.036 & 0.052 & 0.123 & 0.180 &  & Theor Std & 0.036 & 
0.050 & 0.116 & 0.143 \\
\hline
\hline
\end{tabular}
\end{center}
\caption{\label{table:ML_RealData} Mean values and
standard deviations of the Maximum Likelihood estimates of the parameter $\xi$
for the distribution of maximum (cf. equation~\ref{eq:Max}) when data are
clustered in samples of size $20, 40, 200$ and $400$ and for the
Generalized Pareto Distribution
(\ref{eq:GPD}) for thresholds $u$ corresponding to quantiles $90\%, 95\%, 99\%$
ans $99.5\%$. In panel (a),  are presented the results for the Dow
Jones, in panel (b)
for the Nasdaq for raw data and in panel (c) the Nasdaq corrected for
the ``lunch effect''.}
\end{table}

\clearpage

\begin{table}
\begin{tabular}{lrrrrclrrrr}
\hline
\hline
(a)&\multicolumn{10}{c}{Dow Jones} \\
\hline
\multicolumn{5}{c}{Negative Tail} &  & \multicolumn{5}{c}{Positive Tail} \\
quantile & 0.9 & 0.95 & 0.99 & 0.995 &  & quantile & 0.9 & 0.95 & 
0.99 & 0.995 \\
\cline{1-5} \cline{7-11}
N/k & 4 &  &  &  &  & N/k & 4 &  &  &  \\
$\xi$ & 0.2314 & 0.2944 & -0.1115 & 0.3314 &  & $\xi$ & 0.2419 & 
0.4051 & -0.3752 & 0.5516 \\
emp. Std & 0.1073 & 0.1550 & 0.3897 & 0.6712 &  & emp. Std & 0.0915 & 
0.1274 & 0.3474 & 0.5416 \\
th. Std & 0.1176 & 0.1680 & 0.3563 & 0.5344 &  & th. Std & 0.1178 & 
0.1712 & 0.3497 & 0.5562 \\
  &  &  &  &  &  &  &  &  &  &  \\
N/k & 10 &  &  &  &  & N/k & 10 &  &  &  \\
mean & 0.3119 & 0.0890 & -0.3452 & 0.9413 &  & $\xi$ & 0.3462 & 
0.3215 & 0.9111 & -0.3873 \\
emp. Std & 0.1523 & 0.2219 & 0.8294 & 1.1352 &  & emp. Std & 0.1766 & 
0.1929 & 0.6983 & 1.6038 \\
th. Std & 0.1883 & 0.2577 & 0.5537 & 0.9549 &  & th. Std & 0.1894 & 
0.2668 & 0.6706 & 0.7816 \\
  &  &  &  &  &  &  &  &  &  &  \\
\hline
(b) & \multicolumn{10}{c}{Nasdaq (Raw data)} \\
\hline
\multicolumn{5}{c}{Negative Tail} &  & \multicolumn{5}{c}{Positive Tail} \\
quantile & 0.9 & 0.95 & 0.99 & 0.995 &  & quantile & 0.9 & 0.95 & 
0.99 & 0.995 \\
N/k & 4 &  &  &  &  & N/k & 4 &  &  &  \\
\cline{1-5} \cline{7-11}
$\xi$ & 0.0493 & 0.0539 & -0.0095 & 0.4559 &  & $\xi$ & 0.0238 & 
0.1511 & 0.1745 & 1.1052 \\
emp. Std & 0.1129 & 0.1928 & 0.4393 & 0.6205 &  & emp. Std & 0.1003 & 
0.1599 & 0.4980 & 0.6180 \\
th. Std & 0.1147 & 0.1623 & 0.3601 & 0.5462 &  & th. Std & 0.1143 & 
0.1644 & 0.3688 & 0.6272 \\
  &  &  &  &  &  &  &  &  &  &  \\
N/k & 10 &  &  &  &  & N/k & 10 &  &  &  \\
$\xi$ & 0.2623 & 0.1583 & -0.8781 & 0.8855 &  & $\xi$ & 0.2885 & 
0.1435 & 1.3734 & -0.8395 \\
emp. Std & 0.1940 & 0.3085 & 0.9126 & 1.5711 &  & emp. Std & 0.2166 & 
0.3220 & 0.7359 & 1.5087 \\
th. Std & 0.1868 & 0.2602 & 0.5543 & 0.9430 &  & th. Std & 0.1876 & 
0.2596 & 0.7479 & 0.7824 \\
  &  &  &  &  &  &  &  &  &  &  \\
\hline
(c) & \multicolumn{10}{c}{Nasdaq (Corrected data)} \\
\hline
\multicolumn{5}{c}{Negative Tail} &  & \multicolumn{5}{c}{Positive Tail} \\
quantile & 0.9 & 0.95 & 0.99 & 0.995 &  & quantile & 0.9 & 0.95 & 
0.99 & 0.995 \\
\cline{1-5} \cline{7-11}
N/k & 4 &  &  &  &  & N/k & 4 &  &  &  \\
$\xi$ & 0.2179 & 0.0265 & 0.3977 & 0.1073 &  & $\xi$ & 0.2545 & 
-0.0402 & -0.0912 & 1.3915 \\
emp. Std & 0.1211 & 0.1491 & 0.4585 & 0.7206 &  & emp. Std & 0.1082 & 
0.1643 & 0.4317 & 0.6220 \\
th. Std & 0.1174 & 0.1617 & 0.3822 & 0.5167 &  & th. Std & 0.1180 & 
0.1605 & 0.3570 & 0.6720 \\
  &  &  &  &  &  &  &  &  &  &  \\
N/k & 10 &  &  &  &  & N/k & 10 &  &  &  \\
$\xi$ & -0.0878 & 0.4619 & 0.0329 & 0.3742 &  & $\xi$ & 0.0877 & 
0.3907 & 1.4680 & 0.1098 \\
emp. Std & 0.1882 & 0.2728 & 0.7561 & 1.1948 &  & emp. Std & 0.1935 & 
0.2495 & 0.8045 & 1.2345 \\
th. Std & 0.1786 & 0.2734 & 0.5722 & 0.8512 &  & th. Std & 0.1822 & 
0.2699 & 0.7655 & 0.8172 \\
\hline
\hline
\end{tabular}
\caption{\label{tableRealData} Pickands estimates (\ref{eq:Pickands}) of the
parameter $\xi$ for the Generalized Pareto Distribution (\ref{eq:GPD}) for
thresholds $u$ corresponding to quantiles $90\%, 95\%, 99\%$ ans $99.5\%$ and
two different values of the ratio $N/k$ respectiveley equal to $4$ and $10$. In
panel (a), are presented the results for the Dow Jones, in panel (b) for the
Nasdaq for raw data and in panel (c) the Nasdaq corrected for the ``lunch
effect''.}
\end{table}

\clearpage

\begin{table}
\begin{center}
\begin{tabular}{lrrrrrrrrr}
\hline
\hline
      & \multicolumn{4}{c}{Nasdaq} &  & \multicolumn{4}{c}{Dow Jones} \\
      & \multicolumn{2}{c}{{Pos. Tail}} & \multicolumn{2}{c}{{Neg. Tail}}
& {} & \multicolumn{2}{c}{{Pos. Tail}} & \multicolumn{2}{c}{{Neg.
Tail}} \\
\hline
{{\textit     q}} & $10^3 u$ & $n_u$ & $10^3 u$ & $n_u$ & {} &  $10^2
u$ & $n_u$ & $10^2 u$ & $n_u$\\
{$q_1$=0} & {0.0053} & {11241} & {0.0053} & {10751} & {} & {0.0032} & {14949} &
{0.0028} & {13464} \\
{$q_2$=0.1} & {0.0573} & {10117} & {0.0571} & {9676} & {} &
{0.0976} & {13454} & {0.0862} & {12118} \\ {$q_3$=0.2} & {0.1124} &
{8993} & {0.1129}
& {8601} & {} & {0.1833} & {11959} & {0.1739} & {10771} \\
{$q_4$=0.3} & {0.1729} &
{7869} & {0.1723} & {7526} & {} & {0.2783} & {10464} & {0.263} &
{9425} \\ {$q_5$=0.4}
& {0.238} & {6745} & {0.2365} & {6451} & {} & {0.3872} & {8969} & {0.3697} &
{8078} \\ {$q_6$=0.5} & {0.3157} & {5620} & {0.3147} & {5376} & {} &
{0.5055} & {7475}
& {0.4963} & {6732} \\ {$q_7$=0.6} & {0.406} & {4496} & {0.412} & {4300} & {} &
{0.6426} & {5980} & {0.6492} & {5386} \\ {$q_8$=0.7} & {0.5211} &
{3372} & {0.5374} &
{3225} & {} & {0.8225} & {4485} & {0.8376} & {4039} \\ {$q_9$=0.8} &
{0.6901} & {2248}
& {0.7188} & {2150} & {} & {1.0545} & {2990} & {1.1057} & {2693} \\
{$q_{10}$=0.9} &
{0.973} & {1124} & {1.0494} & {1075} & {} & {1.4919} & {1495} & {1.6223} &
{1346} \\ {$q_{11}$=0.925} & {1.1016} & {843} & {1.1833} & {806} & {}
& {1.6956} & {1121}
& {1.8637} & {1010} \\ {$q_{12}$=0.95} & {1.2926} & {562} & {1.3888}
& {538} & {} &
{1.9846} & {747} & {2.2285} & {673} \\ {$q_{13}$=0.96} & {1.3859} &
{450} & {1.4955} &
{430} & {} & {2.1734} & {598} & {2.4197} & {539} \\ {$q_{14}$=0.97} &
{1.53} & {337} &
{1.639} & {323} & {} & {2.413} & {448} & {2.7218} & {404} \\
{$q_{15}$=0.98} & {1.713} &
{225} & {1.8557} & {215} & {} & {2.7949} & {299} & {3.1647} & {269}
\\ {$q_{16}$=0.99} &
{2.1188} & {112} & {1.8855} & {108} & {} & {3.5704} & {149} & {4.1025} & {135}
\\ {$q_{17}$=0.9925} & {2.3176} & {84} & {2.4451} & {81} & {} &
{3.9701} & {112} &
{4.3781} & {101} \\ {$q_{18}$=0.995} & {3.0508} & {56} & {2.7623} &
{54} & {} & {4.5746}
& {75} & {5.0944} & {67} \\ \hline \hline \end{tabular}

\end{center}
\caption{\label{table1} Significance levels $q_{k}$ and their corresponding
lower thresholds $u_{k}$ for the four different samples. The number $n_u$
provides the size of the sub-sample beyond the threshold $u_k$.} \end{table}

\clearpage

\begin{table}
\begin{center}
\begin{tabular}{lrlrlrlrl}
\hline
\hline
\multicolumn{9}{c}{Mean AD-statistic for u$_{1}$ -- u$_{9}$ }  \\
\hline
&
\multicolumn{2}{c}{N-pos}&
\multicolumn{2}{c}{N-neg}  &
\multicolumn{2}{c}{DJ-pos}&
\multicolumn{2}{c}{DJ-neg}\\
\hline
Weibull&
1.37& {\tiny(79.97\%)}&
.851 &{\tiny(54.70\%)}&
4.96 &{\tiny(99.71\%)}&
3.86 &{\tiny(98.92\%)}\\

Gen. Pareto&
3.37&{\tiny(98.21\%)}&
2.28&{\tiny(93.49\%)}&
7.21&{\tiny(99.996\%)}&
3.90&{\tiny(98.97\%)}\\

Gamma&
3.04 &{\tiny(97.39\%)}&
2.36 &{\tiny(94.13\%)}&
5.44 &{\tiny(99.82\%)}&
4.73 &{\tiny(99.62\%)} \\

Exponential&
5.41 &{\tiny(99.81\%)}&
3.33 &{\tiny(98.13\%)}&
16.48 &{\tiny(99.996\%)}&
10.30 &{\tiny(99.996\%)}\\

Pareto&
475.0 &{\tiny(99.996\%)}&
441.4 &{\tiny(99.996\%)}&
691.3 &{\tiny(99.996\%)}&
607.3 &{\tiny(99.996\%)}\\

Log-Weibull&
35.90 &{\tiny(99.996\%)}&
30.92 &{\tiny(99.996\%)}&
32.30 &{\tiny(99.996\%)}&
28.27 &{\tiny(99.996\%)}\\

\hline
\multicolumn{9}{c}{Mean AD-statistic for u$_{10}$ -- u$_{18}$}  \\
\hline

Weibull&
.674&{\tiny(42.11\%)}&
.498&{\tiny(29.13\%)}&
.377&{\tiny(20.55\%)}&
.349&{\tiny(18.65\%)}\\

Gen. Pareto&
2.29&{\tiny(93.57\%)}&
1.88&{\tiny(89.52\%)}&
1.95&{\tiny(90.28\%)}&
1.36&{\tiny(79.67\%)}\\

Gamma&
2.49&{\tiny(95.00\%)}&
1.90&{\tiny(89.74\%)}&
2.12&{\tiny(92.01\%)}&
1.63&{\tiny(86.02\%)}\\

Exponential&
3.06&{\tiny(97.45\%)}&
1.97&{\tiny(90.48\%)}&
3.06&{\tiny(97.45\%)}&
1.89&{\tiny(89.63\%)}\\

Pareto&
1.30&{\tiny(77.73\%)}&
1.33&{\tiny(78.33\%)}&
.775&{\tiny(49.42\%)}&
1.26&{\tiny(76.30\%)} \\

Log-Weibull&
.459 &{\tiny(28.90\%)}&
.490 &{\tiny(29.51\%)}&
.375 &{\tiny(20.52\%)}&
.685 &{\tiny(43.45\%)}\\

\hline
\hline
\end{tabular}
\end{center}
\caption{\label{table2} Mean Anderson-Darling distances in the range
of thresholds
$u_1$-$u_{9}$ and in the range $u_{10}$-$u_{18}$. The figures within
parenthesis
characterize the goodness of fit: they represent the significance levels with
which the considered model can be rejected. Note that these significance levels
are only lower bounds since one or two parameters are fitted.}
\end{table}

\clearpage
\begin{table}
\begin{center}
\begin{tabular}{rccccccccc}
\hline
\hline
      & \multicolumn{4}{c}{Nasdaq} &  & \multicolumn{4}{c}{Dow Jones} \\
\hline
      & \multicolumn{2}{c}{{Pos. Tail}} & \multicolumn{2}{c}{{Neg. Tail}}
& {} & \multicolumn{2}{c}{{Pos. Tail}} & \multicolumn{2}{c}{{Neg.
Tail}} \\
{} & {MLE} & {ADE} & {MLE} & {ADE} & & {MLE} & {ADE} & {MLE} & {ADE} \\
\cline{2-5}
\cline{7-10}
{1} & {0.256 \small{(0.002)}} & {0.192} & {0.254 \small{(0.002)}} &
{0.191} & {} & {0.204 \small{(0.002)}} & {0.150} & {0.199
\small{(0.002)}} & {0.147} \\
{2} & {0.555 \small{(0.006)}} & {0.443} & {0.548 \small{(0.006)}} &
{0.439} & {} & {0.576 \small{(0.005)}} & {0.461} & {0.538
\small{(0.005)}} & {0.431} \\
{3} & {0.765 \small{(0.008)}} & {0.630} & {0.755 \small{(0.008)}} &
{0.625} & {} & {0.782 \small{(0.007)}} & {0.644} & {0.745
\small{(0.007)}} & {0.617} \\
{4} & {0.970 \small{(0.011)}} & {0.819} & {0.945 \small{(0.011)}} &
{0.800} & {} & {0.989 \small{(0.010)}} & {0.833} & {0.920
\small{(0.009)}} & {0.777} \\
{5} & {1.169 \small{(0.014)}} & {1.004} & {1.122 \small{(0.014)}} &
{0.965} & {} & {1.219 \small{(0.013)}} & {1.053} & {1.114
\small{(0.012)}} & {0.960} \\
{6} & {1.400 \small{(0.019)}} & {1.227} & {1.325 \small{(0.018)}} &
{1.157} & {} & {1.447 \small{(0.017)}} & {1.279} & {1.327
\small{(0.016)}} & {1.169} \\
{7} & {1.639 \small{(0.024)}} & {1.460} & {1.562 \small{(0.024)}} &
{1.386} & {} & {1.685 \small{(0.022)}} & {1.519} & {1.563
\small{(0.021)}} & {1.408} \\
{8} & {1.916 \small{(0.033)}} & {1.733} & {1.838 \small{(0.032)}} &
{1.655} & {} & {1.984 \small{(0.030)}} & {1.840} & {1.804
\small{(0.028)}} & {1.659} \\
{9} & {2.308 \small{(0.049)}} & {2.145} & {2.195 \small{(0.047)}} &
{1.999} & {} & {2.240 \small{(0.041)}} & {2.115} & {2.060
\small{(0.040)}} & {1.921} \\
{10} & {2.759 \small{(0.082)}} & {2.613} & {2.824 \small{(0.086)}} &
{2.651} & {} & {2.575 \small{(0.067)}} & {2.474} & {2.436
\small{(0.066)}} & {2.315} \\
{11} & {2.955 \small{(0.102)}} & {2.839} & {3.008 \small{(0.106)}} &
{2.836} & {} & {2.715 \small{(0.081)}} & {2.648} & {2.581
\small{(0.081)}} & {2.467} \\
{12} & {3.232 \small{(0.136)}} & {3.210} & {3.352 \small{(0.145)}} &
{3.259} & {} & {2.787 \small{(0.102)}} & {2.707} & {2.765
\small{(0.107)}} & {2.655} \\
{13} & {3.231 \small{(0.152)}} & {3.193} & {3.441 \small{(0.166)}} &
{3.352} & {} & {2.877 \small{(0.118)}} & {2.808} & {2.782
\small{(0.120)}} & {2.642} \\
{14} & {3.358 \small{(0.183)}} & {3.390} & {3.551 \small{(0.198)}} &
{3.479} & {} & {2.920 \small{(0.138)}} & {2.841} & {2.903
\small{(0.144)}} & {2.740} \\
{15} & {3.281 \small{(0.219)}} & {3.306} & {3.728 \small{(0.254)}} &
{3.730} & {} & {2.989 \small{(0.173)}} & {2.871} & {3.059
\small{(0.186)}} & {2.870} \\
{16} & {3.327 \small{(0.313)}} & {3.472} & {3.990 \small{(0.384)}} &
{3.983} & {} & {3.226 \small{(0.263)}} & {3.114} & {3.690
\small{(0.318)}} & {3.668} \\
{17} & {3.372 \small{(0.366)}} & {3.636} & {3.917 \small{(0.435)}} &
{3.860} & {} & {3.427 \small{(0.322)}} & {3.351} & {3.518
\small{(0.350)}} & {3.397} \\
{18} & {3.136 \small{(0.415)}} & {3.326} & {4.251 \small{(0.578)}} &
{4.302} & {} & {3.818 \small{(0.441)}} & {3.989} & {4.168
\small{(0.506)}} & {4.395} \\
\hline
\hline
\end{tabular}
\end{center}
\caption{\label{table3} Maximum Likelihood and Anderson-Darling estimates of
the Pareto parameter $b$. Figures within parentheses give the
standard deviation of the Maximum Likelihood estimator.}
\end{table}

\clearpage

\begin{table}
\begin{center}
\begin{tabular}{rccccccccc}
\hline
\hline
      & \multicolumn{4}{c}{Nasdaq} &  & \multicolumn{4}{c}{Dow Jones} \\
\hline
      & \multicolumn{2}{c}{{Pos. Tail}} & \multicolumn{2}{c}{{Neg. Tail}}
& {} & \multicolumn{2}{c}{{Pos. Tail}} & \multicolumn{2}{c}{{Neg.
Tail}} \\
      & {MLE} & {ADE} & {MLE} & {ADE} & & {MLE} & {ADE} & {MLE} & {ADE} \\
\cline{2-5}
\cline{7-10}
{1} & {1.007 \small{(0.008)}} & {1.053} & {0.987 \small{(0.008)}} &
{1.017} & {} & {1.040 \small{(0.007)}} & {1.104} & {0.975
\small{(0.007)}} & {1.026} \\
{2} & {0.983 \small{(0.011)}} & {1.051} & {0.953 \small{(0.011)}} &
{0.993} & {} & {0.973 \small{(0.010)}} & {1.075} & {0.910
\small{(0.010)}} & {0.989} \\
{3} & {0.944 \small{(0.014)}} & {1.031} & {0.912 \small{(0.014)}} &
{0.955} & {} & {0.931 \small{(0.013)}} & {1.064} & {0.856
\small{(0.012)}} & {0.948} \\
{4} & {0.896 \small{(0.018)}} & {0.995} & {0.876 \small{(0.018)}} &
{0.916} & {} & {0.878 \small{(0.015)}} & {1.038} & {0.821
\small{(0.015)}} & {0.933} \\
{5} & {0.857 \small{(0.021)}} & {0.978} & {0.861 \small{(0.021)}} &
{0.912} & {} & {0.792 \small{(0.019)}} & {0.955} & {0.767
\small{(0.018)}} & {0.889} \\
{6} & {0.790 \small{(0.026)}} & {0.916} & {0.833 \small{(0.026)}} &
{0.891} & {} & {0.708 \small{(0.023)}} & {0.873} & {0.698
\small{(0.022)}} & {0.819} \\
{7} & {0.732 \small{(0.033)}} & {0.882} & {0.796 \small{(0.033)}} &
{0.859} & {} & {0.622 \small{(0.028)}} & {0.788} & {0.612
\small{(0.028)}} & {0.713} \\
{8} & {0.661 \small{(0.042)}} & {0.846} & {0.756 \small{(0.042)}} &
{0.834} & {} & {0.480 \small{(0.035)}} & {0.586} & {0.531
\small{(0.035)}} & {0.597} \\
{9} & {0.509 \small{(0.058)}} & {0.676} & {0.715 \small{(0.059)}} &
{0.865} & {} & {0.394 \small{(0.047)}} & {0.461} & {0.478
\small{(0.047)}} & {0.527} \\
{10} & {0.359 \small{(0.092)}} & {0.631} & {0.522 \small{(0.099)}} &
{0.688} & {} & {0.304 \small{(0.074)}} & {0.346} & {0.403
\small{(0.076)}} & {0.387} \\
{11} & {0.252 \small{(0.110)}} & {0.515} & {0.481 \small{(0.120)}} &
{0.697} & {} & {0.231 \small{(0.087)}} & {0.158} & {0.379
\small{(0.091)}} & {0.337} \\
{12} & {0.039 \small{(0.138)}} & {0.177} & {0.273 \small{(0.155)}} &
{0.275} & {} & {0.269 \small{(0.111)}} & {0.207} & {0.357
\small{(0.119)}} & {0.288} \\
{13} & {0.057 \small{(0.155)}} & {0.233} & {0.255 \small{(0.177)}} &
{0.274} & {} & {0.253 \small{(0.127)}} & {0.147} & {0.428
\small{(0.136)}} & {0.465} \\
{14} & {$< 10^{-8}$} & {0} & {0.215 \small{(0.209)}} & {0.194} & {} & {0.290
\small{(0.150)}} & {0.174} & {0.448 \small{(0.164)}} & {0.641} \\
{15} & {$< 10^{-8}$} & {0} & {0.103 \small{(0.260)}} & {0} & {} & {0.379
\small{(0.192)}} & {0.407} & {0.451 \small{(0.210)}} & {0.863} \\
{16} & {$9.6 \cdot 10^{-8}$} & {0} & {0.064 \small{(0.390)}} & {0} &
{} & {0.398
\small{(0.290)}} & {0.382} & {0.022 \small{(0.319)}} & {0.110} \\
{17} & {$< 10^{-8}$} & {0} & {0.158 \small{(0.452)}} & {0.224} & {} & {0.307
\small{(0.346)}} & {0.255} & {0.178 \small{(0.367)}} & {0.703} \\
{18} & {$< 10^{-8}$} & {0} & {$< 10^{-8}$} & {0} & {} & {$ 2 \cdot 10^{-8}$} &
{0} & {$< 10^{-8}$} & {0} \\ \hline
\hline
\end{tabular}
\end{center}
\caption{\label{table5} Maximum Likelihood and Anderson-Darling estimates of
the form parameter $c$ of the Weibull (Stretched-Exponential) distribution.}
\end{table}

\clearpage

\begin{table}
\begin{center}
\begin{tabular}{rccccccccc}
\hline
\hline
      & \multicolumn{4}{c}{Nasdaq} &  & \multicolumn{4}{c}{Dow Jones} \\
\hline
      & \multicolumn{2}{c}{{Pos. Tail}} & \multicolumn{2}{c}{{Neg. Tail}}
& {} & \multicolumn{2}{c}{{Pos. Tail}} & \multicolumn{2}{c}{{Neg.
Tail}} \\
      & {MLE} & {ADE} & {MLE} & {ADE} & & {MLE} & {ADE} & {MLE} & {ADE} \\
\cline{2-5}
\cline{7-10}
{1} & {0.443 \small{(0.004)}} & {0.441} & {0.455 \small{(0.005)}} &
{0.452} & {} & {7.137 \small{(0.060)}} & {7.107} & {7.268
\small{(0.068)}} & {7.127} \\
{2} & {0.429 \small{(0.006)}} & {0.440} & {0.436 \small{(0.006)}} &
{0.443} & {} & {6.639 \small{(0.082)}} & {6.894} & {6.726
\small{(0.094)}} & {6.952} \\
{3} & {0.406 \small{(0.008)}} & {0.432} & {0.410 \small{(0.009)}} &
{0.424} & {} & {6.236 \small{(0.113)}} & {6.841} & {6.108
\small{(0.131)}} & {6.640} \\
{4} & {0.372 \small{(0.011)}} & {0.414} & {0.383 \small{(0.012)}} &
{0.402} & {} & {5.621 \small{(0.155)}} & {6.655} & {5.656
\small{(0.175)}} & {6.515} \\
{5} & {0.341 \small{(0.015)}} & {0.404} & {0.369 \small{(0.016)}} &
{0.399} & {} & {4.515 \small{(0.215)}} & {5.942} & {4.876
\small{(0.235)}} & {6.066} \\
{6} & {0.283 \small{(0.020)}} & {0.364} & {0.345 \small{(0.021)}} &
{0.383} & {} & {3.358 \small{(0.277)}} & {5.081} & {3.801
\small{(0.305)}} & {5.220} \\
{7} & {0.231 \small{(0.026)}} & {0.339} & {0.309 \small{(0.028)}} &
{0.358} & {} & {2.192 \small{(0.326)}} & {4.073} & {2.475
\small{(0.366)}} & {3.764} \\
{8} & {0.166 \small{(0.034)}} & {0.311} & {0.269 \small{(0.039)}} &
{0.336} & {} & {0.682 \small{(0.256)}} & {1.606} & {1.385
\small{(0.389)}} & {2.149} \\
{9} & {0.053 \small{(0.030)}} & {0.164} & {0.225 \small{(0.057)}} &
{0.365} & {} & {0.195 \small{(0.163)}} & {0.510} & {0.810
\small{(0.417)}} & {1.297} \\
{10} & {0.005 \small{(0.010)}} & {0.128} & {0.058 \small{(0.057)}} &
{0.184} & {} & {0.019 \small{(0.048)}} & {0.065} & {0.276
\small{(0.361)}} & {0.207} \\
{11} & {0.000 \small{(0.001)}} & {0.049} & {0.036 \small{(0.053)}} &
{0.194} & {} & {0.001 \small{(0.003)}} & {0.000} & {0.169
\small{(0.316)}} & {0.065} \\
{12} & {0.000 \small{(0.000)}} & {0.000} & {0.000 \small{(0.001)}} &
{0.000} & {} & {0.005 \small{(0.025)}} & {0.000} & {0.103
\small{(0.291)}} & {0.012} \\
{13} & {0.000 \small{(0.000)}} & {0.000} & {0.000 \small{(0.001)}} &
{0.000} & {} & {0.001 \small{(0.010)}} & {0.000} & {0.427
\small{(0.912)}} & {0.729} \\
{14} & {0.000 \small{(0.000)}} & {-} & {0.000 \small{(0.000)}} &
{0.000} & {} & {0.009
\small{(0.055)}} & {0.000} & {0.577 \small{(1.357)}} & {3.509} \\
{15} & {0.000 \small{(0.000)}} & {-} & {0.000 \small{(0.000)}} & {-}
& {} & {0.149
\small{(0.629)}} & {0.282} & {0.613 \small{(1.855)}} & {9.640} \\
{16} & {0.000 \small{(0.000)}} & {-} & {0.000 \small{(0.000)}} & {-}
& {} & {0.145
\small{(0.960)}} & {0.179} & {0.000 \small{(0.000)}} & {0.000} \\
{17} & {0.000 \small{(0.000)}} & {-} & {0.000 \small{(0.000)}} &
{0.000} & {} & {0.007
\small{(0.109)}} & {0.002} & {0.000 \small{(0.000)}} & {5.528} \\
{18} & {0.000 \small{(0.000)}} & {-} & {0.000 \small{(0.000)}} & {-}
& {} & {0.000 \small{(0.000)}} & {-} & {0.000 \small{(0.000)}} & {-}
\\
\hline
\hline
\end{tabular}
\end{center}
\caption{\label{table6} Maximum Likelihood and Anderson-Darling estimates of
the form parameter $d ( \times 10^3)$ of the Weibull (Stretched-Exponential)
distribution.} \end{table}

\clearpage

\begin{table}
\begin{center}
\begin{tabular}{rccccccccc}
\hline
\hline
      & \multicolumn{4}{c}{Nasdaq} &  & \multicolumn{4}{c}{Dow Jones} \\
\hline
      & \multicolumn{2}{c}{{Pos. Tail}} & \multicolumn{2}{c}{{Neg. Tail}}
& {} & \multicolumn{2}{c}{{Pos. Tail}} & \multicolumn{2}{c}{{Neg.
Tail}} \\
      & {MLE} & {ADE} & {MLE} & {ADE} &  & {MLE} & {ADE} & {MLE} & {ADE} \\
\cline{2-5}
\cline{7-10}
{1} & {0.441 \small{(0.004)}} & {0.441} & {0.458 \small{(0.004)}} &
{0.451} & {} & {7.012 \small{(0.057)}} & {7.055} & {7.358
\small{(0.063)}} & {7.135} \\
{2} & {0.435 \small{(0.004)}} & {0.431} & {0.454 \small{(0.005)}} &
{0.444} & {} & {6.793 \small{(0.059)}} & {6.701} & {7.292
\small{(0.066)}} & {6.982} \\
{3} & {0.431 \small{(0.005)}} & {0.424} & {0.452 \small{(0.005)}} &
{0.438} & {} & {6.731 \small{(0.062)}} & {6.575} & {7.275
\small{(0.070)}} & {6.890} \\
{4} & {0.428 \small{(0.005)}} & {0.416} & {0.453 \small{(0.005)}} &
{0.437} & {} & {6.675 \small{(0.065)}} & {6.444} & {7.358
\small{(0.076)}} & {6.938} \\
{5} & {0.429 \small{(0.005)}} & {0.415} & {0.458 \small{(0.006)}} &
{0.443} & {} & {6.607 \small{(0.070)}} & {6.264} & {7.429
\small{(0.083)}} & {6.941} \\
{6} & {0.429 \small{(0.006)}} & {0.411} & {0.464 \small{(0.006)}} &
{0.447} & {} & {6.630 \small{(0.077)}} & {6.186} & {7.529
\small{(0.092)}} & {6.951} \\
{7} & {0.436 \small{(0.006)}} & {0.413} & {0.472 \small{(0.007)}} &
{0.453} & {} & {6.750 \small{(0.087)}} & {6.207} & {7.700
\small{(0.105)}} & {7.005} \\
{8} & {0.447 \small{(0.008)}} & {0.421} & {0.483 \small{(0.009)}} &
{0.463} & {} & {6.920 \small{(0.103)}} & {6.199} & {8.071
\small{(0.127)}} & {7.264} \\
{9} & {0.462 \small{(0.010)}} & {0.425} & {0.503 \small{(0.011)}} &
{0.482} & {} & {7.513 \small{(0.137)}} & {6.662} & {8.797
\small{(0.170)}} & {7.908} \\
{10} & {0.517 \small{(0.015)}} & {0.468} & {0.529 \small{(0.016)}} &
{0.496} & {} & {8.792 \small{(0.227)}} & {7.745} & {10.205
\small{(0.278)}} & {9.175} \\
{11} & {0.540 \small{(0.019)}} & {0.479} & {0.551 \small{(0.019)}} &
{0.514} & {} & {9.349 \small{(0.279)}} & {8.148} & {10.835
\small{(0.341)}} & {9.751} \\
{12} & {0.574 \small{(0.024)}} & {0.489} & {0.570 \small{(0.025)}} &
{0.516} & {} & {10.487 \small{(0.383)}} & {9.265} & {11.796
\small{(0.454)}} & {10.657} \\
{13} & {0.615 \small{(0.029)}} & {0.526} & {0.594 \small{(0.029)}} &
{0.537} & {} & {11.017 \small{(0.451)}} & {9.722} & {12.598
\small{(0.543)}} & {11.581} \\
{14} & {0.653 \small{(0.035)}} & {0.543} & {0.627 \small{(0.035)}} &
{0.564} & {} & {11.920 \small{(0.563)}} & {10.626} & {13.349
\small{(0.664)}} & {12.386} \\
{15} & {0.750 \small{(0.050)}} & {0.625} & {0.671 \small{(0.046)}} &
{0.594} & {} & {13.251 \small{(0.766)}} & {12.062} & {14.462
\small{(0.880)}} & {13.521} \\
{16} & {0.917 \small{(0.086)}} & {0.741} & {0.760 \small{(0.073)}} &
{0.674} & {} & {15.264 \small{(1.246)}} & {13.943} & {15.294
\small{(1.316)}} & {13.285} \\
{17} & {0.991 \small{(0.107)}} & {0.783} & {0.827 \small{(0.092)}} &
{0.744} & {} & {15.766 \small{(1.483)}} & {14.210} & {17.140
\small{(1.705)}} & {15.327} \\
{18} & {1.178 \small{(0.156)}} & {0.978} & {0.857 \small{(0.117)}} &
{0.742} & {} & {16.207 \small{(1.871)}} & {13.697} & {16.883
\small{(2.047)}} & {13.476} \\
\hline
\hline
\end{tabular}
\end{center}
\caption{\label{table4} Maximum Likelihood- and Anderson-Darling estimates of
the scale parameter $d=10^{ - 3}d'$ of the Exponential
distribution.Figures within parentheses give the
standard deviation of the Maximum Likelihood estimator. }
\end{table}

\clearpage

\begin{table}
\begin{center}
\begin{tabular}{rccccccccc}
\hline
\hline
      & \multicolumn{4}{c}{Nasdaq} &  & \multicolumn{4}{c}{Dow Jones} \\
\hline
      & \multicolumn{2}{c}{{Pos. Tail}} & \multicolumn{2}{c}{{Neg. Tail}}
& {} & \multicolumn{2}{c}{{Pos. Tail}} & \multicolumn{2}{c}{{Neg.
Tail}} \\
      & {MLE} & {ADE} & {MLE} & {ADE} &  & {MLE} & {ADE} & {MLE} & {ADE} \\
\cline{2-5}
\cline{7-10}
{1} & {-1.03} & {-1.09} & {-1.00} & {-1.03} & {} & {-1.12} & {-1.18}
& {-0.100} & {-1.05} \\
{2} & {-1.02} & {-1.13} & {-0.934} & {-1.01} & {} & {-1.01} & {-1.19}
& {-0.862} & {-1.01} \\
{3} & {-0.931} & {-1.13} & {-0.821} & {-0.955} & {} & {-0.921} &
{-1.23} & {-0.710} & {-0.943} \\
{4} & {-0.787} & {-1.09} & {-0.701} & {-0.887} & {} & {-0.766} &
{-1.24} & {-0.594} & {-0.944} \\
{5} & {-0.655} & {-1.12} & {-0.636} & {-0.914} & {} & {-0.458} &
{-1.09} & {-0.397} & {-0.870} \\
{6} & {-0.395} & {-1.01} & {-0.518} & {-0.911} & {} & {-0.119} &
{-0.929} & {-0.118} & {-0.715} \\
{7} & {-0.142} & {-1.03} & {-0.351} & {-0.906} & {} & {0.261} &
{-0.763} & {0.251} & {-0.462} \\
{8} & {0.206} & {-1.09} & {-0.149} & {-0.97} & {} & {0.881} &
{-0.202} & {0.619} & {-0.160} \\
{9} & {0.971} & {-0.754} & {0.101} & {-1.35} & {} & {1.31} & {0.127}
& {0.930} & {-0.018} \\
{10} & {1.83} & {-1.04} & {1.17} & {-1.33} & {} & {1.82} & {0.408} &
{1.40} & {0.435} \\
{11} & {2.34} & {-0.441} & {1.45} & {-1.53} & {} & {2.10} & {0.949} &
{1.59} & {0.420} \\
{12} & {3.12} & {-0.445} & {2.52} & {-0.435} & {} & {2.04} & {0.733}
& {1.78} & {0.403} \\
{13} & {3.10} & {-0.444} & {2.63} & {-0.402} & {} & {2.16} & {0.886}
& {1.57} & {-0.375} \\
{14} & {3.35} & {1.43} & {2.89} & {-0.419} & {} & {2.07} & {0.786} &
{1.58} & {-0.425} \\
{15} & {3.27} & {1.57} & {3.36} & {1.35} & {} & {1.82} & {-0.282} &
{1.64} & {-2.75} \\
{16} & {3.30} & {2.97} & {3.80} & {-0.411} & {} & {1.88} & {-0.129} &
{3.60} & {-0.428} \\
{17} & {3.34} & {3.19} & {3.46} & {-0.412} & {} & {2.35} & {-0.317} &
{3.19} & {-0.433} \\{18} & {2.74} & {2.90} & {4.22} & {-0.408} & {} &
{
3.73} & {3.27} & {4.11} & {0.374} \\
\hline
\hline
\end{tabular}
\end{center}
\caption{\label{tableXX} Maximum Likelihood- and Anderson-Darling estimates of
the form parameter $b$ of the Incomplete Gamma
distribution. }
\end{table}

\clearpage

\begin{table}
\begin{center}
\begin{tabular}{cccccccccc}
\hline
\hline
   & \multicolumn{4}{c}{Nasdaq Positive Tail} &  &
\multicolumn{4}{c}{Nasdaq Negative Tail} \\
\hline
   & \multicolumn{2}{c}{MLE} & \multicolumn{2}{c}{ADE}   &  &
\multicolumn{2}{c}{MLE} & \multicolumn{2}{c}{ADE} \\
   & c & b & c & b & & c & b & c & b \\
\cline{2-5} \cline{7-10}
1 & 3.835  {\small (0.006)} & 0.004  {\small (0.000)} & 4.310 & 0.002
&  & 3.872  {\small (0.006)} & 0.003  {\small (0.000)} & 4.220 &
0.002 \\
2 & 2.175  {\small (0.010)} & 0.217  {\small (0.002)} & 2.280 & 0.198
&  & 2.126  {\small (0.010)} & 0.219  {\small (0.002)} & 2.220 &
0.202 \\
3 & 1.797  {\small (0.012)} & 0.508  {\small (0.005)} & 1.860 & 0.493
&  & 1.753  {\small (0.012)} & 0.506  {\small (0.005)} & 1.790 &
0.495 \\
4 & 1.590  {\small (0.013)} & 0.812  {\small (0.009)} & 1.620 & 0.800
&  & 1.558  {\small (0.013)} & 0.785  {\small (0.009)} & 1.580 &
0.775 \\
5 & 1.479  {\small (0.014)} & 1.096  {\small (0.013)} & 1.500 & 1.092
&  & 1.472  {\small (0.014)} & 1.032  {\small (0.013)} & 1.480 &
1.030 \\
6 & 1.363  {\small (0.015)} & 1.412  {\small (0.019)} & 1.380 & 1.412
&  & 1.385  {\small (0.015)} & 1.312  {\small (0.018)} & 1.390 &
1.311 \\
7 & 1.301  {\small (0.015)} & 1.723  {\small (0.026)} & 1.310 & 1.724
&  & 1.310  {\small (0.016)} & 1.622  {\small (0.025)} & 1.310 &
1.623 \\
8 & 1.243  {\small (0.017)} & 2.065  {\small (0.036)} & 1.250 & 2.070
&  & 1.250  {\small (0.017)} & 1.968  {\small (0.035)} & 1.250 &
1.969 \\
9 & 1.152  {\small (0.018)} & 2.479  {\small (0.052)} & 1.160 & 2.488
&  & 1.228  {\small (0.020)} & 2.425  {\small (0.052)} & 1.230 &
2.427 \\
10 & 1.124  {\small (0.023)} & 2.981  {\small (0.089)} & 1.130 &
3.003 &  & 1.148  {\small (0.024)} & 3.113  {\small (0.095)} & 1.140
& 3.106 \\
11 & 1.090  {\small (0.025)} & 3.141  {\small (0.108)} & 1.100 &
3.175 &  & 1.148  {\small (0.027)} & 3.343  {\small (0.118)} & 1.150
& 3.344 \\
12 & 1.000  {\small (0.028)} & 3.226  {\small (0.136)} & 1.020 &
3.268 &  & 1.037  {\small (0.030)} & 3.448  {\small (0.149)} & 1.040
& 3.460 \\
13 & 1.042  {\small (0.033)} & 3.327  {\small (0.157)} & 1.050 &
3.356 &  & 1.051  {\small (0.033)} & 3.582  {\small (0.173)} & 1.050
& 3.584 \\
14 & 1.020  {\small (0.036)} & 3.401  {\small (0.185)} & 1.020 &
3.390 &  & 1.064  {\small (0.038)} & 3.738  {\small (0.208)} & 1.040
& 3.676 \\
15 & 1.037  {\small (0.046)} & 3.359  {\small (0.224)} & 1.020 &
3.333 &  & 0.967  {\small (0.043)} & 3.601  {\small (0.245)} & 0.941
& 3.521 \\
16 & 0.961  {\small (0.061)} & 3.202  {\small (0.301)} & 0.959 &
3.195 &  & 1.020  {\small (0.061)} & 4.030  {\small (0.388)} & 0.991
& 3.953 \\
17 & 0.888  {\small (0.067)} & 3.064  {\small (0.332)} & 0.861 &
3.003 &  & 1.015  {\small (0.071)} & 3.924  {\small (0.436)} & 1.010
& 3.937 \\
18 & 0.864  {\small (0.083)} & 2.807  {\small (0.372)} & 0.816 &
2.710 &  & 0.999  {\small (0.084)} & 4.168  {\small (0.567)} & 1.010
& 4.255 \\
\\
\hline
& \multicolumn{4}{c}{Dow Jones Positive Tail} &  &
\multicolumn{4}{c}{Dow jones Negative Tail} \\
\hline
& \multicolumn{2}{c}{MLE} & \multicolumn{2}{c}{ADE}   &  &
\multicolumn{2}{c}{MLE} & \multicolumn{2}{c}{ADE}   \\
& c & b & c & b & & c & b & c & b \\
\cline{2-5} \cline{7-10}
1 &5.262  {\small (0.005)} & 0.000  {\small (0.000)} & 5.55 & 0.000 &
& 5.085  {\small (0.005)} & 0.000  {\small (0.000)} & 5.320 & 0.000 \\
2 &2.140  {\small (0.009)} & 0.241  {\small (0.002)} & 2.25 & 0.220 &
& 2.125  {\small (0.009)} & 0.211  {\small (0.002)} & 2.240 & 0.191 \\
3 &1.790  {\small (0.010)} & 0.531  {\small (0.005)} & 1.87 & 0.510 &
& 1.751  {\small (0.010)} & 0.495  {\small (0.005)} & 1.800 & 0.481 \\
4 &1.616  {\small (0.012)} & 0.830  {\small (0.008)} & 1.65 & 0.820 &
& 1.593  {\small (0.012)} & 0.744  {\small (0.008)} & 1.630 & 0.735 \\
5 &1.447  {\small (0.012)} & 1.165  {\small (0.012)} & 1.47 & 1.160 &
& 1.459  {\small (0.013)} & 1.022  {\small (0.011)} & 1.480 & 1.015 \\
6 &1.339  {\small (0.012)} & 1.472  {\small (0.017)} & 1.36 & 1.473 &
& 1.353  {\small (0.013)} & 1.311  {\small (0.016)} & 1.370 & 1.311 \\
7 &1.259  {\small (0.013)} & 1.768  {\small (0.023)} & 1.28 & 1.773 &
& 1.269  {\small (0.014)} & 1.609  {\small (0.022)} & 1.270 & 1.610 \\
8 &1.173  {\small (0.013)} & 2.097  {\small (0.031)} & 1.17 & 2.096 &
& 1.188  {\small (0.015)} & 1.885  {\small (0.030)} & 1.190 & 1.887 \\
9 &1.125  {\small (0.015)} & 2.362  {\small (0.043)} & 1.12 & 2.358 &
& 1.158  {\small (0.017)} & 2.178  {\small (0.042)} & 1.150 & 2.174 \\
10 & 1.090  {\small (0.020)} & 2.705  {\small (0.070)} & 1.08 & 2.695
&  & 1.087  {\small (0.022)} & 2.545  {\small (0.069)} & 1.090 &
2.545 \\
11 & 1.035  {\small (0.022)} & 2.771  {\small (0.083)} & 1.03 & 2.762
&  & 1.074  {\small (0.024)} & 2.688  {\small (0.085)} & 1.070 &
2.681 \\
12 & 1.047  {\small (0.027)} & 2.867  {\small (0.105)} & 1.04 & 2.857
&  & 1.068  {\small (0.029)} & 2.880  {\small (0.111)} & 1.050 &
2.857 \\
13 & 1.046  {\small (0.030)} & 2.960  {\small (0.121)} & 1.03 & 2.933
&  & 1.067  {\small (0.032)} & 2.900  {\small (0.125)} & 1.080 &
2.924 \\
14 & 1.044  {\small (0.034)} & 3.000  {\small (0.142)} & 1.03 & 2.976
&  & 1.132  {\small (0.038)} & 3.171  {\small (0.158)} & 1.120 &
3.155 \\
15 & 1.090  {\small (0.043)} & 3.174  {\small (0.184)} & 1.09 & 3.165
&  & 1.163  {\small (0.047)} & 3.439  {\small (0.209)} & 1.180 &
3.472 \\
16 & 1.085  {\small (0.059)} & 3.424  {\small (0.280)} & 1.09 & 3.425
&  & 1.025  {\small (0.056)} & 3.745  {\small (0.322)} & 1.010 &
3.731 \\
17 & 1.093  {\small (0.066)} & 3.666  {\small (0.345)} & 1.09 & 3.650
&  & 1.108  {\small (0.069)} & 3.822  {\small (0.380)} & 1.120 &
3.891 \\
18 & 0.935  {\small (0.071)} & 3.556  {\small (0.411)} & 0.902 &
3.484 &  & 0.921  {\small (0.071)} & 3.804  {\small (0.461)} & 0.933
& 3.846 \\

\hline
\hline
\end{tabular}
\end{center}
\caption{\label{tableSL} Maximum Likelihood- and Anderson-Darling estimates of
the parameters $b$ and $c$ of the log-Weibull distribution.  Numbers in
parenthesis give the standard deviations of the estimates.}
\end{table}

\clearpage

\begin{table}
\begin{center}
\begin{tabular}{rrlrlrrlrl}
\hline
\hline
    & \multicolumn{4}{c}{Nasdaq} &  & \multicolumn{4}{c}{Dow Jones} \\
    \hline
    & \multicolumn{2}{c}{Pos. Tail} & \multicolumn{2}{c}{Neg. Tail} &
     & \multicolumn{2}{c}{Pos. Tail} & \multicolumn{2}{c}{Neg. Tail} \\
     \cline{2-5}
     \cline{7-10}
1 & 19335  & (100\%)          & 18201  & (100\%)          &  & 28910
& (100\%)          & 24749  & (100\%)          \\
2 & 7378    & (100\%) & 6815    & (100\%) &  & 9336    & (100\%) &
8377    & (100\%) \\
3 & 4162    & (100\%) & 3795    & (100\%) &  & 5356    & (100\%) &
4536    & (100\%) \\
4 & 2461    & (100\%) & 2311    & (100\%) &  & 3172    & (100\%) &
2832    & (100\%) \\
5 & 1532    & (100\%) & 1520    & (100\%) &  & 1734    & (100\%) &
1681    & (100\%) \\
6 & 853      & (100\%) & 933      & (100\%) &  & 930      & (100\%) &
933      & (100\%) \\
7 & 491      & (100\%)  & 555      & (100\%)  &  & 483      & (100\%)
& 466      & (100\%)  \\
8 & 248      & (100\%) & 301      & (100\%) &  & 177      & (100\%) &
218      & (100\%) \\
9 & 78.6     & (100\%) & 141     & (100\%) &  & 68.0     & (100\%) &
98.0     & (100\%) \\
10 & 16.1  & (99.99\%) & 28       & (100\%) &  & 16    & (99.99\%) &
27        & (100\% ) \\
11 & 5.70    & (98.3\%) & 16       & (98.6\%) &  & 6.69    & (99.0\%)
& 16     & (99.99\%) \\
12 & .102    & (24.8\%) & 3.03    & (91.7\%) &  & 5.71    & (98.3\%)
& 9.0       & (99.7\%) \\
13 & .141    & (30.1\%) & 2.17    & (86.2\%) &  & 3.70    & (94.5\%)
& 9.9       & (99.8\%) \\
14 & 9e-6    & (7e-3\%) & 1.04    & (68.3\%) &  & 3.48    & (93.2\%)
& 7.9       & (99.5\%) \\
15 & 5e-6    & (3e-3\%) & .149    & (30.1\%) &  & 3.73    & (94.5\%)
& 5.4       & (97.8\%) \\
16 & 2e-7    & (1e-3\%) & .028    & (13.8\%) &  & 1.77    & (82.5\%)
& .007& (6.00\%) \\
17 & 2e-6    & (1e-2\%) & .127    & (27.5\%) &  & .729    & (41.1\%)
& .30       & (41.6\%) \\
18 & 3e-7    & (2e-3\%) & 7e-7    & (4e-3\%) &  & 1e-6 & (1e-3\%) &
2e-6    & (1e-2\%) \\
\hline
\hline
\end{tabular}

\caption{\label{table:ParSE} Wilks' test for the Pareto distribution versus the
Stretched-Exponentail distribution. The p-value (figures within
parentheses) gives the significance with which one can reject the null
hypothesis that the Pareto distribution is sufficient to accurately
describe the
data.} \end{center}
\end{table}

\clearpage

\begin{table}
\begin{center}
\begin{tabular}{rrlrlrrlrl}
\hline
\hline
    & \multicolumn{4}{c}{Nasdaq} &  & \multicolumn{4}{c}{Dow Jones} \\
    \hline
    & \multicolumn{2}{c}{Pos. Tail} & \multicolumn{2}{c}{Neg. Tail} &
     & \multicolumn{2}{c}{Pos. Tail} & \multicolumn{2}{c}{Neg. Tail} \\
     \cline{2-5}
     \cline{7-10}
1 & 17235   & (100\%)  & 16689   & (100\%) &  & 27632  & (100\%)   &
23670   & (100\%) \\
2 & 6426    & (100\%)  & 5834    & (100\%) &  & 8262    & (100\%)  &
7313    & (100\%) \\
3 & 3533    & (100\%)  & 3134    & (100\%) &  & 4680    & (100\%)  &
3933    & (100\%) \\
4 & 2051    & (100\%)  & 1795    & (100\%) &  & 2959    & (100\%)  &
2497    & (100\%) \\
5 & 1308    & (100\%)  & 1209    & (100\%) &  & 1587    & (100\%)  &
1482    & (100\%) \\
6 & 698     & (100\%)  & 730     & (100\%) &  & 853      & (100\%) &
817     & (100\%) \\
7 & 426     & (100\%)  & 421     & (100\%) &  & 442      & (100\%) &
414     & (100\%)  \\
8 & 226     & (100\%)  & 222     & (100\%) &  & 164      & (100\%) &
172     & (100\%) \\
9 & 57.9    & (100\%)  & 127     & (100\%) &  & 62.4     & (100\%) &
84.9    & (100\%) \\
10 & 22.9   & (99.99\%)& 30.5    & (100\%) &  & 15.8    & (100\%)  &
14.0      & (99.99\% ) \\
11 & 9.77   & (99.8\%) & 22.9    & (100\%) &  & 2.09    & (85.2\%) &
6.91      & (99.15\%) \\
12 & 0.008  & (7.1\%)  & 0.506   & (52.3\%) & & 2.48    & (88.5\%) &
4.35     & (96.3\%) \\
13 & 0.675  & (58.9\%) & 1.56    & (78.8\%) & & 2.05    & (84.8\%) &
2.40     & (87.9\%) \\
14 & 0.185  & (33.3\%) & 0.892   & (65.5\%) & & 1.25    & (73.7\%) &
7.88     & (99.5\%) \\
15 & 0.073  & (21.3\%) & 0.599   & (56.1\%) & & 2.89    & (91.1\%) &
9.12     & (99.75\%) \\
16 & 0.308  & (42.2\%) & 0.103   & (25.2\%) & & 1.53    & (78.4\%) &
0.00062    & (2\%) \\
17 & 2.21   & (86.3\%) & 0.309   & (42.2\%) & & 1.14    & (71.5\%) &
0.909    & (66.0\%) \\
18 & 2.17   & (85.9\%) & 0.032   & (14.2\%) & & 1.03    & (69.0\%) &
0.848    & (64.3\%) \\
\hline
\hline
\end{tabular}
\caption{\label{table:ParSWeib} Wilks' test for the Pareto
distribution versus the
log-Weibull distribution. The p-value (figures within
parentheses) gives the significance with which one can reject the null
hypothesis that the Pareto distribution is sufficient to accurately
describe the
data.} \end{center}
\end{table}

\clearpage

\begin{table}
\begin{center}
\begin{tabular}{lcccc}
\hline
\hline
\hspace{2cm} Sample&& $c$ & $d$ &  $c (u_{12}/d)^c$ \\
\hline
ND positive returns  && 0.039 {\small(0.138)} & 4.54 $\cdot$
10$^{-52}$ {\small(2.17$\cdot$ 10$^{-49}$)} & $3.03$ \\
ND negative returns && 0.273  {\small(0.155)}& 1.90 $\cdot$ 10$^{-7}$
{\small(1.38$\cdot$ 10$^{-6}$)}  & $3.10$\\
DJ positive returns && 0.274  {\small(0.111)}&  4.81 $\cdot$
10$^{-6}$ {\small(2.49$\cdot$ 10$^{-5}$)}  & $2.68$\\
DJ negative returns && 0.362  {\small(0.119)}&1.02 $\cdot$ 10$^{-4}$
{\small(2.87$\cdot$ 10$^{-4}$)}  & $2.57$\\ \hline
\hline
\end{tabular}
\end{center}
\caption{\label{sumse} Best parameters $c$ and $d$ of the Stretched Exponential
model estimated up to quantile $q_{12}=95\%$. The apparent Pareto exponent
$c (u_{12}/d)^c$ (see expression (\ref{jgjlke})) is also shown.
$u_{12}$ are the lower thresholds corresponding to the significance
levels $q_{12}$
given in table \ref{table1}. }
\end{table}

\clearpage

\clearpage

\clearpage

\clearpage

\clearpage

\clearpage

\clearpage

\clearpage

\clearpage

\clearpage

\end{document}